\documentclass[twocolumn, trackchanges]{aastex631}
\usepackage{amsmath} 
\usepackage{booktabs}  
\usepackage{tabularx}  
\usepackage{graphicx}  
\usepackage{rotating}
\usepackage{xcolor}
\usepackage{verbatim}

\newcommand{\lcdm}{$\Lambda$CDM}

\begin{document}

\title{Searching for Dark Structures: A Comparison of Weak-lensing Convergence Maps and Lensing-weighted Galaxy Density Maps}

\correspondingauthor{Ho Seong Hwang}
\email{hhwang@astro.snu.ac.kr}

\author[0009-0008-3074-9473]{Soojin Kim}
\affiliation{Astronomy Program, Department of Physics and Astronomy, Seoul National University, 1 Gwanak-ro, Gwanak-gu, Seoul 08826, Republic of Korea}

\author[0000-0003-3428-7612]{Ho Seong Hwang}
\affiliation{Astronomy Program, Department of Physics and Astronomy, Seoul National University, 1 Gwanak-ro, Gwanak-gu, Seoul 08826, Republic of Korea}
\affiliation{SNU Astronomy Research Center, Seoul National University, 1 Gwanak-ro, Gwanak-gu, Seoul 08826, Republic of Korea}
\affiliation{Australian Astronomical Optics - Macquarie University, 105 Delhi Road, North Ryde, NSW 2113, Australia}

\author[0000-0003-2927-1800]{Niall Jeffrey}
\affiliation{Department of Physics and Astronomy, University College London, Gower Street, London WC1E 6BT, UK}



\begin{abstract}
We present the result of a comparison between the dark matter distribution inferred from weak gravitational lensing and the observed galaxy distribution to identify dark structures with a high dark matter-to-galaxy density ratio. To do this, we use weak-lensing convergence maps from the Dark Energy Survey Year 3 data, and construct corresponding galaxy convergence maps at z$\lesssim$1.0, representing projected galaxy number density fluctuations weighted by lensing efficiency. The two maps show overall agreement. However, we could identify 22 regions where the dark matter density exhibits an excess compared to the galaxy density. After carefully examining the survey depths and proximity to survey boundaries, we select the seven most probable candidates for dark structures. This sample provides valuable test beds for further investigations into dark matter mapping. Moreover, our method will be very useful for future studies of dark structures as large-scale weak-lensing surveys become available, such as the Euclid mission, the Vera C. Rubin Observatory’s Legacy Survey of Space and Time, and the Nancy Grace Roman Space Telescope.

\end{abstract}



\section{Introduction} \label{sec:intro}
Dark matter constitutes approximately 84\% of the total matter in the Universe, and forms dark matter halos hosting galaxies in them \citep{DMreview12, DMreview24}. As the primary driver of gravity, dark matter creates deep gravitational potential wells, within which baryons accumulate and form stars, shining as galaxies \citep{White&Rees78}. In this way, the Lambda cold dark matter (\lcdm) cosmological model predicts the hierarchical clustering of matter \citep{DMreview24}. Numerous cosmological simulations based on this framework successfully reproduce the observed large-scale structures in the Universe \citep{TNGclustering18, FLAMINGO23}. These simulations also indicate that the spatial distribution of dark matter is well correlated with that of galaxies. However, directly confirming this correlation in the real Universe remains challenging, as dark matter is not detectable by light. Various methods have been developed to probe dark matter \citep{DMreview24}, with gravitational lensing being the most direct approach, as it measures the distortion of spacetime caused by mass without a hydrostatic equilibrium assumption \citep{Bartelmann&Schneider, WL&cosmologyreview08,  Mandelbaum18}.

To study dark matter itself, it is good to probe dark matter without any contamination by visible matter including galaxies. However, galaxy distribution is closely related to the distribution of dark matter, which is known as galaxy bias \citep{Kaiser84, LSgalaxybias18}. Studies on galaxy bias have addressed the systematic differences between the spatial distribution of galaxies and the underlying distribution of total matter. \citet{Pujol16bias} and \citet{Chang16bias} have developed statistical methods to examine the redshift evolution of galaxy bias. Local deviations in galaxy bias have been investigated in the context of dark structures (e.g. dark galaxies, dark cores). A dark galaxy is an object composed primarily of dark matter with little or no visible stellar content, predicted by most cosmological simulations \citep{Jimenez97, Benitez-Llambay17, Jimenez&Heavens20, Gain24}. Observational efforts to identify dark galaxies have focused on detecting HI clouds without optical counterparts \citep{Davies06, Cannon15, Kwon25}. A search for such structures larger than galaxies has been conducted focusing on the scales of galaxy clusters (e.g. \citealt{Geller14, Hwang14, Liu18, Shin22, Kang25}). In particular, \citet{Jee12} used weak gravitational lensing (WL) and claimed the detection of a dark matter-dominated structure in the galaxy cluster A520. By comparing a weak-lensing convergence map with the spatial distribution of cluster member galaxies, they discovered significant concentrations of dark matter associated with a small number of galaxies. They referred to this peculiar structure as the dark core. However, \citet{Clowe12} challenged the finding through an independent weak-lensing analysis. Later, \citet{Jee14} reaffirmed its presence using improved imaging data. There are even larger dark structures dominated by dark matter. For example, the Great Attractor is the region where local velocity flows converge, which is considered to be the region dominated by dark matter \citep{Dressler87, Courtois13}. Unfortunately, it lies in the direction of the Zone of Avoidance, making it difficult to study the nature of its structure; a recent study by \citet{Dupuy25} has tackled this issue using the machine learning technique developed by \cite{Hong21}. In this study, we continue such searches for dark structures that are dominated by dark matter through the comparison between the weak-lensing maps and the galaxy distribution maps.

To do this, we devise a novel method to compare the dark matter distribution reconstructed from WL with the observed galaxy distribution for a study of their similarities and discrepancies. We identify regions with abundant dark matter but relatively few galaxies. We refer to these regions as candidates for “dark structures” and focus on their detection. A dark structure is defined as a concentration of matter spanning a few Mpc predominantly composed of dark matter with a relatively low galaxy content. Such structures are not readily expected from the standard \lcdm$\ $paradigm, where galaxies are good tracers of dark matter in large scales. This means that the identification of such structures can provide new insights into unresolved questions in cosmology, such as the nature of dark matter and dark energy \citep{DMDEreview}, the discrepancy in Hubble constant estimates between the cosmic microwave background (CMB) and distance ladder methods \citep{HubbleTensionreview}, and the mechanisms governing structure formation across different scales \citep{Bullock17}. Specifically, by investigating why galaxies fail to form sufficiently in dark structures and what conditions might give rise to these structures, we can gain new understanding about the nature of dark matter and its interaction with baryons.

For this analysis, we use a weak-lensing convergence map derived from the Dark Energy Survey Year 3 (DES Y3) data \citep{DESWLmap}, covering an area of approximately 4000 deg$^2$. WL occurs when light from background galaxies passes through an inhomogeneous foreground matter distribution, which is observed by the distorted shapes of background galaxies. We use the WL convergence map as a proxy for the true dark matter distribution to examine its correlation with the galaxy distribution. The WL convergence map represents the two-dimensional projected surface mass density along the line of sight between the observer and background galaxies. The efficiency of the lensing effect depends on the relative distances between the observer, the background galaxies, and the intervening lensing mass. Thus, the contribution to the WL convergence map varies as a function of the redshift of the lensing mass. This dependency is characterized by the lensing weight function, which we apply to the observed galaxy distribution to construct the “lensing-weighted galaxy density” maps. Lensing-weighted galaxy density maps represent the expected convergence inferred from galaxies alone \citep{Vikram15}. In this work, we introduce a new methodology to identify dark structure candidates, with the following objectives: (1) to find the overall correlation between dark matter and galaxies by comparing the two maps, (2) to construct a catalog of dark structure candidates, including their precise locations, sizes, and reliability estimates, and (3) to conduct further analyses of these candidate regions to assess their reliability. 

For the full analysis, we adopt the cosmological parameter $h=0.7,\ \Omega_m=0.279$ and a flat-\lcdm$\ $assumption which is consistent with Wilkinson Microwave Anisotropy Probe 9 yr result \citep{WMAP9yr} and used for the simulation test in \cite{DESWLmap}. In Section \ref{sec:theory}, we briefly summarize the theory of our study. In Section \ref{sec:Data}, we introduce the DES Y3 data products. In Section \ref{sec:method}, we explain the detailed steps to construct lensing-weighted galaxy density maps. In Section \ref{sec:result}, we present our main results. We discuss our results in Section \ref{sec:dis}, and conclude in Section \ref{sec:summary}.

\section{Theoretical Background} \label{sec:theory}
\subsection{Weak Gravitational Lensing Convergence Map}
A comprehensive review of WL can be found in many papers including \cite{Bartelmann&Schneider} and \cite{Mandelbaum18}. Here we briefly describe some basics that are necessary for our work. 

The fluctuation of the mass distribution in the Universe is described by $\delta=(\rho-\bar{\rho})/\bar{\rho}$, where $\rho$ is the matter density, $\bar{\rho}$ is the mean matter density, and $\delta$ represents the underdense and overdense regions in our Universe. Using the Limber approximation, the gravitational potential $\Phi$ is related to the matter fluctuation $\delta$ by the Poisson equation,
\begin{equation}
    \nabla^2\Phi = \frac{3H_0^2\Omega_m}{2a}\delta
\end{equation}
where $H_0$ is the present value of the Hubble parameter, $a$ is the scale factor, and $\Omega_m$ is the matter density parameter. To use the analogy of the Poisson equation in two-dimensional projected space, the lensing potential ($\Psi$) is defined as follows:

\begin{equation}
    \Psi(\vec{\theta}, r) = 2\int_{0}^{r} \frac{r-r'}{rr'}\Phi(\vec{\theta}, r')\, dr' 
\end{equation}
where $r$ is the comoving distance to the source, $\vec{\theta}$ is the two-dimensional angular coordinate of the source, and $r'$ is the comoving distance to the lens. $\Psi$ is the weighted integral of the gravitational potential along the line of sight from the source to the observer. The weight is determined by the relative distance between the source, lens, and observer. 

The gravitational lensing effect is the mapping between the source plane and the lens plane. The coordinate in the source plane is converted to the coordinate in the lens plane; then the source shape appears as a distorted image in the lens plane. The two-dimensional coordinate in the source plane ($\vec{\beta}$) is mapped to the two-dimensional coordinate in the lens plane ($\vec{\theta}$) by the lens equation \citep{Bartelmann&Schneider}:
\begin{equation}
    \vec{\beta}-\vec{\beta_0} = A(\vec{\theta})(\vec{\theta}-\vec{\theta_0})
\end{equation}
where $A$ is the Jacobian matrix containing the amplitude and direction of the gravitational lensing. $A$ is written in terms of two weak-lensing parameters, reduced shear ($g$) and convergence ($\kappa$):
\begin{equation}
    A_{ij}(\vec{\theta}, r) = \delta_{ij} - \frac{\partial^2 \Psi}{\partial x_i \partial x_j} = (1-\kappa)\begin{pmatrix}
        1-g_1 & -g_2 \\
        -g_2 & 1+g_1
    \end{pmatrix}
\end{equation}
The reduced shear, $g_i=\gamma_i/(1-\kappa)$, is the combination of shear ($\gamma$) and convergence ($\kappa$). The $(1-\kappa)$ term describes the magnification, the difference in image size and brightness. The matrix part describes the direction of the distortion in the image. In WL, we can assume $\kappa \ll 1$ so that the reduced shear is considered as shear, $g\approx\gamma$. According to the above formula, convergence and shear are the second-order derivatives of the lensing potential. The convergence can be reconstructed by Fourier transform of the shear field \citep{KS93}.
\begin{equation}
    \kappa=\frac{1}{2}\nabla^2\Psi
\end{equation}
\begin{equation}
    \gamma=\gamma_1+i\gamma_2=\frac{1}{2}(\Psi_{,11}-\Psi_{,22})+i\Psi_{,12}
\end{equation}
The convergence is the dimensionless surface mass density. If there are multiple sources and lenses along the line of sight, the convergence is written as follows:
\begin{equation}
    \kappa(\vec{\theta}, p_s) = \int_0^{\infty} q(r', p_s) \delta(\vec{\theta}, r')\, dr' 
\end{equation}
\begin{equation}
    q(r', p_s) = \frac{3H_0^2 \Omega_m r'}{2c^2a(r')}\int_{r'}^{\infty} \frac{r-r'}{r}p_s(r)\, dr
\end{equation}
where $\kappa(\vec{\theta}, p_s)$ is the convergence at a fixed angular position $\vec{\theta}$ with respect to a given source redshift probability density $p_s$ and $q(r', p_s)$ is the lensing weight or lensing efficiency for a fixed lens plane $r'$ and source distribution $p_s$. The convergence is the weighted integral of the density fluctuation along the line of sight.

\subsection{Lensing-Weighted Galaxy Density Map}
To compare the matter distribution inferred from the weak-lensing convergence maps with the galaxy distribution, we need to make the galaxy distribution maps comparable to the weak-lensing convergence maps. To do this, we construct lensing-weighted galaxy density maps as a `predicted convergence' from the foreground galaxies. We adopt a parameter $\kappa_g$ introduced in \cite{Vikram15} as the lensing-weighted galaxy density, and follow their method to compute it. The $\kappa_g$ is an analog of $\kappa$ but with the galaxy density fluctuation $\delta_g$ in the three-dimensional spatial pixel, 
\begin{equation}
    \kappa_g(\vec{\theta}, p_s) = \int_0^{\infty} q(r', p_s) \delta_g(\vec{\theta}, r')\, dr' 
\end{equation}
In this work, we call $\kappa_g$ galaxy convergence. It is the weighted integral of the galaxy density fluctuation $\delta_g$ along the line of sight. The radial selection function $\phi$ is introduced to account for the "partial galaxy convergence," which is the line-of-sight slice of the galaxy convergence between the lens distance of $r'_{min} < r' < r'_{max}$.
\begin{equation}
    \phi(r') = 
    \begin{cases}
    1 & r'_{min} < r' < r'_{max} \\
    0 & \text{otherwise}
    \end{cases}
\end{equation}
For \textbf{N} lens redshift bins, the galaxy convergence ($\kappa_g$) is the sum of the \textbf{N} partial galaxy convergences ($\kappa_g'$) derived for each redshift bin:
\begin{equation}
    \kappa_g = \sum_{i=0}^N \kappa_g' = \sum_{i=0}^N \int_0^{\infty} q(r', p_s) \delta_g(\vec{\theta}, r')\phi(r')\, dr' 
    \label{equ:kappa_g}
\end{equation}
With some steps to simplify the above formula and the assumption of that the lensing weight is constant for the small width of the lens redshift bin, the partial galaxy convergence can be written as follows:
\begin{equation}
    \kappa_g'(\vec{\theta}, \phi, p_s) = q(r'_{m}, p_s)(r'_{max}-r'_{min})\delta^{2D}_g(\vec{\theta})
    \label{eq:partialk}
\end{equation}
where $r'_{m}$ is the midpoint between $r'_{min}$ and $r'_{max}$,  $q(r'_{m}, p_s)$ is the mean lensing weight defined as the lensing weight at $r'_{m}$, and $\delta^{2D}_g$ is the two-dimensional projected galaxy matter density fluctuation at the given lens redshift bin. For \textbf{M} source redshift bins, the comoving distance to the \textbf{i}-th source bin is given as $r_i$. To express the lensing weight term as a discrete sum rather than an integral, 
\begin{equation}\label{equ:q_lensweight}
    q(r'_{m}, p_s) = \frac{3H_0^2\Omega_m r'_{m}}{2c^2a(r'_{m})} \sum\limits_{\substack{r_i \geq r'_{m}}}^{M} p_s(r_i)\frac{r_i-r'_{m}}{r_i}
\end{equation}
The maps of galaxy convergence correspond to the "lensing-weighted galaxy density maps." They are used for comparison with weak-lensing convergence maps. 

\section{Data} \label{sec:Data}

\begin{figure*}[htb!]
    \centering
    \includegraphics[width=1\linewidth, trim=0 70 0 0, clip]{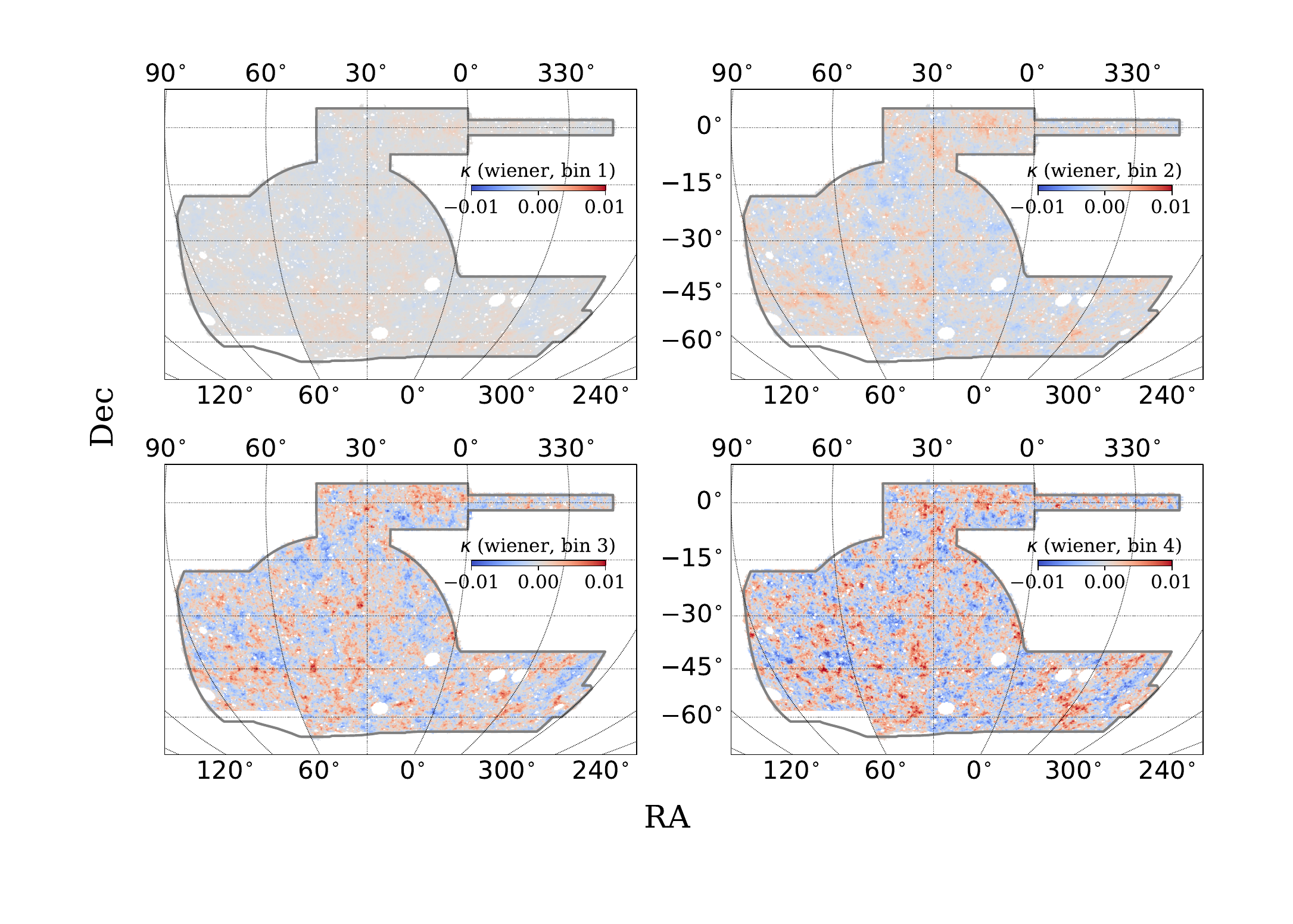} 
    \caption{The Wiener weak-lensing convergence maps from \cite{DESWLmap} before smoothing are shown for each tomographic bin. Gray boundaries show the DES Y3 survey footprint.}
    \label{fig:kappamaps_fig1}
\end{figure*}

\begin{figure*}[htb!]
    \centering
    \includegraphics[width=1\linewidth]{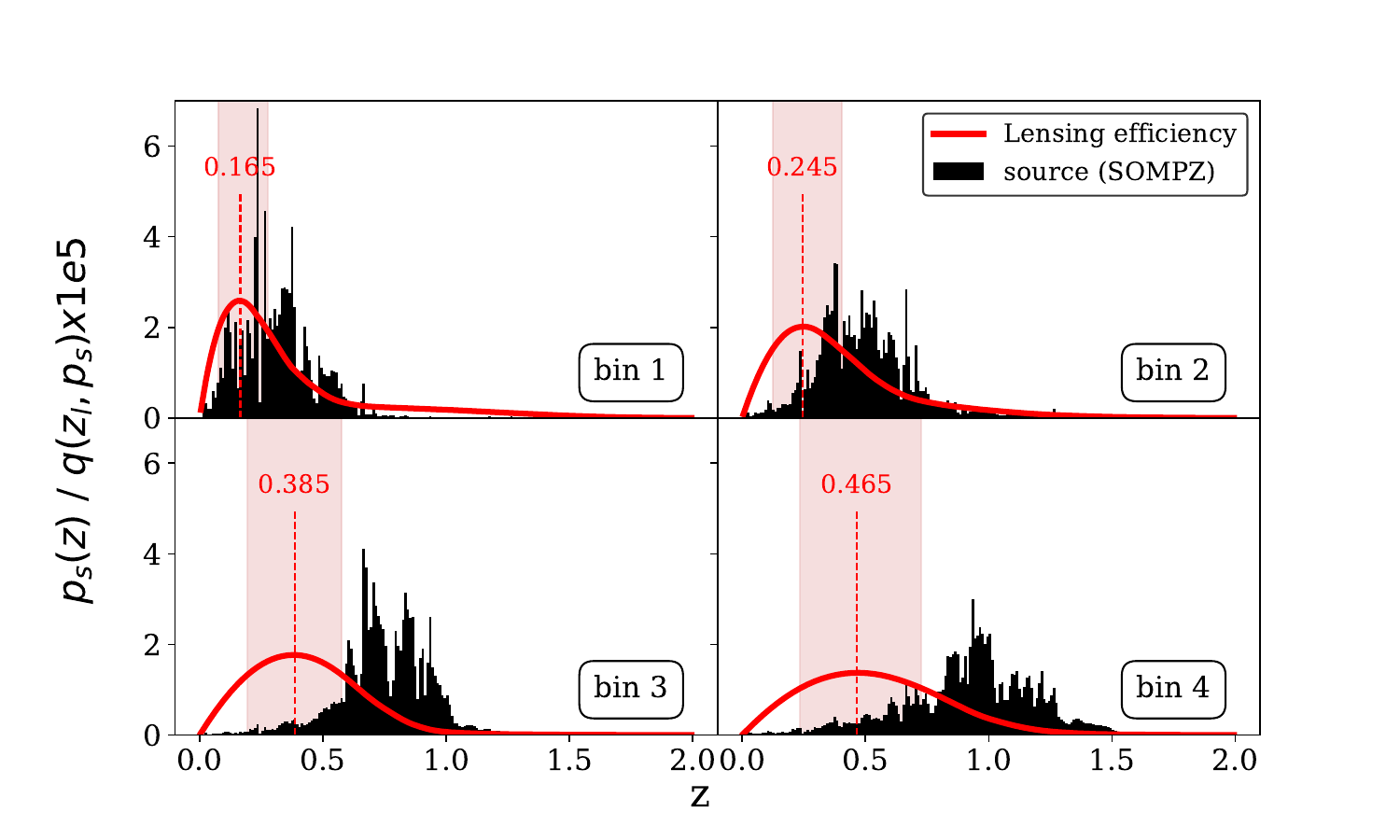}
    \caption{For each tomographic bin, the black histograms are the source redshift distributions $p_s(z)$ from \cite{SOMPZ} and the red solid lines are the lensing efficiency curves $q(z_l, p_s)$.  The histograms are normalized so that the total area under each curve is equal to 1. The red vertical dashed lines represent the peak of the lensing efficiency curve. The red shaded areas indicate the interval of lens redshifts where the lensing efficiency is larger than 75\% of its peak value.}
    \label{fig:lensweight_fig2}
\end{figure*}

\subsection{DES Y3 Gold Catalog}

The DES \citep{DESoverview} is a six-year photometric survey of the southern part of the sky, covering the 5000 deg$^2$. The DES uses the Dark Energy Camera (DECam) mounted on the 4 m Blanco telescope of the Cerro Tololo Inter-American Observatory (CTIO) in Chile. DECam is a 570 megapixel CCD camera that images the sky in the \textit{grizY} filters. In this work, we use the data products of the first three years (DES Y3). In particular, the photometric catalog of \cite{DESgold}, called the DES Y3 Gold catalog, is used as a complete dataset of galaxies. The DES Y3 Gold catalog includes 390 million objects, with signal-to-noise ratios $S/N \gtrsim10$ for extended sources with $i_{AB} \lesssim 23.0$. The catalog provides position, magnitude, morphological star-galaxy classification flag, some photometric properties and quality flags, and photometric redshifts. The single-object and multiobject fit pipelines are applied independently; we use only the parameters derived from single-object fits (SOFs). The photometric redshifts are obtained by three different methods, Bayesian photometric redshifts \citep{BPZ}, Directional Neighborhood Fitting (DNF) photometric redshifts \citep{DNF}, and machine learning for photometric redshifts (ANNz2; \citealt{ANNz2}). \cite{DESgold} performed an accuracy test among these methods, and found that DNF provides the best performance through the comparison with the spectroscopic redshifts. In this work, we use DNF photometric redshifts to produce galaxy convergence maps. For a detailed description of the photometry and photometric redshift estimates, see \cite{DESgold}.

\subsection{DES Y3 Weak-Lensing Convergence Map}

\cite{DESWLmap} presented weak-lensing convergence maps reconstructed from the DES Y3 shear catalog \citep{DESshear}. They used four reconstruction methods with different priors in a maximum a posteriori estimation: a uniform prior \citep{KS93}, a null B-mode prior, a Gaussian random-field prior (Wiener filtering), and a sparsity prior (GLIMPSE; \citealt{GLIMPSE}). The performance of each prior was evaluated in \cite{DESWLmap} using a realistic mock galaxy catalog; the Gaussian and sparsity priors were found to perform best in the presence of observational masks and shape noise. 

The Gaussian random-field prior assumes a specific E-mode power spectrum while setting the B-mode power to zero. Under this prior, the maximum a posteriori estimate corresponds to a Wiener filter reconstruction \citep{Wiener}. This prior effectively suppresses the noise and is particularly suitable for recovering large-scale structures, which are less affected by nonlinear structure formation. 

The sparsity prior is based on a wavelet transform approach that incorporates both a `halo' prior and a zero B-mode prior. The structure formation of the Universe expects the late-time matter distribution to be dominated by quasi-spherical halos of bound matter. The starlet basis \citep{starlet}, coefficients of isotropic undecimated wavelets in two dimensions, has been shown to effectively reproduce the observed convergence. The GLIMPSE algorithm solves the optimization problem of the maximum a posteriori estimate with the sparsity prior. It represents a physical model in which the matter distribution is a sum of spherically symmetric dark matter halos. The sparsity prior suppresses the noise and enhances the matter concentration peaks. Thus the convergence maps contain much more localized clumps than Gaussian prior-based convergence maps. From now on, the convergence map reconstructed from the Gaussian prior is called the Wiener map, and the map from the sparsity prior is called the GLIMPSE map.

In this work, we use the Wiener and GLIMPSE convergence maps independently to find similarities and differences with the galaxy convergence maps. The four convergence maps from each reconstruction method are shown in Figure 10 of \cite{DESWLmap}.
Both publicly released weak-lensing convergence maps and the galaxy convergence maps are NSIDE = 1024 HEALPix maps.
To facilitate interpretation, we apply a Gaussian smoothing kernel with an FWHM of 10 arcmin to the galaxy convergence maps, which are derived from discrete galaxy positions. The weak-lensing convergence maps are smoothed with the same kernel for consistency, providing additional smoothing beyond that intrinsic to the prior in the reconstruction. We smooth these maps after filling the masked regions with zeros, while smoothing the mask map in the same manner. The final smoothed map is obtained by dividing the smoothed convergence map by the smoothed mask map.

The shear catalog of source galaxies is divided into four tomographic bins, each corresponding to a different redshift range. \cite{DESWLmap} presented five convergence maps for each reconstruction method: four maps corresponding to the individual tomographic bins and one map reconstructed using the full sample of source galaxies. The photometric redshifts for the source galaxies were derived from the combined information from three independent likelihood functions: self-organizing map $p(z)$ (SOMPZ), clustering redshifts, and shear ratios (SR) \citep{SOMPZ}. Unlike the methods that assign individual redshift estimates to each galaxy, \cite{SOMPZ} used a neural network to group similar galaxies and determined the corresponding redshift distribution of each group. The tomographic bins were constructed by aggregating smaller groups, ensuring that each bin contained approximately the same number of galaxies. In this work, we use the tomographic bin assignments for each source galaxy and the source redshift distributions as provided in \cite{SOMPZ}. We construct four galaxy convergence maps each corresponding to one of the four tomographic bins, to compare with their weak-lensing convergence map counterparts. 

Figure \ref{fig:kappamaps_fig1} represents the weak-lensing convergence maps reconstructed using the Wiener method before smoothing is applied. As the tomographic bin number increases, more distant source galaxies are incorporated into the reconstruction. The amplitude of the projected matter density grows with increasing cosmic volume between the observer and the source galaxies. Consequently, the convergence maps in Figure \ref{fig:kappamaps_fig1} exhibit higher amplitudes for the farthest tomographic bins. 

Figure \ref{fig:lensweight_fig2} represents the source redshift probability distribution for each tomographic bin $p_s(z)$, shown in black histograms, as provided by \cite{SOMPZ}. The source redshift distribution was given for the entire sky without taking into account spatial variations. The red solid lines represent the lensing efficiency curves $q(z_l, p_s)$, which are computed using the corresponding source redshift distributions $p_s$ and the lens redshifts $z_l$ on the x-axis. The red vertical dashed lines indicate the peak of each lensing efficiency curve. The red shaded regions mark the range of lens redshifts where the lensing efficiency exceeds 75\% of its peak value, representing the redshift intervals that contribute most significantly to each weak-lensing convergence map.

\section{Method}\label{sec:method}
\subsection{Building Lensing-Weighted Galaxy Density Maps}
The galaxy convergence, $\kappa_g$, is computed by summing the partial galaxy convergence, $\kappa_g'$, for each lens redshift bin. The partial galaxy convergence is obtained by multiplying the lensing weight, $q(z_l, p_s)$, and the two-dimensional galaxy matter density fluctuation, $\delta^{2D}_g$. These are the four steps to construct the galaxy convergence map ($\kappa_g$):
\begin{enumerate}
    \item For each lens redshift bin, we compute the two-dimensional galaxy matter density fluctuation ($\delta^{2D}_g$) by counting the number of galaxies in each HEALPix pixel.
    \item We derive the source redshift distribution ($p_s$) for each HEALPix pixel.
    \item We compute the lensing weight, $q(z_l, p_s)$, for each lens redshift bin and corresponding sky position.
    \item By multiplying the lensing weight and the 2-dimensional galaxy matter density fluctuation, we obtain the partial galaxy convergence and sum them across all lens redshift bins.
\end{enumerate}
This procedure is repeated for four different source galaxy catalogs, which are used to reconstruct weak-lensing convergence maps corresponding to four tomographic redshift bins. 

Before performing the full procedure, we construct a mask map of the DES Y3 survey area. The mask is applied to exclude regions outside the survey footprint, areas obscured by large foreground objects, and anomalous regions where image processing or photometry may be unreliable. We follow the general guidelines from \cite{DESgold} to use the SOF-based galaxy photometric catalog.\footnote{The condition to obtain the nonmasked HEALPix pixels is as follows: $FLAGS\_FOREGROUND=0$ AND $FRACDET\_I>-1000$ AND $FRACDET\_G>-1000$ AND $FRACDET\_R>-1000$ AND $FRACDET\_Y>-1000$ AND $FRACDET\_Z>-1000$ AND $COVERAGE\_GRIZY>-1000$ AND $COVERAGE\_GRIZ>-1000$ AND $FLAGS\_BADREGIONS<4$} 
The mask map is generated using a HEALPix grid with NSIDE=4096, and galaxies within the masked regions are excluded from all subsequent analysis. For the weak-lensing convergence maps, the Wiener and GLIMPSE reconstructions adopt different masking schemes. The Wiener map contains no masked regions, as its reconstruction algorithm interpolates over areas with no source galaxy detections. Although there are no source galaxies, it is valuable to retain the weak-lensing signal in these regions, as they represent reasonable interpolations from nearby pixels and may correspond to dark-matter-dominated structures with few or no galaxies. The GLIMPSE map is generated through a patch-based reconstruction that incorporates the mask into the final convergence map. Accordingly, for the GLIMPSE comparison, we use the combined mask of the GLIMPSE and galaxy convergence maps, whereas for the Wiener comparison, we apply only the mask from the galaxy convergence map.

To build the galaxy convergence maps, we select the galaxies by applying various criteria: (1) the SOF \textit{i}-band magnitude should be less than 22.7, (2) the star-galaxy identifier should give the highest probability of being a galaxy, and (3) the object should not be located in the masked region. To avoid the contamination from the galaxies with large errors in photometry or photometric redshift, additional quality cuts are applied: the magnitude cut is set slightly shallower than the median survey depth. Including the objects that are too faint will increase the effect of observational systematics in the galaxy convergence maps.

\subsubsection{Computation of the Two-dimensional Galaxy Matter Density Fluctuation Maps}

After selecting the galaxies from the DES Y3 Gold catalog, we divide them into 15 redshift bins spanning $z=0$ to $z=1.5$ with a bin width of $\Delta_z=0.1$. For each redshift bin, we construct galaxy maps by counting the number of galaxies within each HEALPix pixel, using a resolution of NSIDE=1024. To obtain the two-dimensional galaxy matter density fluctuation maps, $\delta_g^{2D}$, we normalize the galaxy number counts by their mean value at each redshift bin. This process results in 15 $\delta_g^{2D}$ maps for different lens redshift bins.
\begin{figure*}[htp!]
    \centering
    \includegraphics[width=1\linewidth]{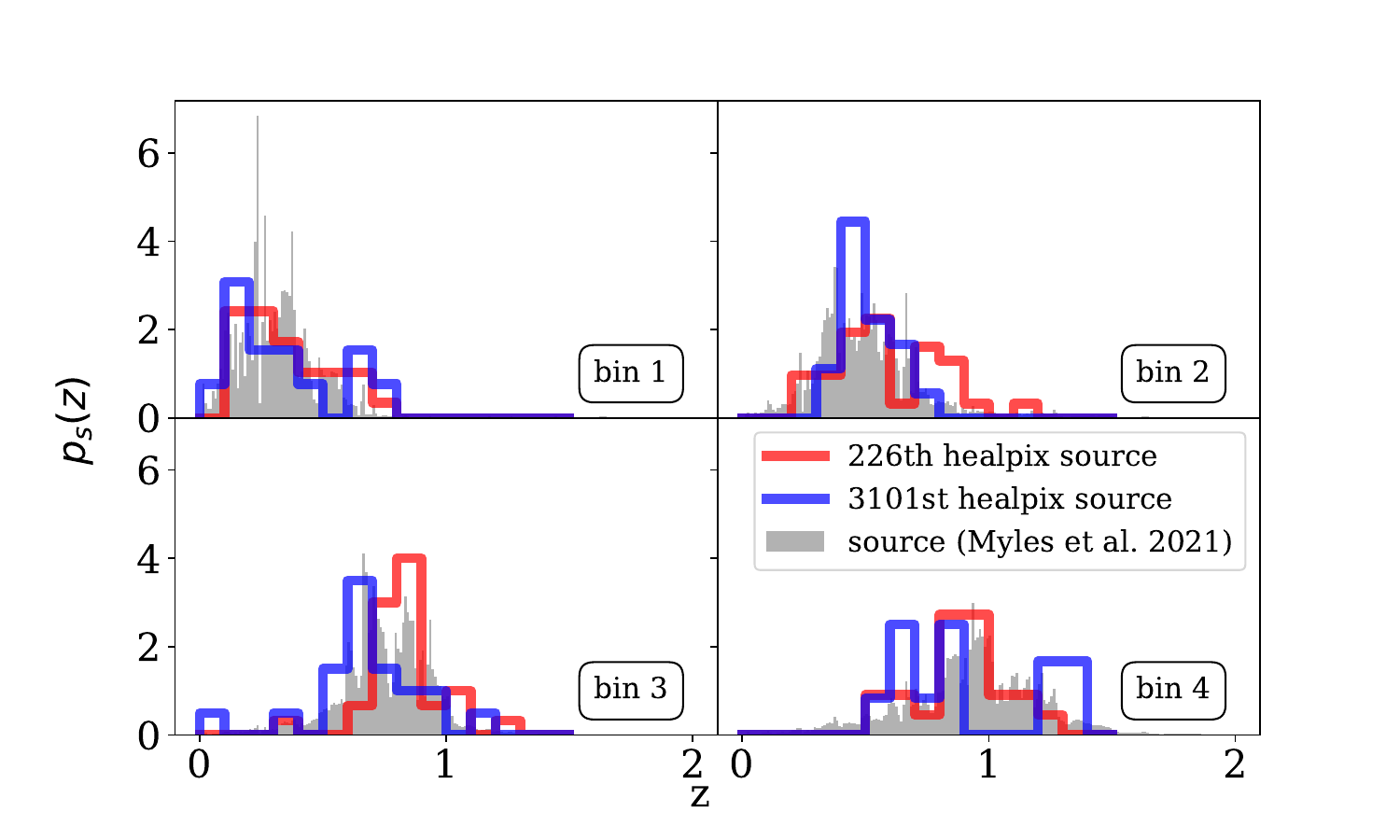}
    \caption{The source redshift distributions $p_s(z)$ for each tomographic bin. Red and blue solid lines show the specific source redshift distribution for the 226th and 3101st HEALPix pixels. The background gray histograms are the fiducial source redshift distributions as presented previously in Figure \ref{fig:lensweight_fig2}. The gray histograms are differ from the solid lines as they show the marginalized source redshift distribution regardless of the sky position.}
    \label{fig:sourcedist_fig3}
\end{figure*}

 \begin{figure*}[htp!]
    \centering
    \includegraphics[width=1\linewidth, trim=0 70 0 0, clip]{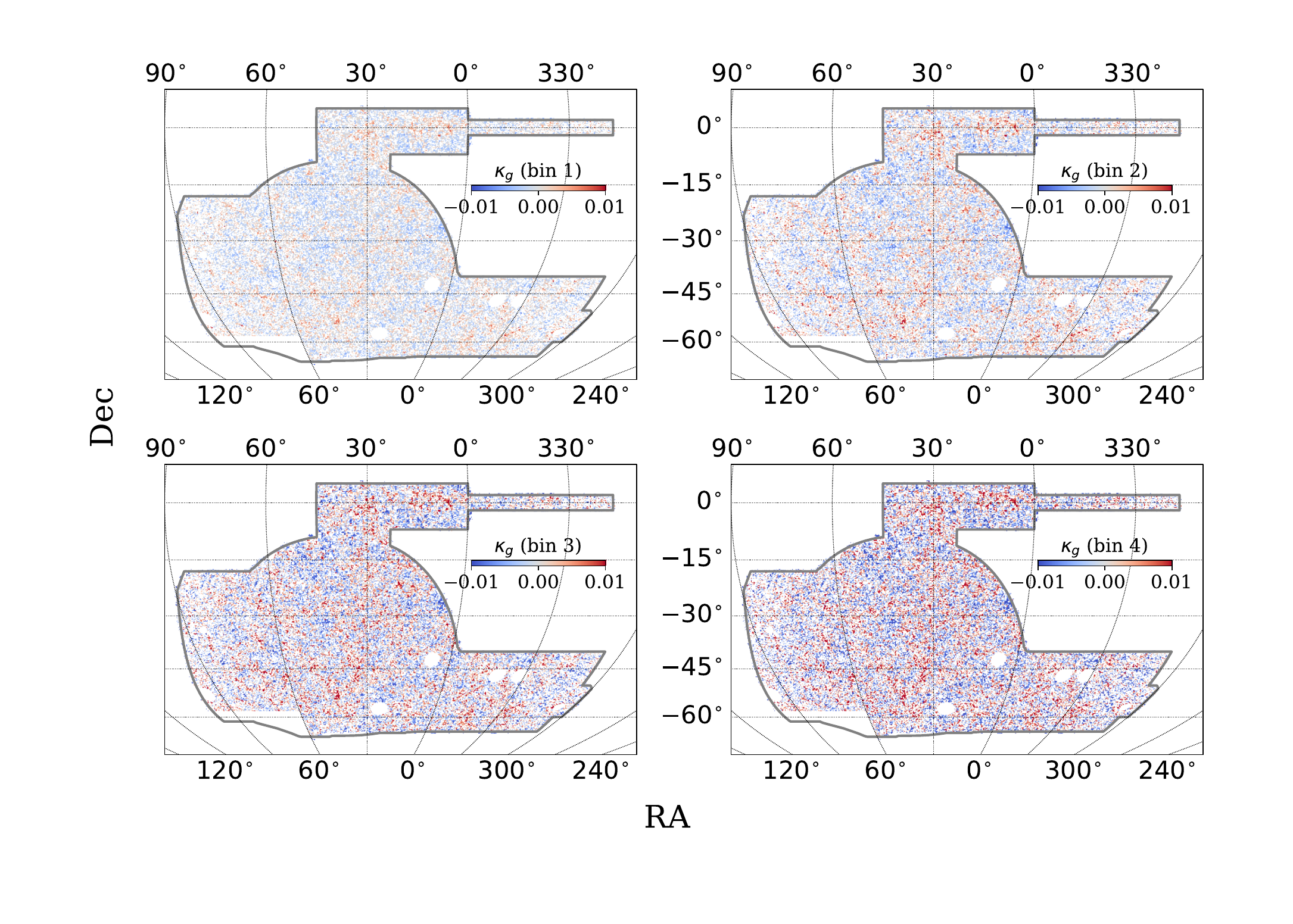}
    \caption{The galaxy convergence maps, $\kappa_g$, for each tomographic bin are shown. Each map is smoothed with a 10 arcmin FWHM Gaussian kernel.}
    \label{fig:kappag_maps_fig4}
\end{figure*}
\subsubsection{Derivation of the Source Redshift Distributions}\label{subsubsec:sourceredshift}

Next, we obtain the source redshift distribution, $p_s$, at each position on the sky. We match the unique object IDs between the DES Y3 Gold catalog and the SOMPZ redshift information catalog. We classify these source galaxies into four samples based on their assigned tomographic bins. This results in four source galaxy catalogs, identical to those used in the reconstruction of the weak-lensing convergence maps in \cite{DESWLmap}.

For each catalog, we further divide the galaxies into 15 source redshift bins ranging from $z=0$ to $z=1.5$ with a bin width of $\Delta_z=0.1$. In other words, the four broad redshift tomographic bins are subdivided into 15 narrower redshift bins. For each of these 15 bins, we construct the galaxy maps by counting the number of galaxies at each HEALPix pixel at a resolution of NSIDE=1024. Within each HEALPix pixel, the source redshift distribution for each tomographic bin is obtained by constructing a histogram of the galaxy counts across the 15 source redshift bins.

Figure \ref{fig:sourcedist_fig3} shows an example of the source redshift distributions at a given HEALPix pixel. The red and blue solid lines represent the source redshift distributions at two different HEALPix pixels, while the gray histograms in the background are the fiducial source redshift distributions derived from the SOMPZ redshift information catalog \citep{SOMPZ}, as in Figure \ref{fig:lensweight_fig2}. The fiducial distributions assume a single redshift distribution for the entire sky, neglecting spatial variations. However, the differences between the red and blue solid lines illustrate the spatial variation of the source redshift distribution depending on the sky position. While the overall shapes of the solid lines closely match the fiducial distributions, small variations between them reflect spatial variations in the source redshift distribution. In this work, we calculate the lensing weight by considering the subtle variations in the source redshift distributions across different regions of the sky.

\subsubsection{Computation of the Lensing Weight}

We estimate the lensing weight, $q(r_m', p_s)$, at the midpoint of each lens redshift bin, $r_m'$, using the previously obtained source redshift distribution, $p_s$. For each HEALPix pixel and tomographic bin, the source redshift distribution is used to compute the Equation (\ref{equ:q_lensweight}) as a function of $r_m'$.

\subsubsection{Construction of the Galaxy Convergence Maps}

Next we use Equation (\ref{eq:partialk}), to compute 15 partial galaxy convergence maps, $\kappa_g'$, corresponding to different lens planes at $r_m'$. Both the two-dimensional galaxy matter density fluctuation, $\delta_g^{2D}$, and the lensing weight, $q(r_m', p_s)$, are determined for each of the 15 lens planes using the procedures described above. The final galaxy convergence maps, $\kappa_g$, are obtained by summing the 15 partial galaxy convergence maps as in Equation (\ref{equ:kappa_g}). We construct four separate galaxy convergence maps, each corresponding to one of the four tomographic bins, using different sets of source galaxies initially categorized in Section \ref{subsubsec:sourceredshift}. 

Figure \ref{fig:kappag_maps_fig4} represents the final galaxy convergence maps for each tomographic bin. All maps are smoothed with a Gaussian kernel with an FWHM of 10 arcmin. Some regions appear masked because of the presence of foreground objects and areas affected by poor photometry. The amplitudes of the galaxy convergence maps are increased in higher tomographic bins, as shown in Figure \ref{fig:kappamaps_fig1}, because the projected mass of lens galaxies accumulates as more distant source galaxies are used.

\subsection{Estimation of the Uncertainties of Weak-lensing Convergence and Galaxy Convergence}

We estimate the uncertainty of the galaxy convergence, which arises from the usage of photometric redshifts. The DES Y3 Gold catalog provides photometric redshift uncertainties based on the DNF method. We generate 50 perturbed realizations of the galaxy catalog by randomly shifting the redshifts of lens galaxies according to their reported uncertainties using a Gaussian probability distribution.

In this process, only the redshifts of the lens galaxies are perturbed, while the source redshift distribution, $p_s$, remains unchanged for each HEALPix pixel. From these 50 realizations, we compute the standard deviation across the maps to obtain the uncertainty map, $\sigma_{\kappa_{g}}$. 

Due to the high computational cost of generating each realization, we limit the number to 50. To validate this choice, we perform a test with a small patch of 15$^\circ$$\times$15$^\circ$ size up to 100 realizations. The result shows that the uncertainty derived from 50 realizations does not introduce any systematic bias.

We do not take into account the Poisson noise from galaxy counts in the above process. It should be noted that the combined noise from the weak-lensing convergence and galaxy convergence will be dominated by the shape noise of the weak-lensing convergence.

We estimate the contribution of the uncertainties in the WL convergence map to the statistical confidence of our analysis. The Wiener map presented in \cite{DESWLmap} is one that which maximizes the posterior probability of the map given the observed data $\boldsymbol{\gamma} $,
\begin{equation} \label{eq:posterior_optimization}
\hat{\boldsymbol{\kappa}} = \underset{\boldsymbol{\kappa}}{\rm arg \ max} \  \log p(\boldsymbol{\gamma} | \boldsymbol{\kappa} ) + \log p(\boldsymbol{\kappa} ) \ ,
\end{equation}
\noindent using a Gaussian form for the prior $\log p(\boldsymbol{\kappa} ) $ with a chosen fiducial power spectrum. As the likelihood is also Gaussian, this maximum posterior estimate $\hat{\boldsymbol{\kappa}}$ is also the mean of the posterior. In this work, we instead use a set of samples from the posterior:
\begin{equation} 
\{ \boldsymbol{\kappa}_i \} \sim p(\boldsymbol{\kappa} | \boldsymbol{\gamma})  .
\end{equation}
\noindent These 29 posterior samples are ``constrained realizations'' (previously used in~\citealt{des_supervoid}) that are drawn using the \texttt{dante} algorithm~\citep{dante}. These samples are constructed using the entire source galaxy sample without division into tomographic bins. The per-pixel standard deviation serves as our convergence uncertainty map. It provides a measure of the reliability of the weak-lensing mass reconstruction in different regions. Using these uncertainty maps, we proceed to identify discrepancies between the convergence maps and galaxy convergence maps by defining peaks in their residual maps and assessing their significance in relation to the derived uncertainty estimates.

\begin{figure*}[htp!]
    \centering
    \includegraphics[width=1\linewidth, trim=70 70 70 0, clip]{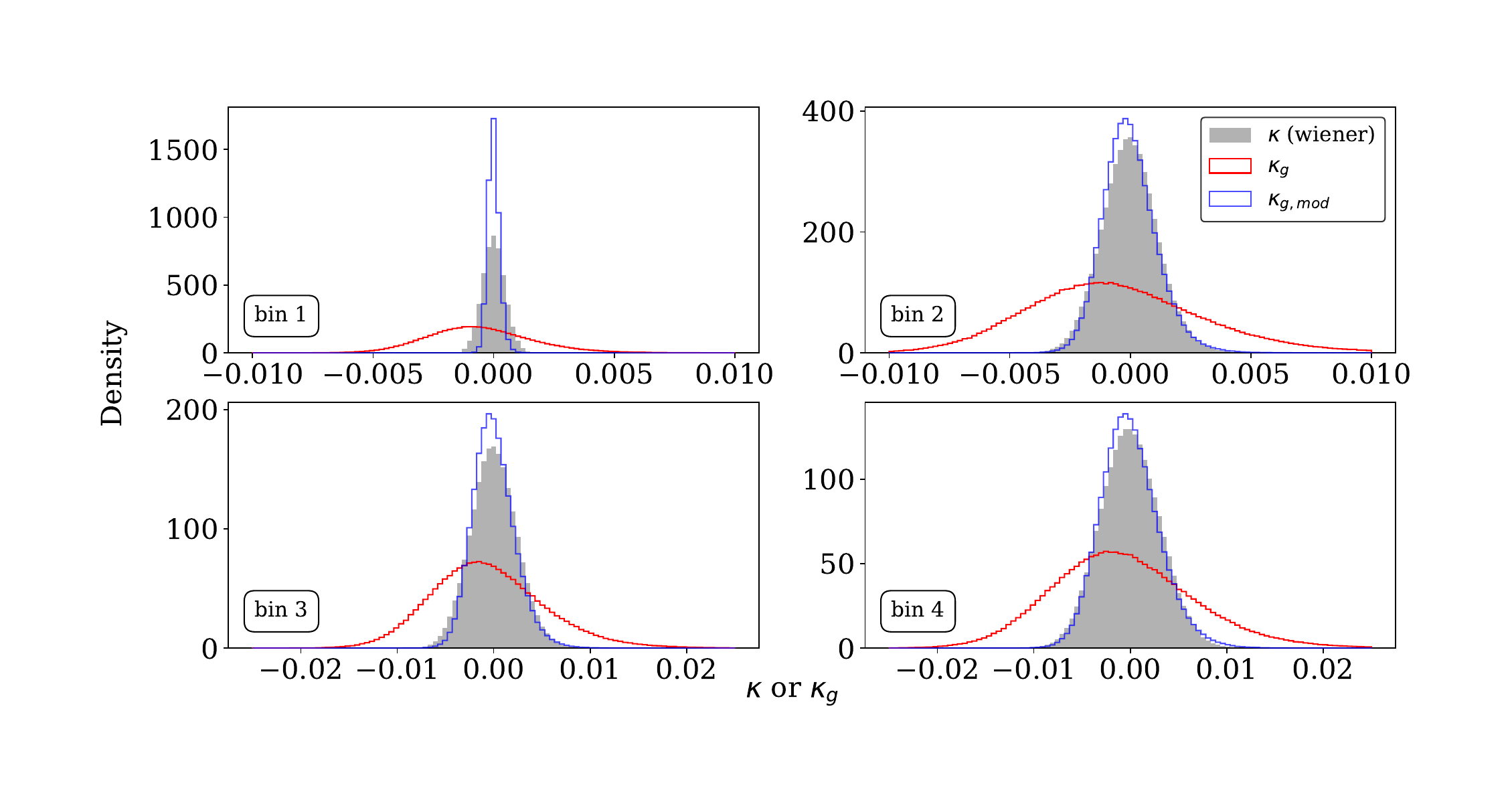}
    \caption{The histograms of the Wiener weak-lensing convergence in gray, the galaxy convergence in red lines, and the scaled galaxy convergence in blue lines are shown for each tomographic bin. All histograms are normalized such that the area under each curve is equal to 1.}
    \label{fig:scaling_w_fig5}
\end{figure*}
\subsection{Scaling between the Weak-lensing Convergence and the Galaxy Convergence}\label{sec:scaling}

To identify any discrepancy between the two maps, we subtract the galaxy convergence maps from the weak-lensing convergence maps. Before this, we first compare the histograms of weak-lensing convergence ($\kappa$) and galaxy convergence ($\kappa_g$). \cite{Clerkin16} noted that the weak-lensing convergence distribution is well modeled by a lognormal distribution convolved with Gaussian shape noise. However, in cases where the reconstruction is affected by large shape noise, \cite{Clerkin16} mentioned that a simple Gaussian distribution provides a sufficient approximation. In this regard, we can characterize both distributions as approximately Gaussian (to leading order).

Figure \ref{fig:scaling_w_fig5} shows the histograms of the weak-lensing convergence and galaxy convergence distributions. The gray bar histogram represents the distribution of weak-lensing convergence, reconstructed using the Wiener method in Figure \ref{fig:kappamaps_fig1}. The red solid line shows the original distribution of galaxy convergence ($\kappa_g$). The histograms from the GLIMPSE reconstruction are in Appendix A. 
While the gray and red histograms are expected to be similar, noticeable differences are observed in the distributions. The weak-lensing convergence exhibits a nearly symmetric distribution, whereas the galaxy convergence shows a slightly skewed distribution with a positive tail, which becomes more pronounced in the farther tomographic bins. This discrepancy can be interpreted as follows.

First, the power spectrum of the Wiener WL map is expected to be biased low. This is an expected feature: the variance of the mean map (e.g. from posterior samples) is not equal to the mean posterior of the map variance (or, equivalently, the power spectrum). The samples themselves have unbiased power spectra, which can be jointly inferred if necessary \citep[e.g.,][]{alsing_maps}. In this work, this difference is completely accounted for via rescaling (see below).

Secondly, even if we had access to the noise-free (true) convergence map, there may be a mismatch between $\kappa$ and $\kappa_g$. \citet{Waerbeke13} compared the galaxy convergence with the weak-lensing convergence at the identified peak regions and reported that the former is 2-3 times larger. Moreover, the scatter in Figure 9 of \citet{Waerbeke13} indicates a wider distribution for the galaxy convergence than for the weak-lensing convergence, in agreement with our findings. They attributed this discrepancy to the assumption that all galaxies host dark matter halos of identical mass. In this work, we adopt a similar assumption while counting the galaxies to obtain two-dimensional galaxy matter density fluctuations. 

One way to account for the mass of dark matter halos associated with galaxies is to use their intrinsic luminosities as a proxy for mass. We attempt to construct luminosity-weighted galaxy convergence maps; however, the resulting distributions appear more skewed than those of the unweighted maps. The reason for this behavior remains unclear at this stage—whether it arises from systematics in the data or assumptions in the weighting scheme or whether it is an expected outcome. Given these ambiguities, the true distributions of the weak-lensing and galaxy convergence remain uncertain.

To ensure a fair comparison using the available datasets, we adopt a simple approach: we construct the galaxy convergence maps from galaxy number counts and apply a scaling adjustment to the galaxy convergence ($\kappa_g$) so that its Gaussian-fitted distribution better matches that of the weak-lensing convergence ($\kappa$). This number count-based approach minimizes potential sources of systematic bias compared to the luminosity-weighted approach, which is affected by several factors: the use of photometric redshifts to estimate intrinsic luminosities, the fact that luminosity traces only the stellar rather than the total mass, and the nonconstant galaxy mass-to-light ratio. Moreover, the unweighted galaxy convergence yields a distribution that more closely matches the shape of the reconstructed weak-lensing convergence. After the scaling, the weak-lensing convergence (gray histogram) and the scaled galaxy convergence (blue solid line) show good agreement.

\section{Results} \label{sec:result}

From the comparison between the weak-lensing convergence map and the scaled galaxy convergence map, we examine the similarities between the two in Section \ref{sec:corr} and the discrepancies in Sections \ref{sec:dscfinding} and \ref{sec:dscexamination}.

\subsection{Correlation between the Weak-lensing Convergence and the Galaxy Convergence}\label{sec:corr}

\begin{figure*}[htp!]
    \centering
    \includegraphics[width=1\linewidth, trim=70 60 70 30, clip]{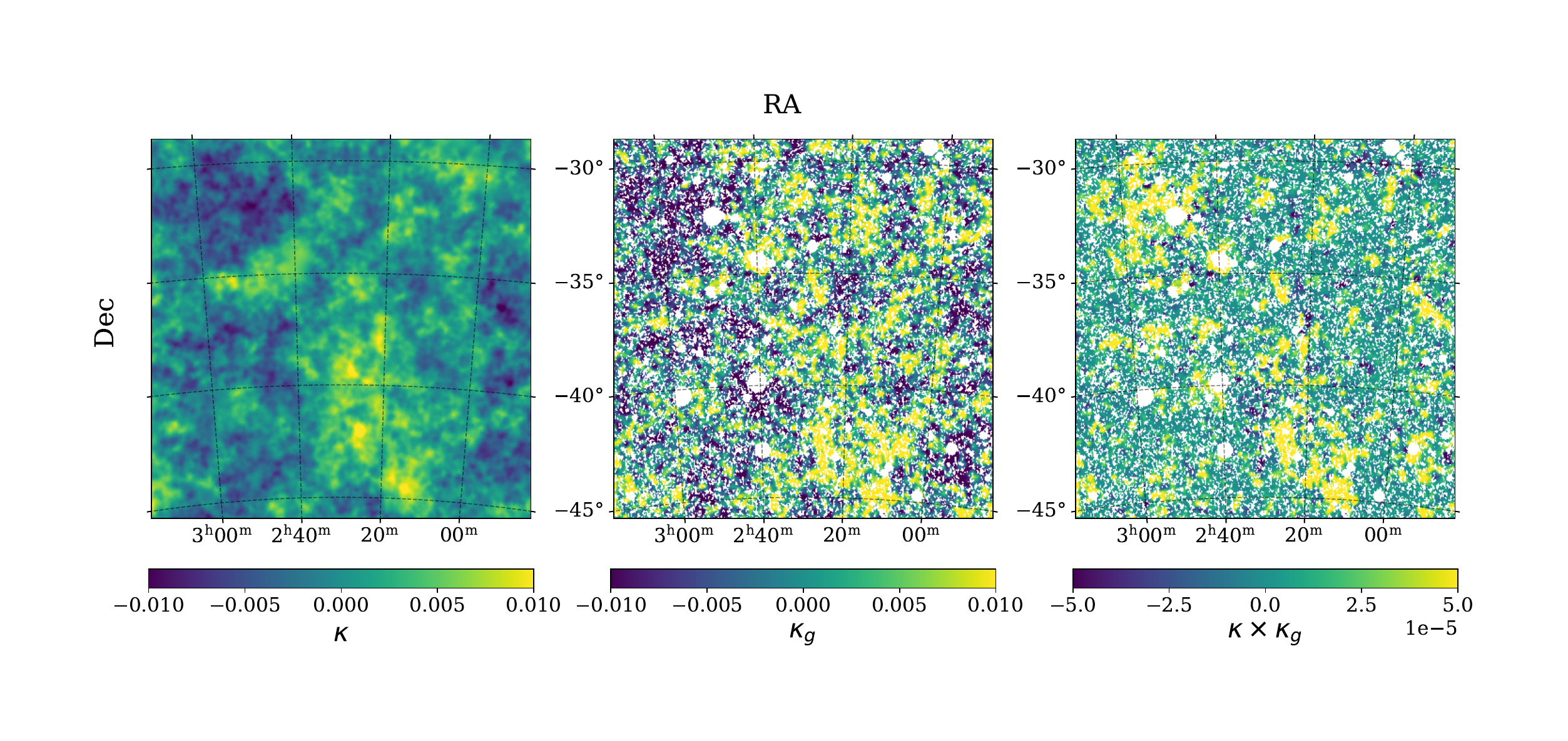}
    \caption{(Left) Patch of the Wiener-reconstructed weak-lensing convergence map for the fourth tomographic bin. (Middle) Patch of the corresponding galaxy convergence map. All maps are smoothed with a Gaussian kernel with an FWHM of 10 arcmin. Yellow regions represent overdensities, while blue regions represent underdensities. (Right) The product map, $\kappa \times \kappa_g$, highlighting regions of high correlation (yellow) and low correlation (blue).}
    \label{fig:ktimeskg_fig6}
\end{figure*}

\begin{figure}[htp!]
    \centering
    \includegraphics[width=1\linewidth]{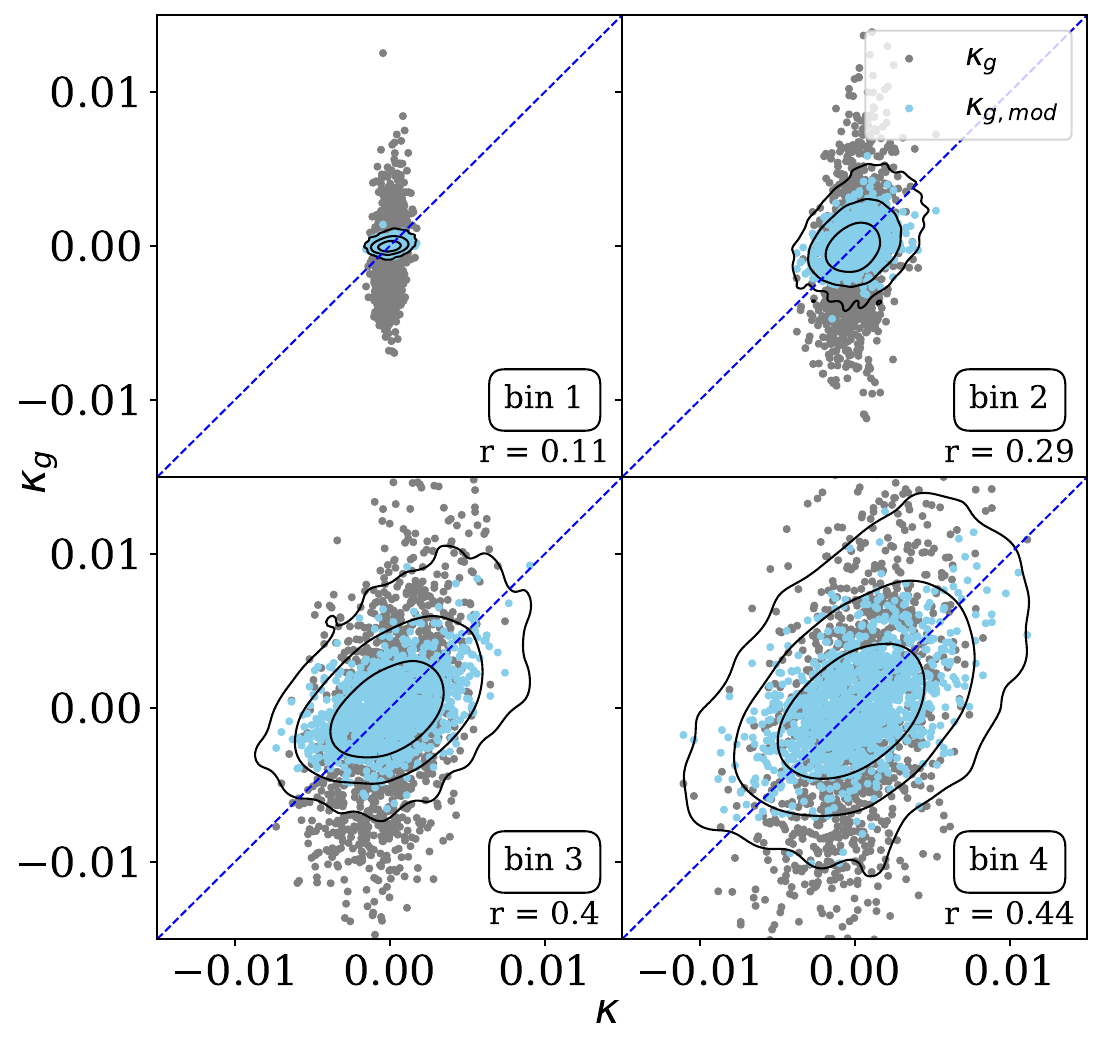}
    \caption{Pixel-by-pixel comparison between the weak-lensing convergence ($\kappa$) reconstructed using the Wiener method and the galaxy convergence ($\kappa_g$). The gray points represent data from the unscaled galaxy convergence maps, while the sky blue points correspond to the scaled galaxy convergence maps adjusted to match the Wiener convergence. For visualization purposes, only 0.01\% of the total data points are displayed as points. The black contour lines represent the 1$\sigma$, 2$\sigma$, and 3$\sigma$ levels of the full distribution of sky blue points. The blue dashed line indicates the one-to-one correspondence in the $\kappa$-$\kappa_g$ plane.}
    \label{fig:kvskg_fig7}
\end{figure}

The weak-lensing convergence ($\kappa$) and the galaxy convergence ($\kappa_g$) are expected to be correlated, because they represent the total matter and galaxy distributions, respectively. Figure \ref{fig:ktimeskg_fig6} shows an example of a strong correlation between these two quantities. Each panel represents a cutout of a particular region of the sky, which clearly shows the large-scale matter distribution. The first panel shows the Wiener-reconstructed weak-lensing convergence map for the fourth tomographic bin, while the second panel shows the corresponding galaxy convergence map. In both maps, yellow regions indicate overdensities, and blue regions indicate underdensities. The third panel shows the product map, $\kappa \times \kappa_g$, obtained by multiplying the first and second maps. In this panel, the yellow regions highlight areas where the convergence and the galaxy convergence are strongly correlated. Notably, there are no significant blue regions with amplitudes or sizes comparable to those of the yellow regions, indicating that the overall distributions of $\kappa$ and $\kappa_g$ are well matched.

To examine the correlation over the entire survey area, Figure \ref{fig:kvskg_fig7} represents a pixel-by-pixel comparison between the weak-lensing convergence ($\kappa$) from Wiener reconstruction and the galaxy convergence ($\kappa_g$) maps. The gray dots represent data points from the unscaled galaxy convergence maps, while the sky blue dots correspond to the scaled galaxy convergence maps ($\kappa_{g,mod}$) to the Wiener convergence maps. The blue dashed line denotes the one-to-one correspondence in the $\kappa$-$\kappa_g$ plane. The contours correspond to the 1$\sigma$, 2$\sigma$, and 3$\sigma$ levels of the data distribution. The contours show a clear linear correlation between the weak-lensing convergence and the scaled galaxy convergence. Additionally, the Pearson correlation coefficient ($r$) between the $\kappa$ and $\kappa_{g,mod}$ maps is computed and displayed in each panel. 
The highest Pearson correlation coefficient is observed in the farthest source tomographic bin, as the increased line-of-sight projection of the lens makes the galaxy distribution smoother and more closely aligned with the total mass distribution.
In Appendix A, the correlation between the GLIMPSE convergence and the galaxy convergence is described. 

\subsection{Finding Dark Structure Candidates in the Residual Maps}\label{sec:dscfinding}

To identify dark structure candidates, we first construct residual maps by subtracting the scaled galaxy convergence ($\kappa_{g,mod}$) from the weak-lensing convergence ($\kappa$) maps for each tomographic bin. We then implement the following steps to identify dark structure candidates.
\begin{enumerate}
    \item We compute the mean and standard deviation of the residual maps and apply a 3$\sigma$-clipping threshold to filter out HEALPix pixels with residuals exceeding this threshold. 
    \item From these pixels, we connect adjacent HEALPix pixels to define peak regions. 
    \item Peaks consisting of fewer than 10 HEALPix pixels are excluded because they are smaller than the smoothing scale of 10 arcmin. Peaks where some of the convergence values of the selected pixels are negative are also discarded.
    \item We exclude peaks whose boundaries are directly connected to the survey footprint boundary to minimize edge effects.
    \item We estimate the integrated S/N of the peaks by incorporating the uncertainties of both the weak-lensing convergence and the galaxy convergence maps. The S/N of the residual peaks serves as a measure of the reliability of the detected residual signals. We compute the integrated S/N for each peak and exclude those with S/N$<3$. 
    \item We crossmatch the detected residual peaks between the Wiener and GLIMPSE convergence maps for each tomographic bin to identify consistent residual signals present in both maps. 
\end{enumerate}

We do not use the shape information of dark structure candidates for their identification. The apparent morphology of residual peaks is highly sensitive to both the pixelization scheme and the specific weak-lensing reconstruction method. Moreover, there is no prior constraint on the expected morphology of dark structures. We therefore rely solely on their size for candidate identification. We adopt a size threshold of 10 HEALPix pixels, which provides a robust criterion for extracting high-significance candidates. This criterion is validated, as no contiguous peak consisting of more than 10 outlier pixels is detected in the null tests.

Following these steps, we identify 49 peaks in the Wiener residual map and 352 peaks in the GLIMPSE residual map. Peaks are defined from the HEALPix maps and their centroids are computed as the mean of the spatial distribution, weighted by the pixel values in the residual map. To construct the final candidate list, we match the peak centroids of Wiener and GLIMPSE peaks if their separation is less than 10 arcmin, corresponding to the smoothing scale. This process results in a final catalog of 22 dark structure candidates.

Table \ref{tab} summarizes the properties of the identified candidates. The first column lists the candidate ID, while the second column indicates the corresponding tomographic redshift bin. The third and fourth columns list the peak centroid coordinates in degrees. The fifth column shows the candidate area size in arcmin$^2$. The sixth column is the integrated S/N of the residual signal for each candidate. The seventh to 10th columns list the information from the GLIMPSE map comparison, which is comparable with the Wiener map comparison results. The 11th column contains a bitwise quality flag that assesses the reliability of each candidate as a dark structure.

The bitwise quality flag is assigned based on a visual inspection of the weak-lensing convergence map, galaxy convergence map, residual map, and survey \textit{i}-band depth map. There are some candidates that have some suspicious features indicating a false signal of a dark structure. Candidates are evaluated using the following quantitative criteria during visual inspection:

\begin{enumerate}
    \item \textit{Bit 1.} In the weak-lensing convergence map, the candidates are associated with a diffuse overdense signal or reside in the outskirts of a weak-lensing mass peak. Because galaxies form at the high matter concentration, a diffuse mass distribution may not be associated with a sufficient number of galaxies. This behavior is consistent with the expectations from \lcdm$\ $cosmology, and thus we mainly focus on candidates that are well matched to the weak-lensing mass peaks.
    \item \textit{Bit 2.} The candidate boundaries are directly connected to the large foreground object masks. We also flag candidates whose 5$^\circ$$\times$5$^\circ$ cutout regions are covered by masks more than 45\%, or include the survey footprint boundary. Nearby masked regions in the survey can introduce significant uncertainties in both the weak-lensing convergence and galaxy convergence maps.
    \item \textit{Bit 4.} The survey depth in the candidate region is lower than 22 mag in the \textit{i}-band. The observational completeness should be high enough to ensure that an apparent underdensity of galaxies is not simply due to incomplete observations.
\end{enumerate}
Based on these criteria, we identify the seven most promising dark structure candidates, which are marked with a zero flag in the 11th column of Table \ref{tab}.

\begin{deluxetable*}{ccccccccccc}
\label{tab}
\tablewidth{0pt}
\tablecaption{Catalog of the dark structure candidates}
\tablehead{
\colhead{ID} & \colhead{Bin} & \colhead{R.A.} & \colhead{Decl.} & \colhead{Size} & \colhead{SNR} & \colhead{R.A.} & \colhead{Decl.} & \colhead{Size} & \colhead{SNR} & \colhead{Flag} \\
& & Wiener & Wiener & Wiener & Wiener & GLIMPSE & GLIMPSE & GLIMPSE & GLIMPSE &  \\
& & (deg) & (deg) & (arcmin$^2$) & & (deg) & (deg) & (arcmin$^2$) & & 
}
\startdata
1 & 2 & 6.26 & -3.23 & 188.8 & 3.5 & 6.30 & -3.24 & 283.3 & 7.1 & 0 \\
2 & 2 & 33.98 & -3.24 & 188.8 & 4.3 & 33.97 & -3.26 & 141.6 & 5.9 & 1 \\
3 & 2 & 36.77 & -33.45 & 236.1 & 4.0 & 36.77 & -33.43 & 177 & 5.6 & 3 \\
4 & 2 & 41.63 & -2.17 & 224.2 & 3.9 & 41.63 & -2.11 & 177 & 5.3 & 5 \\
5 & 2 & 42.52 & -26.76 & 118 & 3.1 & 42.62 & -26.70 & 590.1 & 13.6 & 0 \\
6 & 2 & 54.51 & -12.11 & 141.6 & 3.3 & 54.55 & -12.09 & 177 & 6.1 & 3 \\
7 & 2 & 318.60 & -61.71 & 224.2 & 4.2 & 318.60 & -61.72 & 224.2 & 6.9 & 7 \\
8 & 3 & 2.25 & -36.38 & 141.6 & 6.4 & 2.28 & -36.37 & 224.2 & 9.2 & 6 \\
9 & 3 & 11.30 & -54.84 & 129.8 & 5.7 & 11.33 & -54.68 & 696.4 & 16.5 & 1 \\
10 & 3 & 32.06 & -27.00 & 318.7 & 10 & 31.93 & -27.03 & 944.2 & 19.3 & 1 \\
11 & 3 & 32.60 & -27.95 & 224.2 & 8.1 & 32.57 & -27.95 & 141.6 & 6.4 & 1 \\
12 & 3 & 32.82 & -27.19 & 118 & 5 & 32.82 & -27.15 & 188.8 & 6.7 & 0 \\
13 & 3 & 39.32 & -1.34 & 118 & 5.5 & 39.31 & -1.34 & 165.2 & 6.8 & 3 \\
14 & 3 & 65.08 & -44.78 & 118 & 5.2 & 65.09 & -44.77 & 247.9 & 8.9 & 3 \\
15 & 3 & 69.00 & -34.23 & 129.8 & 5.4 & 69.01 & -34.20 & 224.2 & 8.4 & 0 \\
16 & 3 & 334.66 & -52.52 & 153.4 & 5.9 & 334.61 & -52.51 & 224.2 & 7.4 & 5 \\
17 & 4 & 31.38 & -30.11 & 129.8 & 7.6 & 31.39 & -30.10 & 129.8 & 7.4 & 4 \\
18 & 4 & 32.27 & -2.51 & 141.6 & 7.4 & 32.29 & -2.52 & 129.8 & 6.8 & 1 \\
19 & 4 & 35.15 & -36.99 & 129.8 & 6.7 & 35.16 & -37.00 & 165.2 & 8.2 & 0 \\
20 & 4 & 36.87 & -58.64 & 212.4 & 9.5 & 36.91 & -58.63 & 177 & 8.3 & 1 \\
21 & 4 & 51.95 & -44.22 & 153.4 & 7.5 & 52.01 & -44.13 & 224.2 & 9.1 & 0 \\
22 & 4 & 337.06 & -53.69 & 212.4 & 8.9 & 337.12 & -53.62 & 177 & 8.3 & 0 \\
\enddata
\tablecomments{Column (1): the candidate ID. Column (2): the corresponding redshift tomographic bin. Column (3) and (4): the R.A. and decl. positions of the candidate centroid in degrees. Column (5): the size of the candidate in square arcminutes. Column (6): the integrated S/N of the residual. Column (7)-(10): the centroid position, size, and integrated S/N obtained from the GLIMPSE map comparison. Column (11): the bitwise quality flag. Each tomographic bin corresponds to the following redshift interval: $z=0.075-0.275$ for bin 1, $z=0.125-0.405$ for bin 2, $z=0.195-0.575$ for bin 3, and $z=0.235-0.725$ for bin 4. The most probable candidates based on visual inspection are marked with "0" in column (11). For those with nonzero flags, see Section \ref{sec:dscfinding} for details.}
\end{deluxetable*}

\subsection{Further Examination of Dark Structure Candidates}\label{sec:dscexamination}

\begin{figure*}
    \centering
    \includegraphics[width=1\linewidth, trim=0 0 0 0, clip]{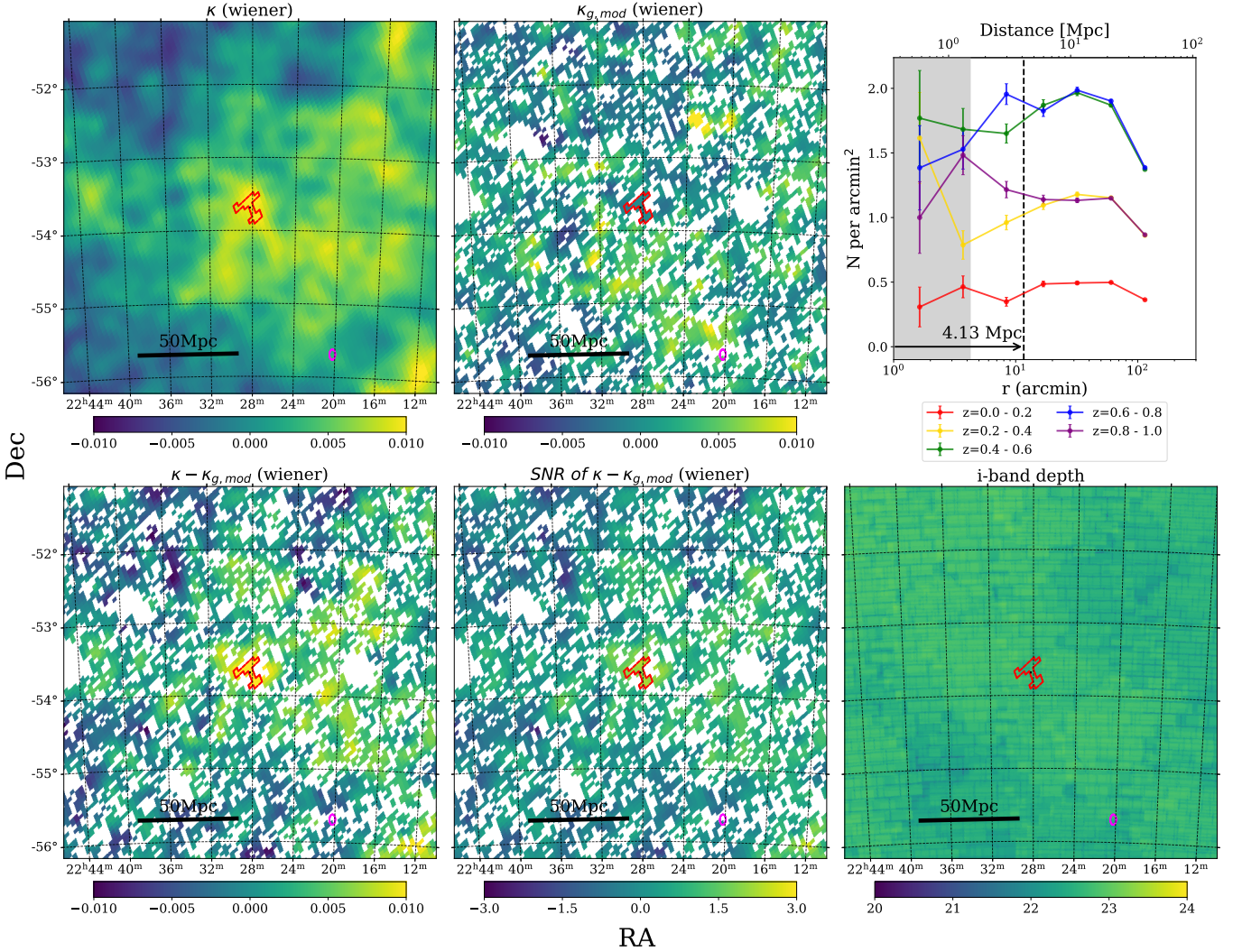}
    \caption{The 5$^\circ$$\times$5$^\circ$ cutout maps of the dark structure candidate ID 22. The red lines outline the the boundary of the candidate in each map. The small magenta circle in the bottom-right corner of each panel represents the 10 arcmin smoothing scale, while the small black dot at the center of each panel marks the centroid of the candidate region. (top-left) Wiener convergence map. (top-center) Galaxy convergence map scaled to the Wiener weak lensing convergence. (top-right) Residual map obtained by subtracting the scaled galaxy convergence from the Wiener convergence. (middle-left) CMB lensing convergence map. (middle-center) Observation depth map for i-band. (middle-right) S/N map of the residual. (bottom-left) Radial galaxy surface number density profile around candidate ID 22 shown as a function of redshift bins and (bottom-right) shown as a function of magnitude bins. The gray shaded region represents the circular radius of the smoothing scale, while the black dashed line indicates the most extended boundary distance from the candidate centroid. The complete figure set for 22 dark structure candidates is available in the online journal. The analogous figure set for GLIMPSE weak lensing convergence is available in Appendix B of the online journal.}
    \label{fig:exzoomin_fig8}
\end{figure*}

Figure \ref{fig:exzoomin_fig8} shows a 5$^\circ$$\times$5$^\circ$ cutout of the convergence, galaxy convergence, residual, and depth maps for dark structure candidate ID 22. This candidate is classified as the most probable dark structure in our sample. The first two top panels display the Wiener convergence map and the galaxy convergence map scaled to match the Wiener convergence. The top-right panel represents the residual map, obtained by subtracting the galaxy convergence map from the convergence map. The middle-right panel shows the S/N map of the residual signal; here the noise is the integrated noise from the weak-lensing convergence and galaxy convergence. We note that the S/N shown in Figure \ref{fig:exzoomin_fig8} is positive for the positive $\kappa-\kappa_{g,mod}$, and negative for the negative $\kappa-\kappa_{g,mod}$. We also note that the amplitude of the S/N map is different from the integrated S/N shown in Table \ref{tab}, because the integrated S/N is obtained from the whole candidate area. The middle-center panel displays the $i$-band survey depth map. The red lines outline the boundary of the candidate in each map. The yellow region inside the red boundary in the top-left panel indicates an excess of dark matter. In contrast, the top-center panel exhibits a blue region within the red boundary, indicating a deficit of galaxies. Consequently, the residual map (top-right panel) and the residual S/N map (middle-right panel) show a yellow color inside the red boundary. Notably, the depth map (middle-center panel) suggests that the observed galaxy underdensity is unlikely to be due to survey systematics, as the magnitude limit remains high enough throughout the region enclosed by the red boundary.

\figsetstart
\figsetnum{8}
\figsettitle{Dark Structure Candidates}

\figsetgrpstart
\figsetgrpnum{8.1}
\figsetgrptitle{Candidate ID 1}
\figsetplot{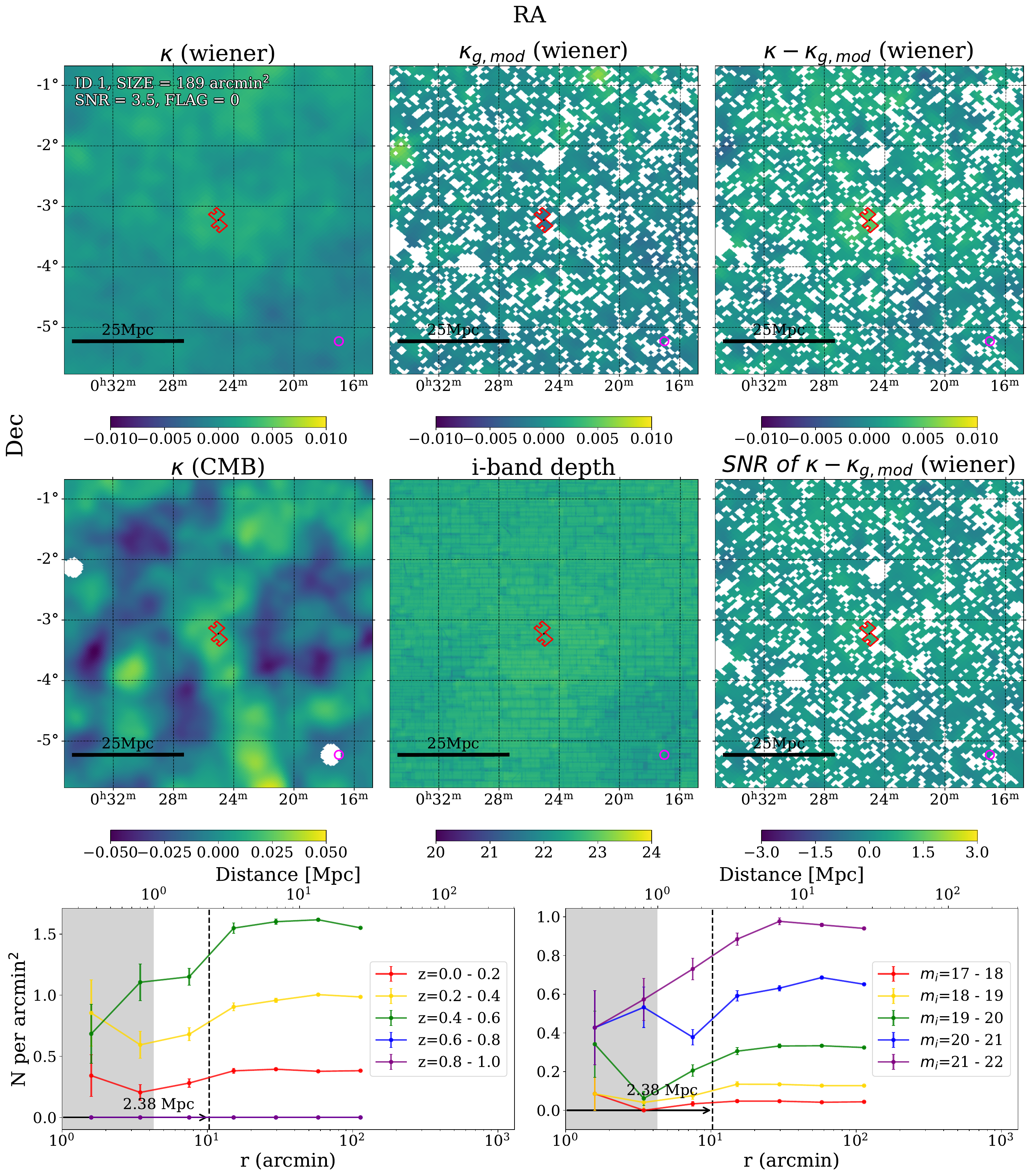}
\figsetgrpnote{The 5$^\circ$$\times$5$^\circ$ cutout maps of the dark structure candidate ID 1. The red lines outline the the boundary of the candidate in each map. The small magenta circle in the bottom-right corner of each panel represents the 10 arcmin smoothing scale, while the small black dot at the center of each panel marks the centroid of the candidate region. (top-left) Wiener convergence map. (top-center) Galaxy convergence map scaled to the Wiener weak lensing convergence. (top-right) Residual map obtained by subtracting the scaled galaxy convergence from the Wiener convergence. (middle-left) CMB lensing convergence map. (middle-center) Observation depth map for i-band. (middle-right) S/N map of the residual. (bottom-left) Radial galaxy surface number density profile around candidate ID 1, shown as a function of redshift bins and (bottom-right) the same profile shown as a function of magnitude bins.}
\figsetgrpend

\figsetgrpstart
\figsetgrpnum{8.2}
\figsetgrptitle{Candidate ID 2}
\figsetplot{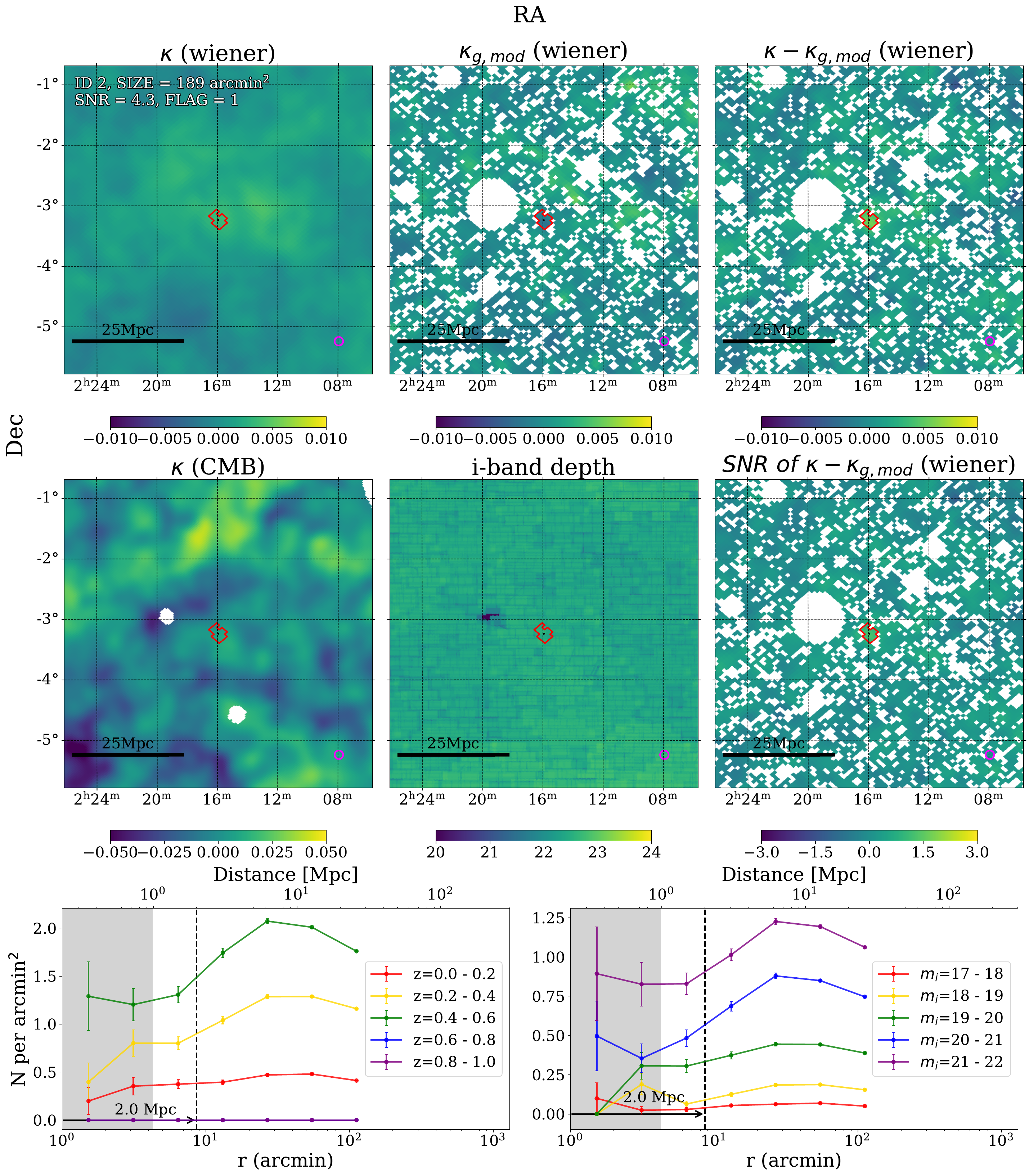}
\figsetgrpnote{The 5$^\circ$$\times$5$^\circ$ cutout maps of the dark structure candidate ID 2. The red lines outline the the boundary of the candidate in each map. The small magenta circle in the bottom-right corner of each panel represents the 10 arcmin smoothing scale, while the small black dot at the center of each panel marks the centroid of the candidate region. (top-left) Wiener convergence map. (top-center) Galaxy convergence map scaled to the Wiener weak lensing convergence. (top-right) Residual map obtained by subtracting the scaled galaxy convergence from the Wiener convergence. (middle-left) CMB lensing convergence map. (middle-center) Observation depth map for i-band. (middle-right) S/N map of the residual. (bottom-left) Radial galaxy surface number density profile around candidate ID 2, shown as a function of redshift bins and (bottom-right) the same profile shown as a function of magnitude bins.}
\figsetgrpend

\figsetgrpstart
\figsetgrpnum{8.3}
\figsetgrptitle{Candidate ID 3}
\figsetplot{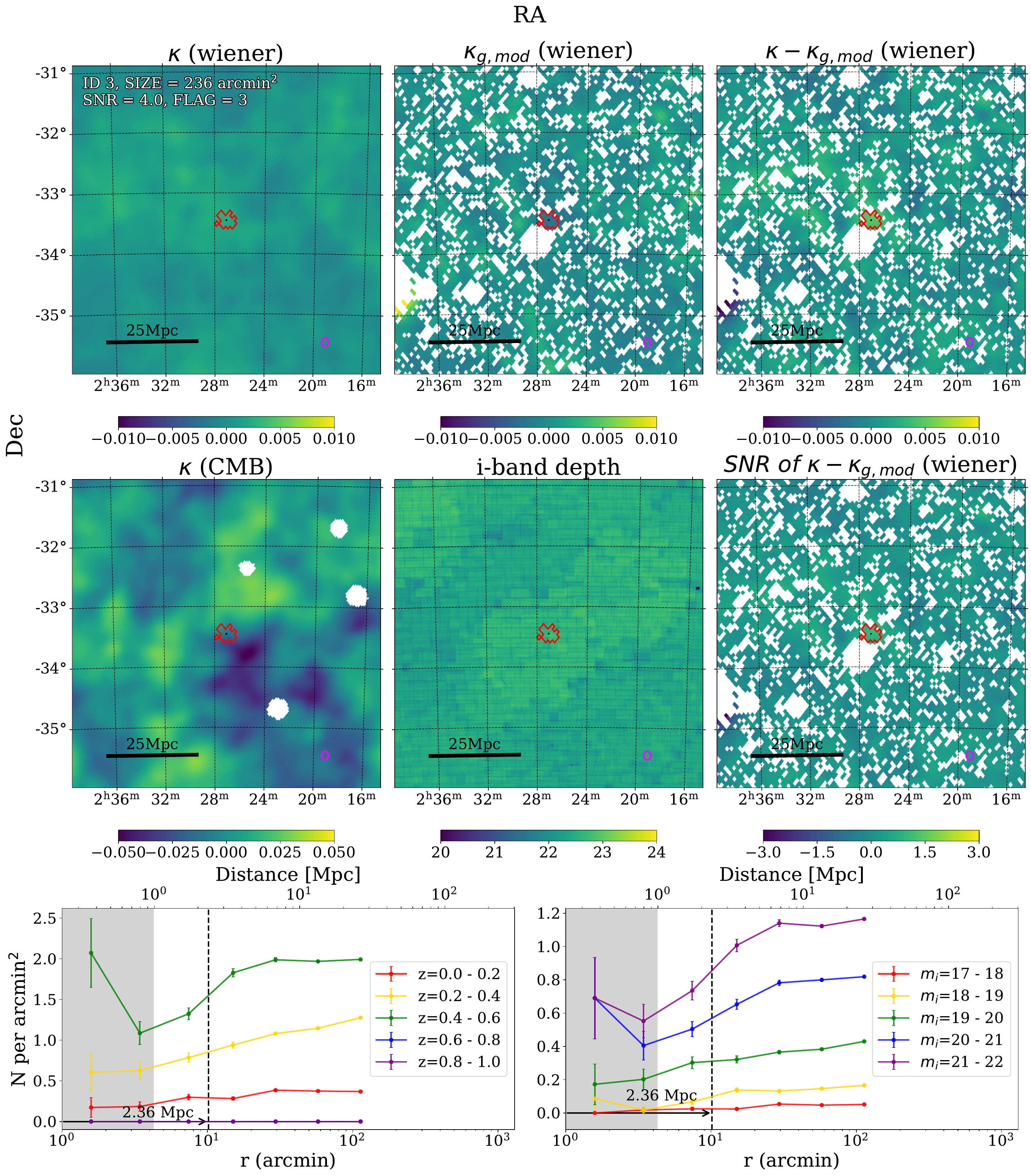}
\figsetgrpnote{The 5$^\circ$$\times$5$^\circ$ cutout maps of the dark structure candidate ID 3. The red lines outline the the boundary of the candidate in each map. The small magenta circle in the bottom-right corner of each panel represents the 10 arcmin smoothing scale, while the small black dot at the center of each panel marks the centroid of the candidate region. (top-left) Wiener convergence map. (top-center) Galaxy convergence map scaled to the Wiener weak lensing convergence. (top-right) Residual map obtained by subtracting the scaled galaxy convergence from the Wiener convergence. (middle-left) CMB lensing convergence map. (middle-center) Observation depth map for i-band. (middle-right) S/N map of the residual. (bottom-left) Radial galaxy surface number density profile around candidate ID 3, shown as a function of redshift bins and (bottom-right) the same profile shown as a function of magnitude bins.}
\figsetgrpend

\figsetgrpstart
\figsetgrpnum{8.4}
\figsetgrptitle{Candidate ID 4}
\figsetplot{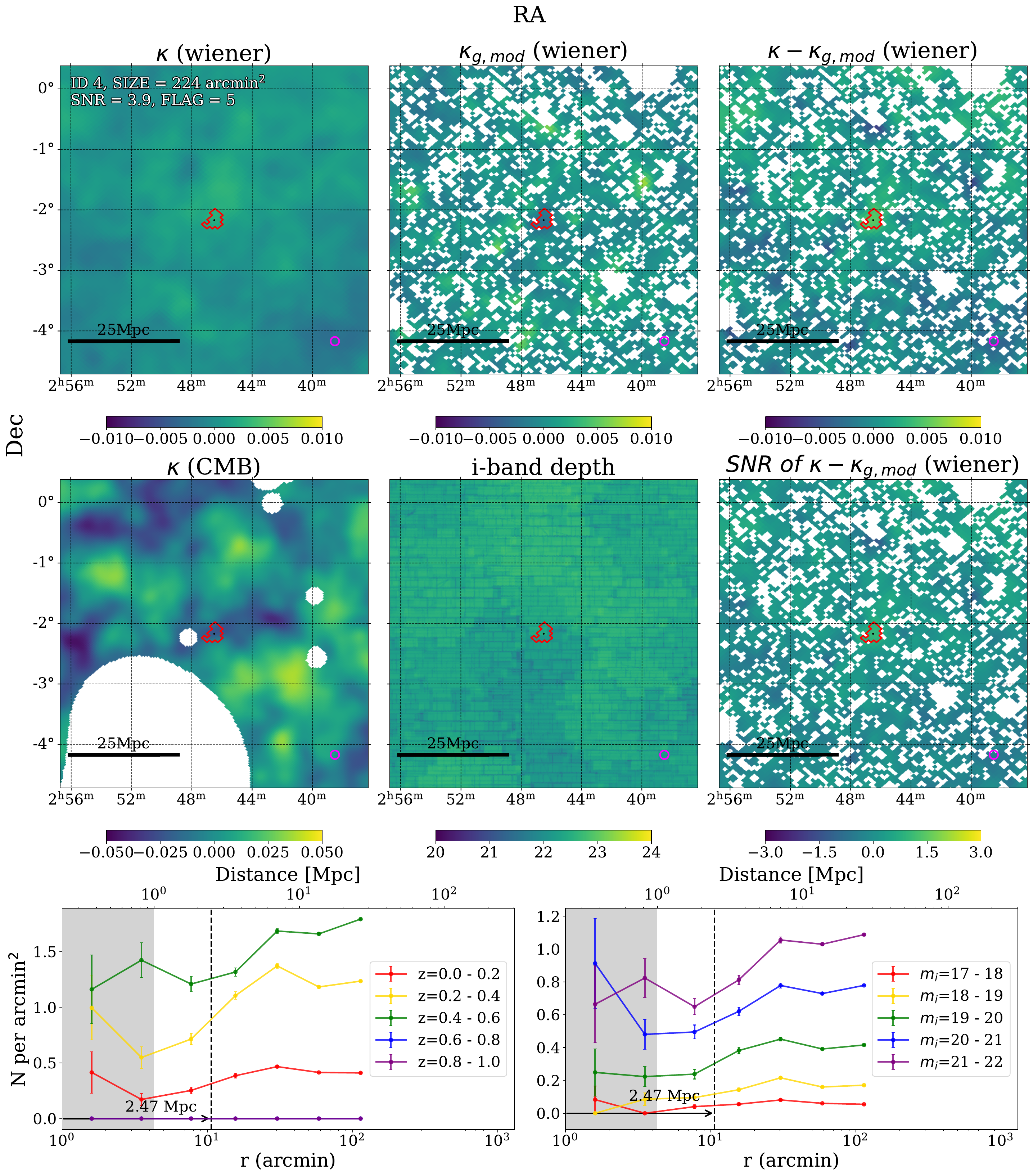}
\figsetgrpnote{The 5$^\circ$$\times$5$^\circ$ cutout maps of the dark structure candidate ID 4. The red lines outline the the boundary of the candidate in each map. The small magenta circle in the bottom-right corner of each panel represents the 10 arcmin smoothing scale, while the small black dot at the center of each panel marks the centroid of the candidate region. (top-left) Wiener convergence map. (top-center) Galaxy convergence map scaled to the Wiener weak lensing convergence. (top-right) Residual map obtained by subtracting the scaled galaxy convergence from the Wiener convergence. (middle-left) CMB lensing convergence map. (middle-center) Observation depth map for i-band. (middle-right) S/N map of the residual. (bottom-left) Radial galaxy surface number density profile around candidate ID 4, shown as a function of redshift bins and (bottom-right) the same profile shown as a function of magnitude bins.}
\figsetgrpend

\figsetgrpstart
\figsetgrpnum{8.5}
\figsetgrptitle{Candidate ID 5}
\figsetplot{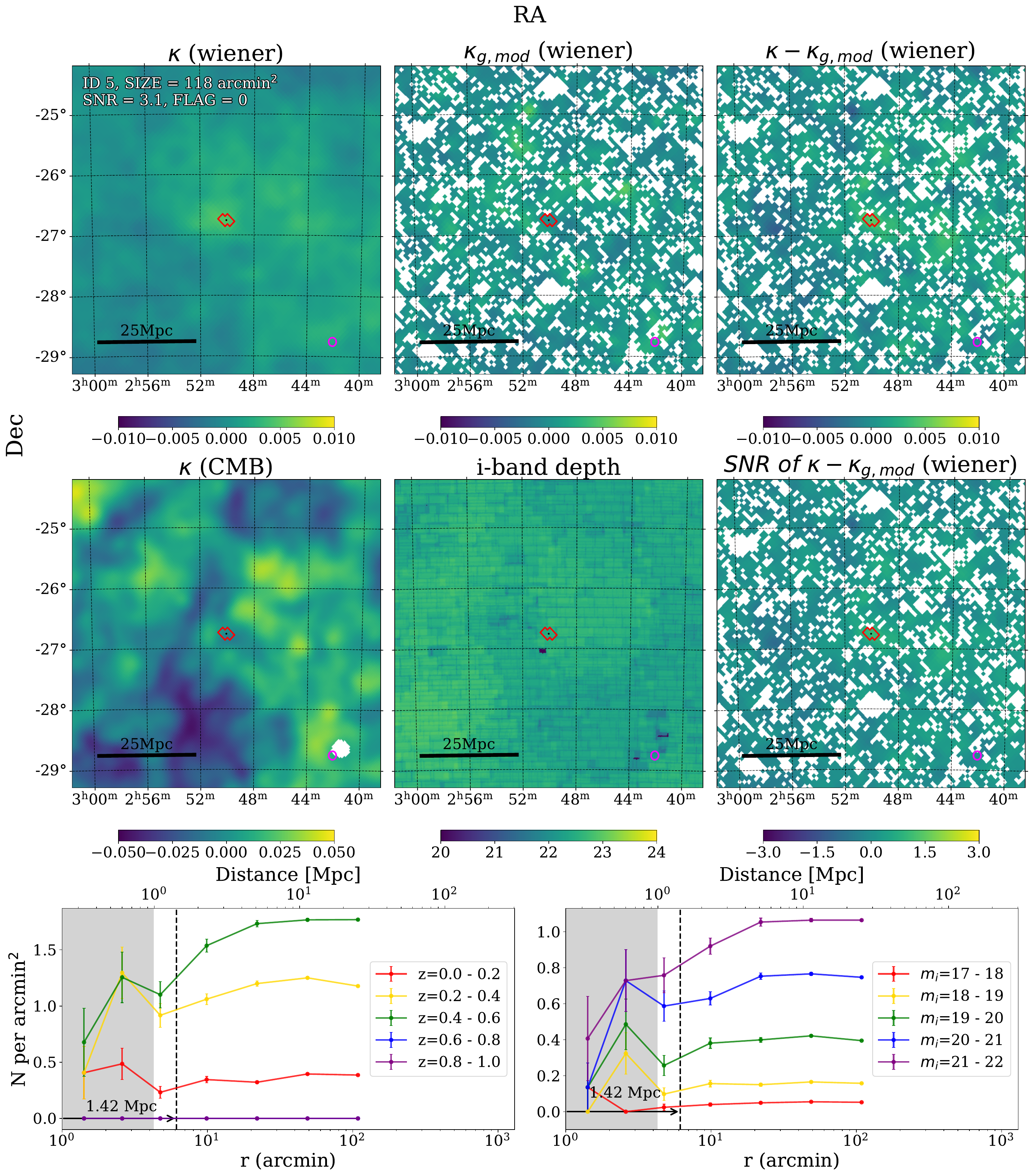}
\figsetgrpnote{The 5$^\circ$$\times$5$^\circ$ cutout maps of the dark structure candidate ID 5. The red lines outline the the boundary of the candidate in each map. The small magenta circle in the bottom-right corner of each panel represents the 10 arcmin smoothing scale, while the small black dot at the center of each panel marks the centroid of the candidate region. (top-left) Wiener convergence map. (top-center) Galaxy convergence map scaled to the Wiener weak lensing convergence. (top-right) Residual map obtained by subtracting the scaled galaxy convergence from the Wiener convergence. (middle-left) CMB lensing convergence map. (middle-center) Observation depth map for i-band. (middle-right) S/N map of the residual. (bottom-left) Radial galaxy surface number density profile around candidate ID 5, shown as a function of redshift bins and (bottom-right) the same profile shown as a function of magnitude bins.}
\figsetgrpend

\figsetgrpstart
\figsetgrpnum{8.6}
\figsetgrptitle{Candidate ID 6}
\figsetplot{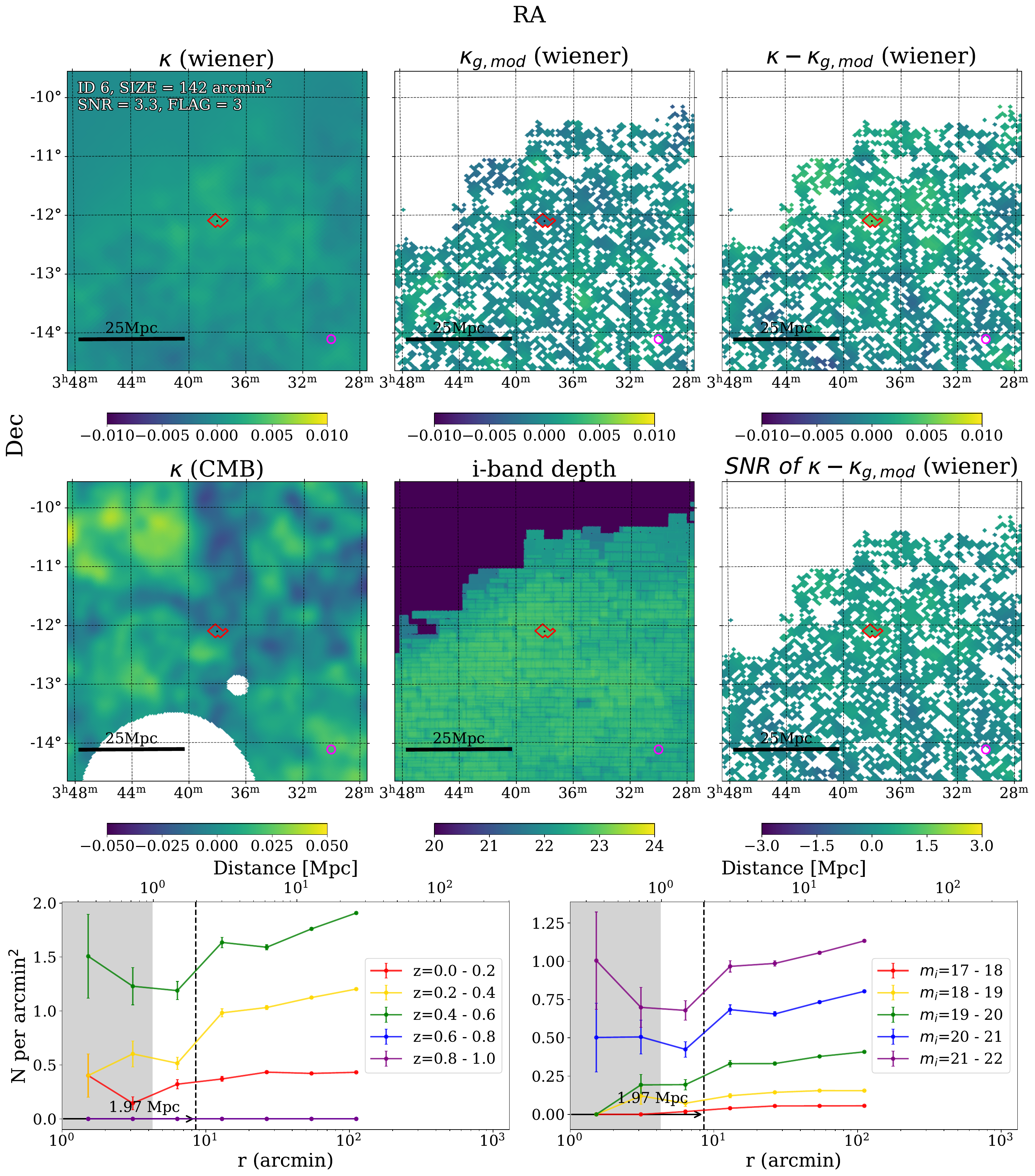}
\figsetgrpnote{The 5$^\circ$$\times$5$^\circ$ cutout maps of the dark structure candidate ID 6. The red lines outline the the boundary of the candidate in each map. The small magenta circle in the bottom-right corner of each panel represents the 10 arcmin smoothing scale, while the small black dot at the center of each panel marks the centroid of the candidate region. (top-left) Wiener convergence map. (top-center) Galaxy convergence map scaled to the Wiener weak lensing convergence. (top-right) Residual map obtained by subtracting the scaled galaxy convergence from the Wiener convergence. (middle-left) CMB lensing convergence map. (middle-center) Observation depth map for i-band. (middle-right) S/N map of the residual. (bottom-left) Radial galaxy surface number density profile around candidate ID 6, shown as a function of redshift bins and (bottom-right) the same profile shown as a function of magnitude bins.}
\figsetgrpend

\figsetgrpstart
\figsetgrpnum{8.7}
\figsetgrptitle{Candidate ID 7}
\figsetplot{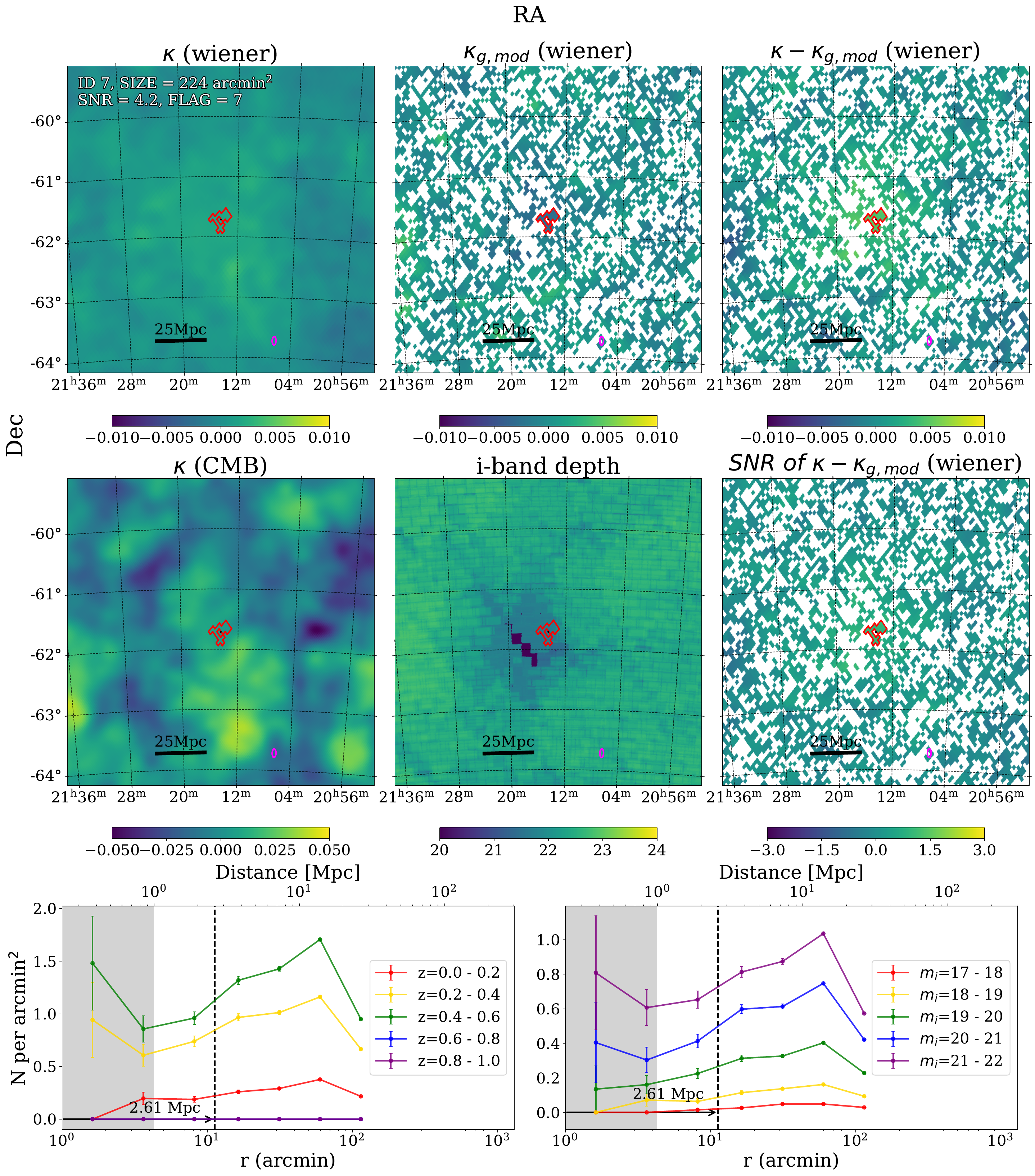}
\figsetgrpnote{The 5$^\circ$$\times$5$^\circ$ cutout maps of the dark structure candidate ID 7. The red lines outline the the boundary of the candidate in each map. The small magenta circle in the bottom-right corner of each panel represents the 10 arcmin smoothing scale, while the small black dot at the center of each panel marks the centroid of the candidate region. (top-left) Wiener convergence map. (top-center) Galaxy convergence map scaled to the Wiener weak lensing convergence. (top-right) Residual map obtained by subtracting the scaled galaxy convergence from the Wiener convergence. (middle-left) CMB lensing convergence map. (middle-center) Observation depth map for i-band. (middle-right) S/N map of the residual. (bottom-left) Radial galaxy surface number density profile around candidate ID 7, shown as a function of redshift bins and (bottom-right) the same profile shown as a function of magnitude bins.}
\figsetgrpend

\figsetgrpstart
\figsetgrpnum{8.8}
\figsetgrptitle{Candidate ID 8}
\figsetplot{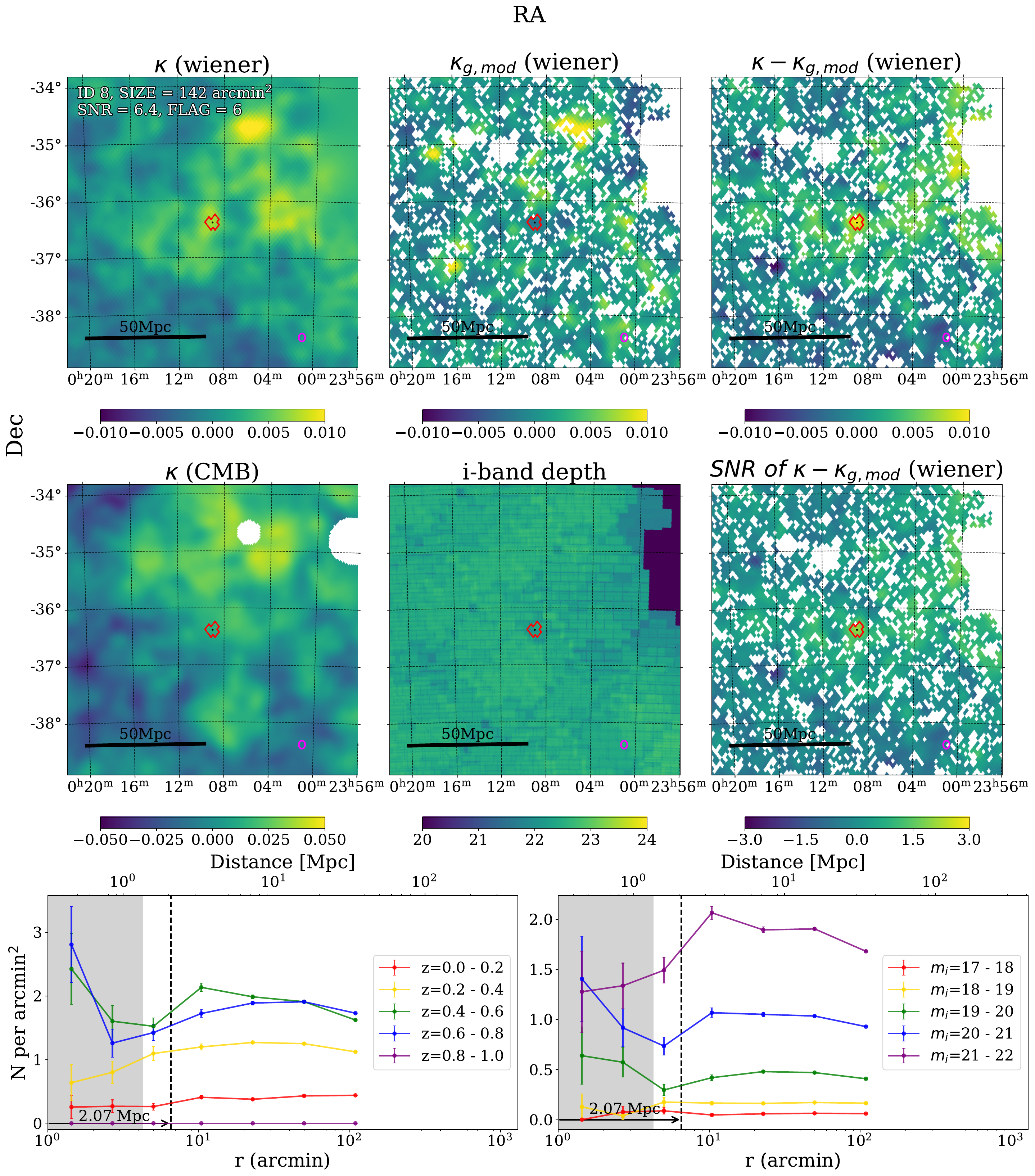}
\figsetgrpnote{The 5$^\circ$$\times$5$^\circ$ cutout maps of the dark structure candidate ID 8. The red lines outline the the boundary of the candidate in each map. The small magenta circle in the bottom-right corner of each panel represents the 10 arcmin smoothing scale, while the small black dot at the center of each panel marks the centroid of the candidate region. (top-left) Wiener convergence map. (top-center) Galaxy convergence map scaled to the Wiener weak lensing convergence. (top-right) Residual map obtained by subtracting the scaled galaxy convergence from the Wiener convergence. (middle-left) CMB lensing convergence map. (middle-center) Observation depth map for i-band. (middle-right) S/N map of the residual. (bottom-left) Radial galaxy surface number density profile around candidate ID 8, shown as a function of redshift bins and (bottom-right) the same profile shown as a function of magnitude bins.}
\figsetgrpend

\figsetgrpstart
\figsetgrpnum{8.9}
\figsetgrptitle{Candidate ID 9}
\figsetplot{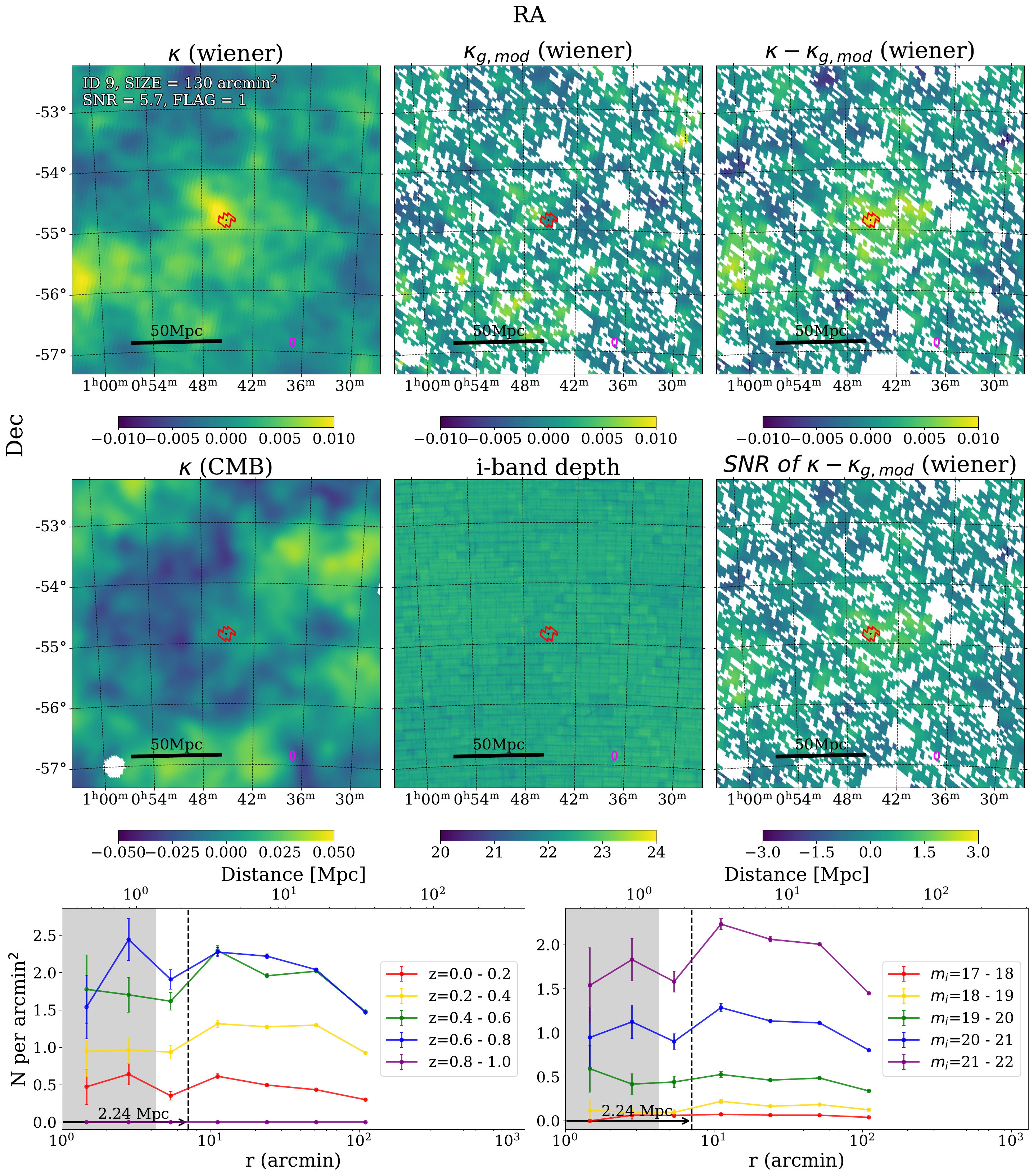}
\figsetgrpnote{The 5$^\circ$$\times$5$^\circ$ cutout maps of the dark structure candidate ID 9. The red lines outline the the boundary of the candidate in each map. The small magenta circle in the bottom-right corner of each panel represents the 10 arcmin smoothing scale, while the small black dot at the center of each panel marks the centroid of the candidate region. (top-left) Wiener convergence map. (top-center) Galaxy convergence map scaled to the Wiener weak lensing convergence. (top-right) Residual map obtained by subtracting the scaled galaxy convergence from the Wiener convergence. (middle-left) CMB lensing convergence map. (middle-center) Observation depth map for i-band. (middle-right) S/N map of the residual. (bottom-left) Radial galaxy surface number density profile around candidate ID 9, shown as a function of redshift bins and (bottom-right) the same profile shown as a function of magnitude bins.}
\figsetgrpend

\figsetgrpstart
\figsetgrpnum{8.10}
\figsetgrptitle{Candidate ID 10}
\figsetplot{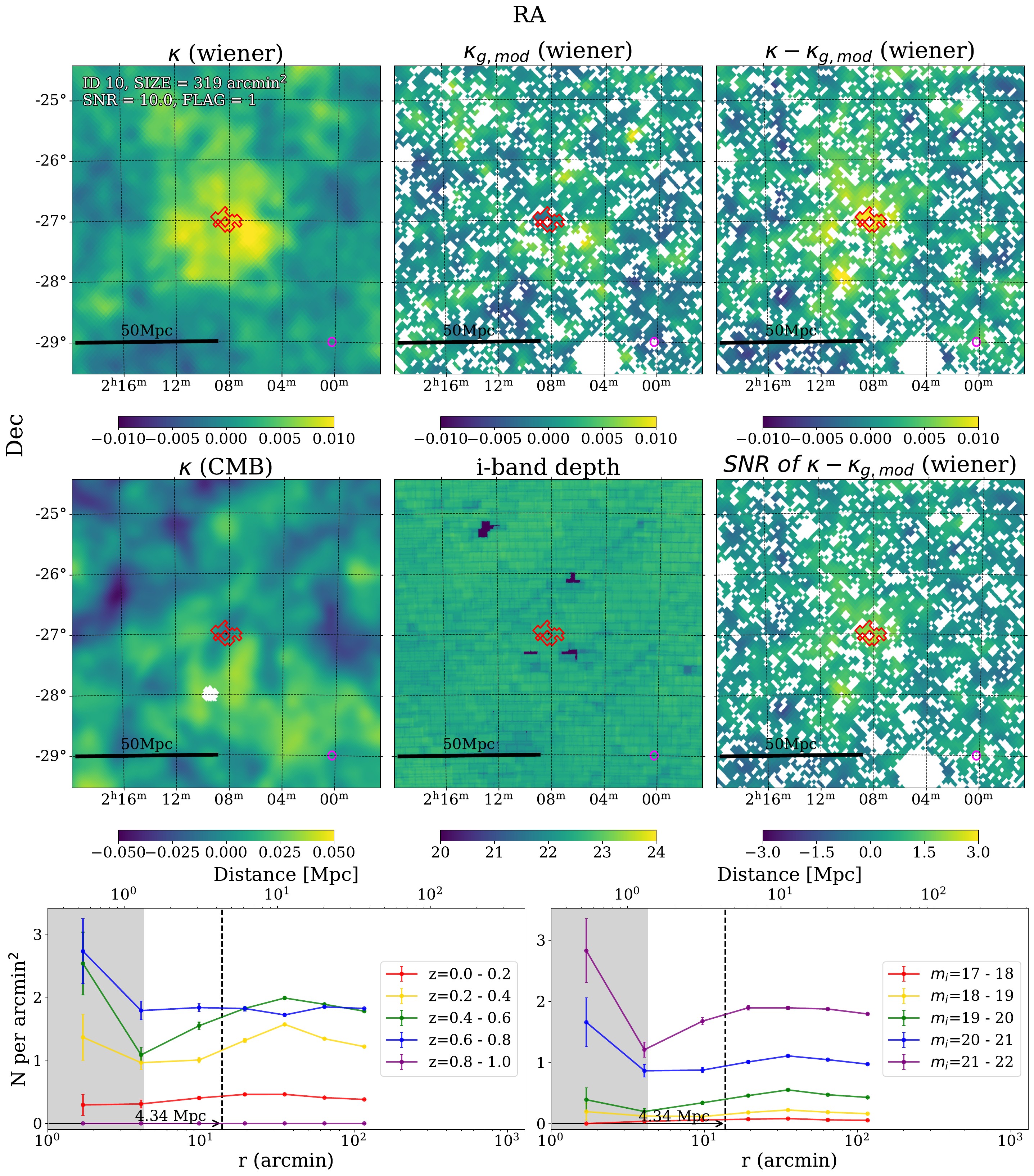}
\figsetgrpnote{The 5$^\circ$$\times$5$^\circ$ cutout maps of the dark structure candidate ID 10. The red lines outline the the boundary of the candidate in each map. The small magenta circle in the bottom-right corner of each panel represents the 10 arcmin smoothing scale, while the small black dot at the center of each panel marks the centroid of the candidate region. (top-left) Wiener convergence map. (top-center) Galaxy convergence map scaled to the Wiener weak lensing convergence. (top-right) Residual map obtained by subtracting the scaled galaxy convergence from the Wiener convergence. (middle-left) CMB lensing convergence map. (middle-center) Observation depth map for i-band. (middle-right) S/N map of the residual. (bottom-left) Radial galaxy surface number density profile around candidate ID 10, shown as a function of redshift bins and (bottom-right) the same profile shown as a function of magnitude bins.}
\figsetgrpend

\figsetgrpstart
\figsetgrpnum{8.11}
\figsetgrptitle{Candidate ID 11}
\figsetplot{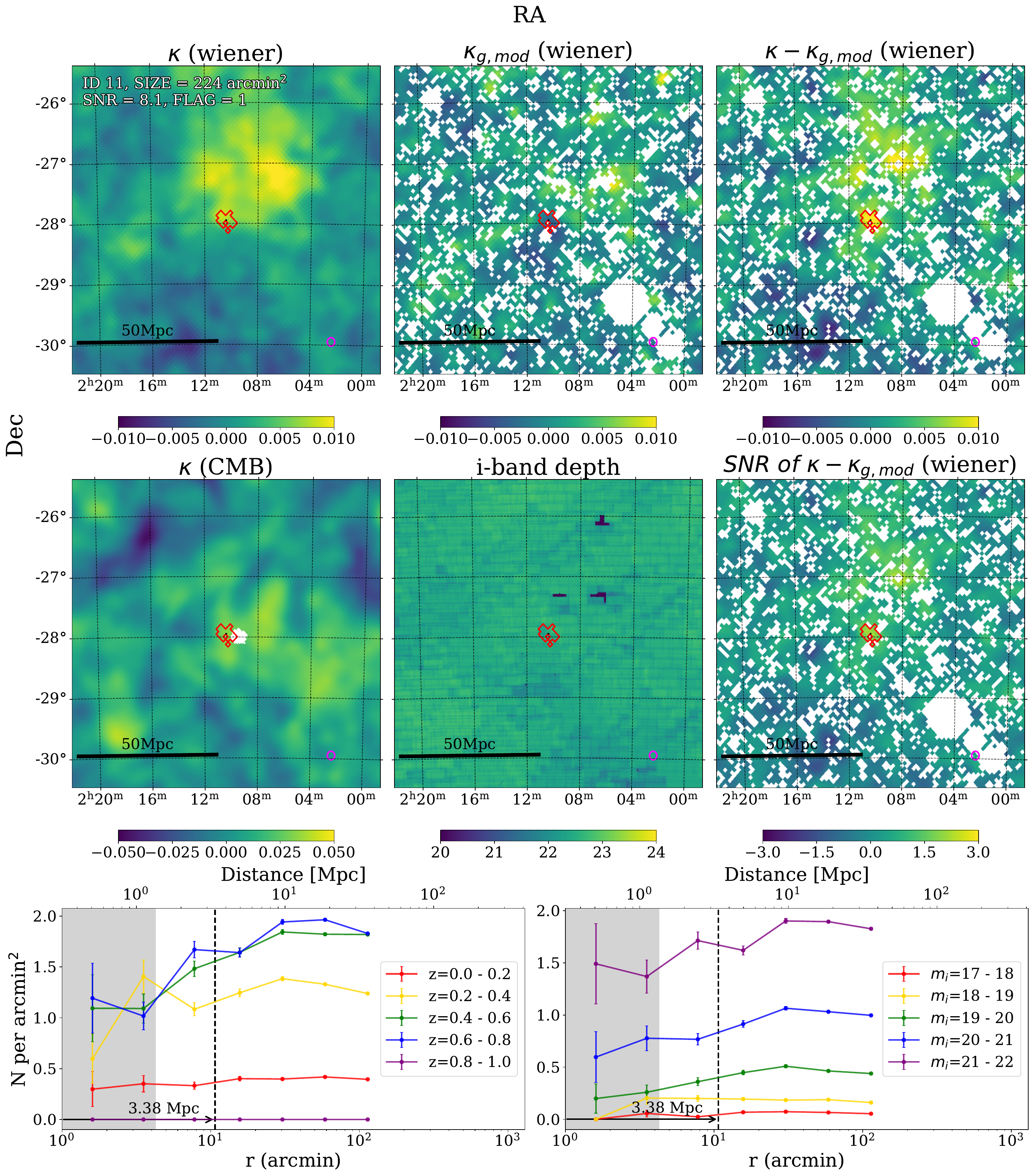}
\figsetgrpnote{The 5$^\circ$$\times$5$^\circ$ cutout maps of the dark structure candidate ID 11. The red lines outline the the boundary of the candidate in each map. The small magenta circle in the bottom-right corner of each panel represents the 10 arcmin smoothing scale, while the small black dot at the center of each panel marks the centroid of the candidate region. (top-left) Wiener convergence map. (top-center) Galaxy convergence map scaled to the Wiener weak lensing convergence. (top-right) Residual map obtained by subtracting the scaled galaxy convergence from the Wiener convergence. (middle-left) CMB lensing convergence map. (middle-center) Observation depth map for i-band. (middle-right) S/N map of the residual. (bottom-left) Radial galaxy surface number density profile around candidate ID 11, shown as a function of redshift bins and (bottom-right) the same profile shown as a function of magnitude bins.}
\figsetgrpend

\figsetgrpstart
\figsetgrpnum{8.12}
\figsetgrptitle{Candidate ID 12}
\figsetplot{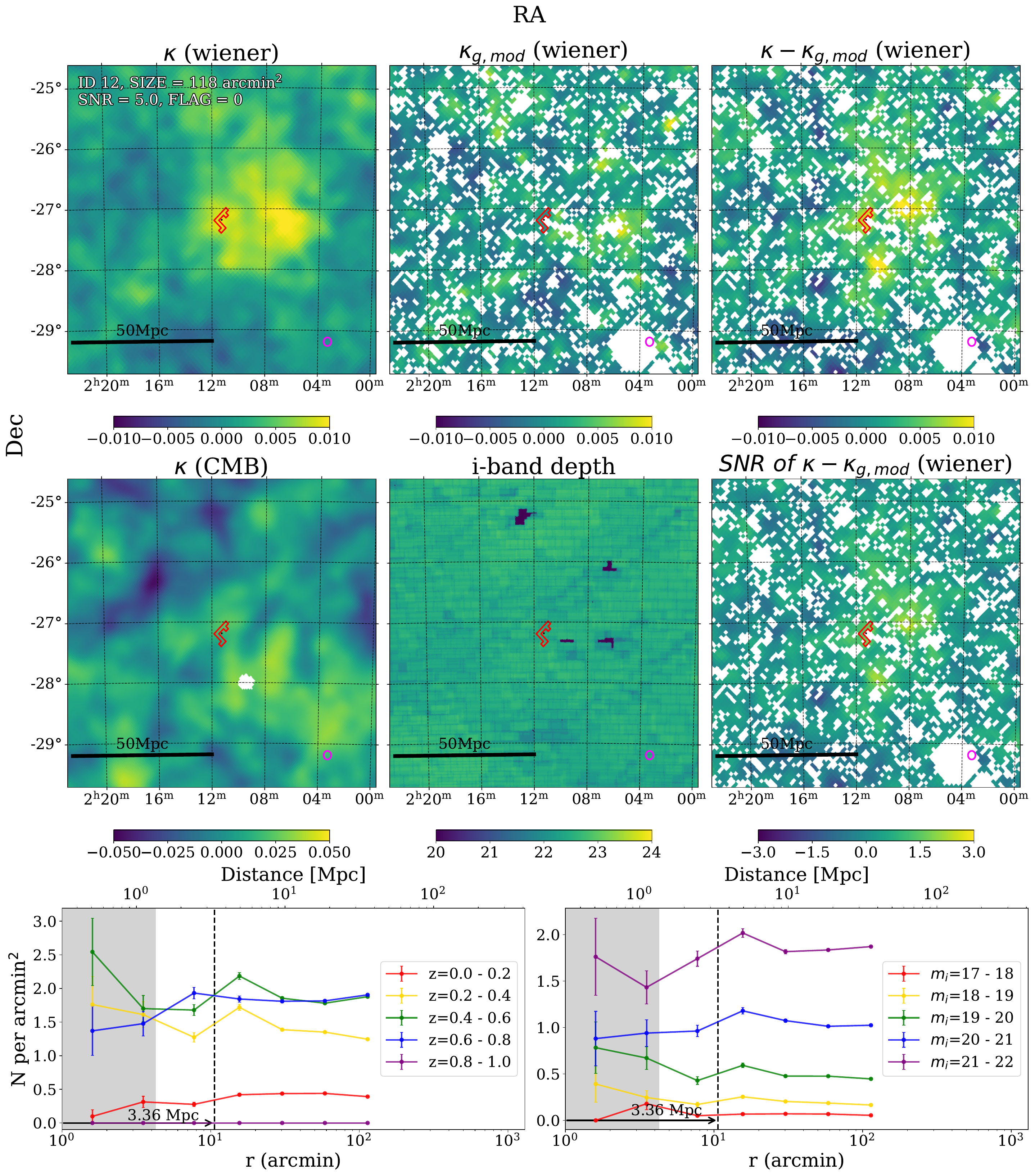}
\figsetgrpnote{The 5$^\circ$$\times$5$^\circ$ cutout maps of the dark structure candidate ID 12. The red lines outline the the boundary of the candidate in each map. The small magenta circle in the bottom-right corner of each panel represents the 10 arcmin smoothing scale, while the small black dot at the center of each panel marks the centroid of the candidate region. (top-left) Wiener convergence map. (top-center) Galaxy convergence map scaled to the Wiener weak lensing convergence. (top-right) Residual map obtained by subtracting the scaled galaxy convergence from the Wiener convergence. (middle-left) CMB lensing convergence map. (middle-center) Observation depth map for i-band. (middle-right) S/N map of the residual. (bottom-left) Radial galaxy surface number density profile around candidate ID 12, shown as a function of redshift bins and (bottom-right) the same profile shown as a function of magnitude bins.}
\figsetgrpend

\figsetgrpstart
\figsetgrpnum{8.13}
\figsetgrptitle{Candidate ID 13}
\figsetplot{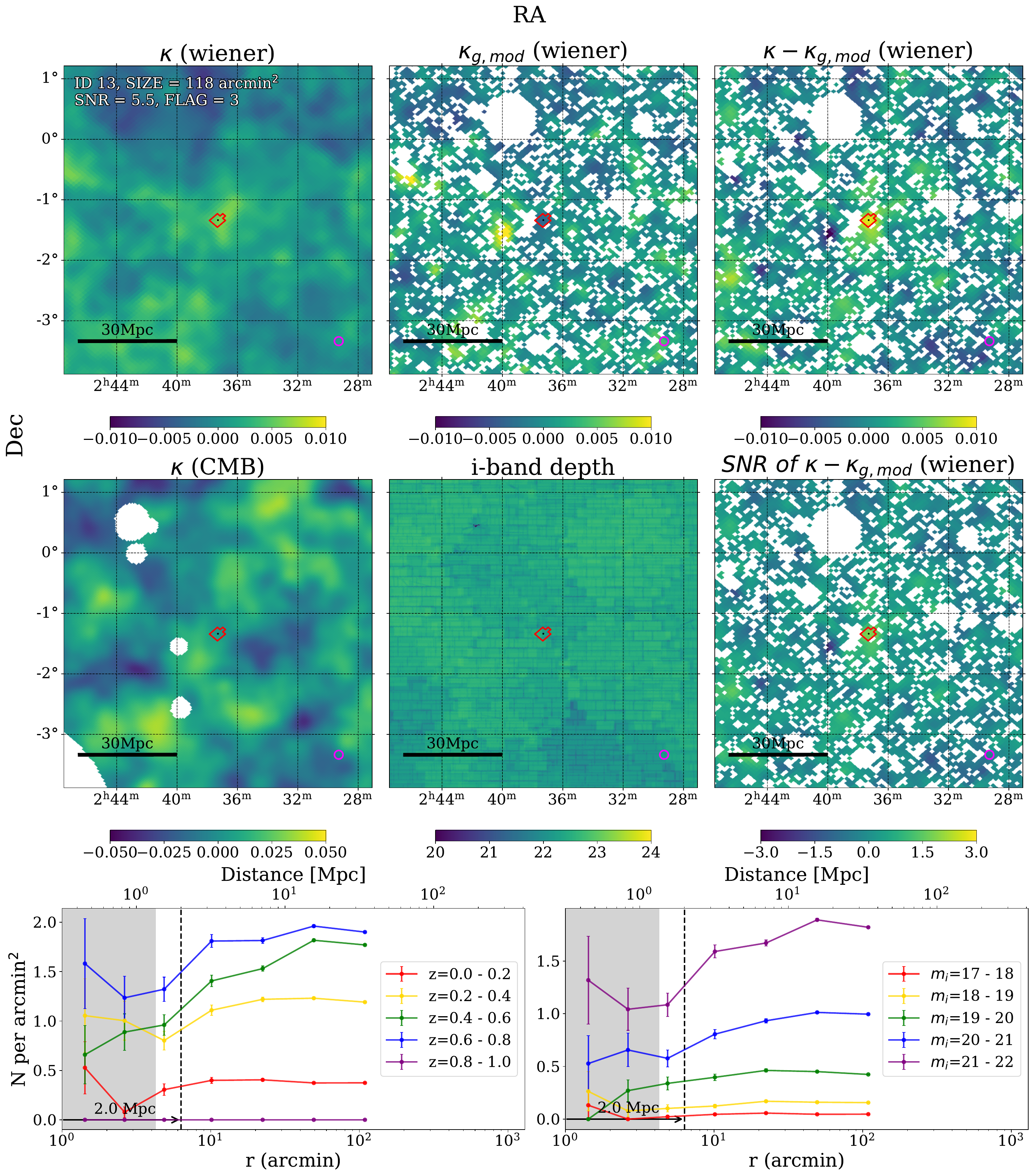}
\figsetgrpnote{The 5$^\circ$$\times$5$^\circ$ cutout maps of the dark structure candidate ID 13. The red lines outline the the boundary of the candidate in each map. The small magenta circle in the bottom-right corner of each panel represents the 10 arcmin smoothing scale, while the small black dot at the center of each panel marks the centroid of the candidate region. (top-left) Wiener convergence map. (top-center) Galaxy convergence map scaled to the Wiener weak lensing convergence. (top-right) Residual map obtained by subtracting the scaled galaxy convergence from the Wiener convergence. (middle-left) CMB lensing convergence map. (middle-center) Observation depth map for i-band. (middle-right) S/N map of the residual. (bottom-left) Radial galaxy surface number density profile around candidate ID 13, shown as a function of redshift bins and (bottom-right) the same profile shown as a function of magnitude bins.}
\figsetgrpend

\figsetgrpstart
\figsetgrpnum{8.14}
\figsetgrptitle{Candidate ID 14}
\figsetplot{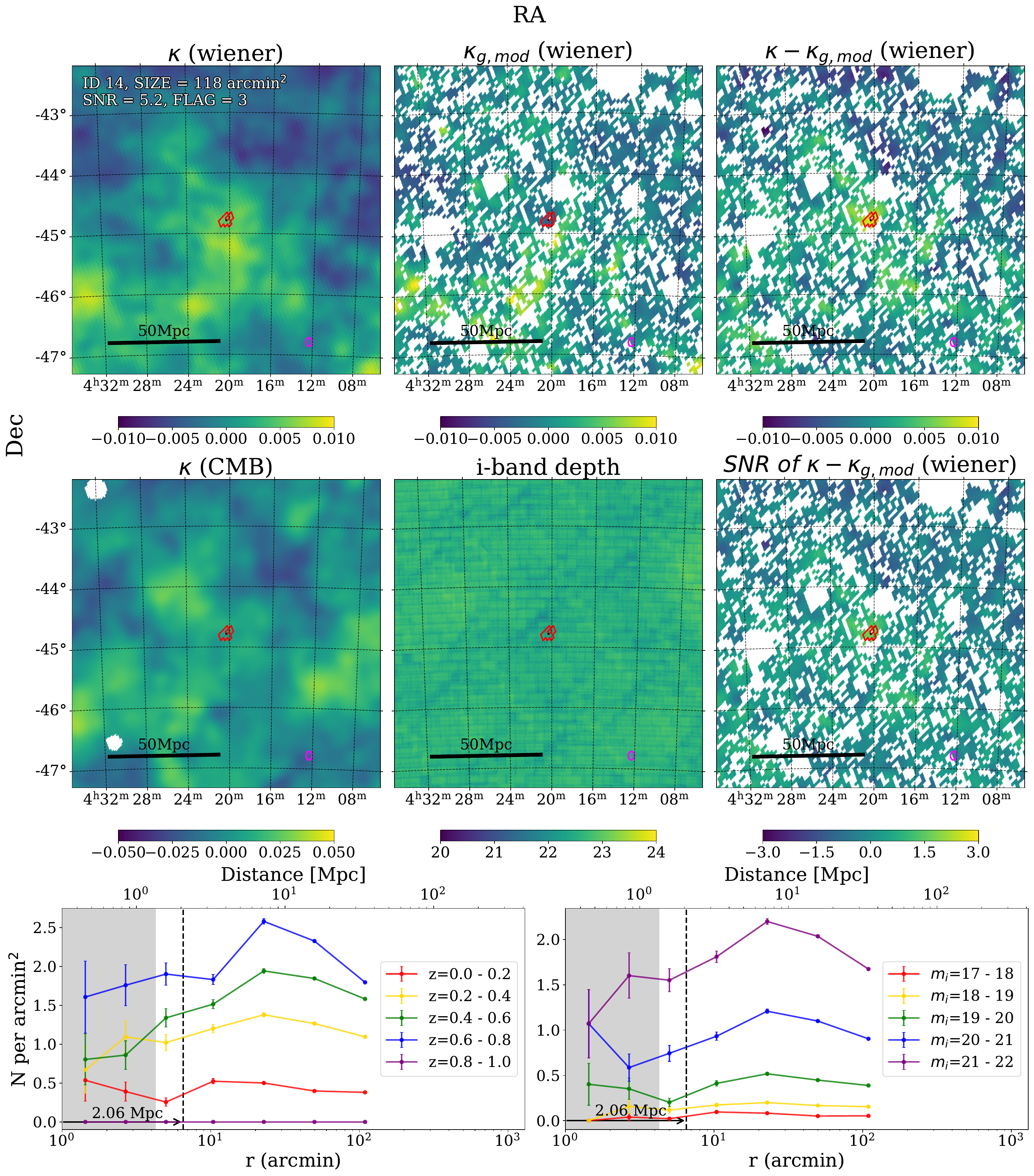}
\figsetgrpnote{The 5$^\circ$$\times$5$^\circ$ cutout maps of the dark structure candidate ID 14. The red lines outline the the boundary of the candidate in each map. The small magenta circle in the bottom-right corner of each panel represents the 10 arcmin smoothing scale, while the small black dot at the center of each panel marks the centroid of the candidate region. (top-left) Wiener convergence map. (top-center) Galaxy convergence map scaled to the Wiener weak lensing convergence. (top-right) Residual map obtained by subtracting the scaled galaxy convergence from the Wiener convergence. (middle-left) CMB lensing convergence map. (middle-center) Observation depth map for i-band. (middle-right) S/N map of the residual. (bottom-left) Radial galaxy surface number density profile around candidate ID 14, shown as a function of redshift bins and (bottom-right) the same profile shown as a function of magnitude bins.}
\figsetgrpend

\figsetgrpstart
\figsetgrpnum{8.15}
\figsetgrptitle{Candidate ID 15}
\figsetplot{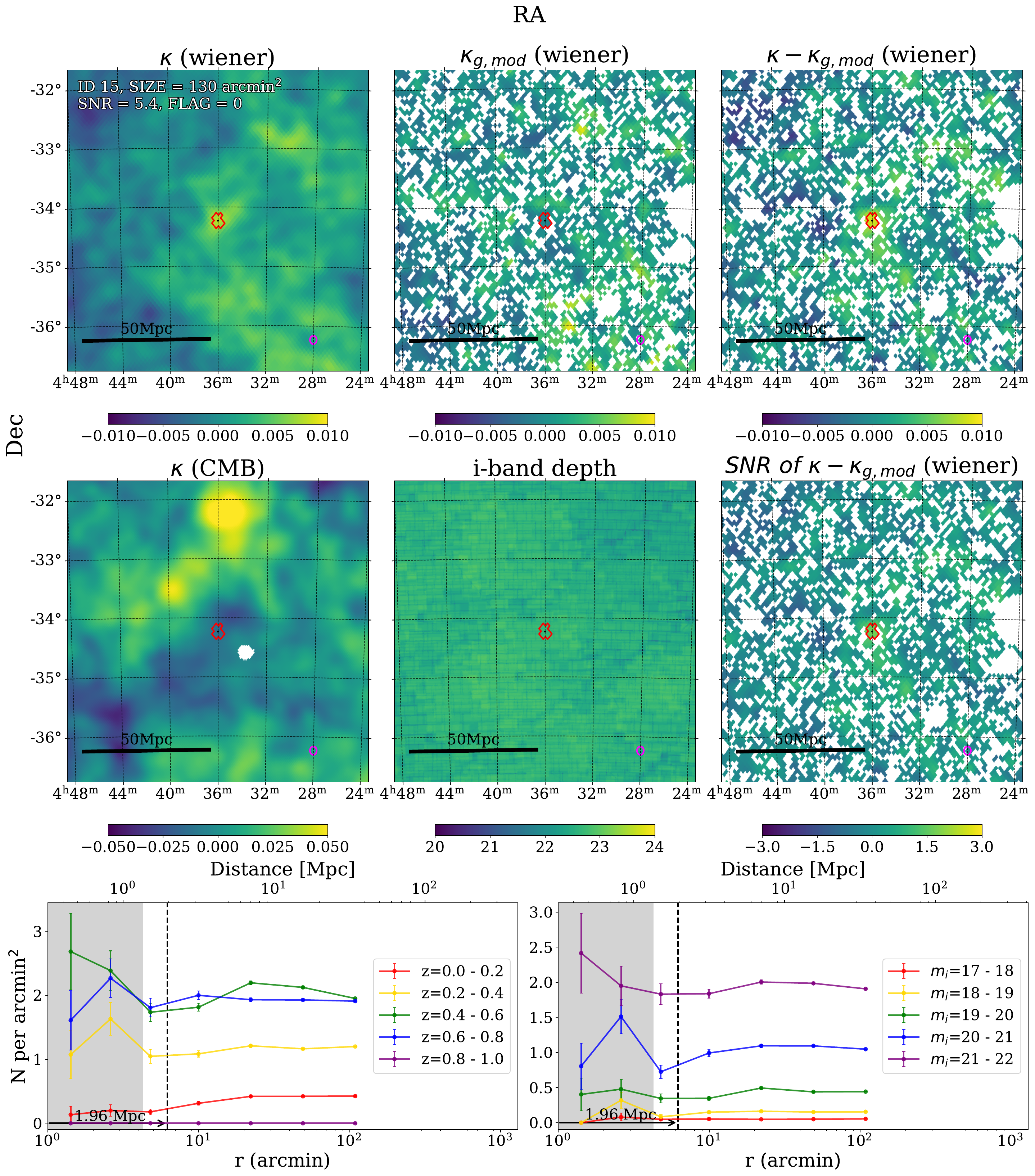}
\figsetgrpnote{The 5$^\circ$$\times$5$^\circ$ cutout maps of the dark structure candidate ID 15. The red lines outline the the boundary of the candidate in each map. The small magenta circle in the bottom-right corner of each panel represents the 10 arcmin smoothing scale, while the small black dot at the center of each panel marks the centroid of the candidate region. (top-left) Wiener convergence map. (top-center) Galaxy convergence map scaled to the Wiener weak lensing convergence. (top-right) Residual map obtained by subtracting the scaled galaxy convergence from the Wiener convergence. (middle-left) CMB lensing convergence map. (middle-center) Observation depth map for i-band. (middle-right) S/N map of the residual. (bottom-left) Radial galaxy surface number density profile around candidate ID 15, shown as a function of redshift bins and (bottom-right) the same profile shown as a function of magnitude bins.}
\figsetgrpend

\figsetgrpstart
\figsetgrpnum{8.16}
\figsetgrptitle{Candidate ID 16}
\figsetplot{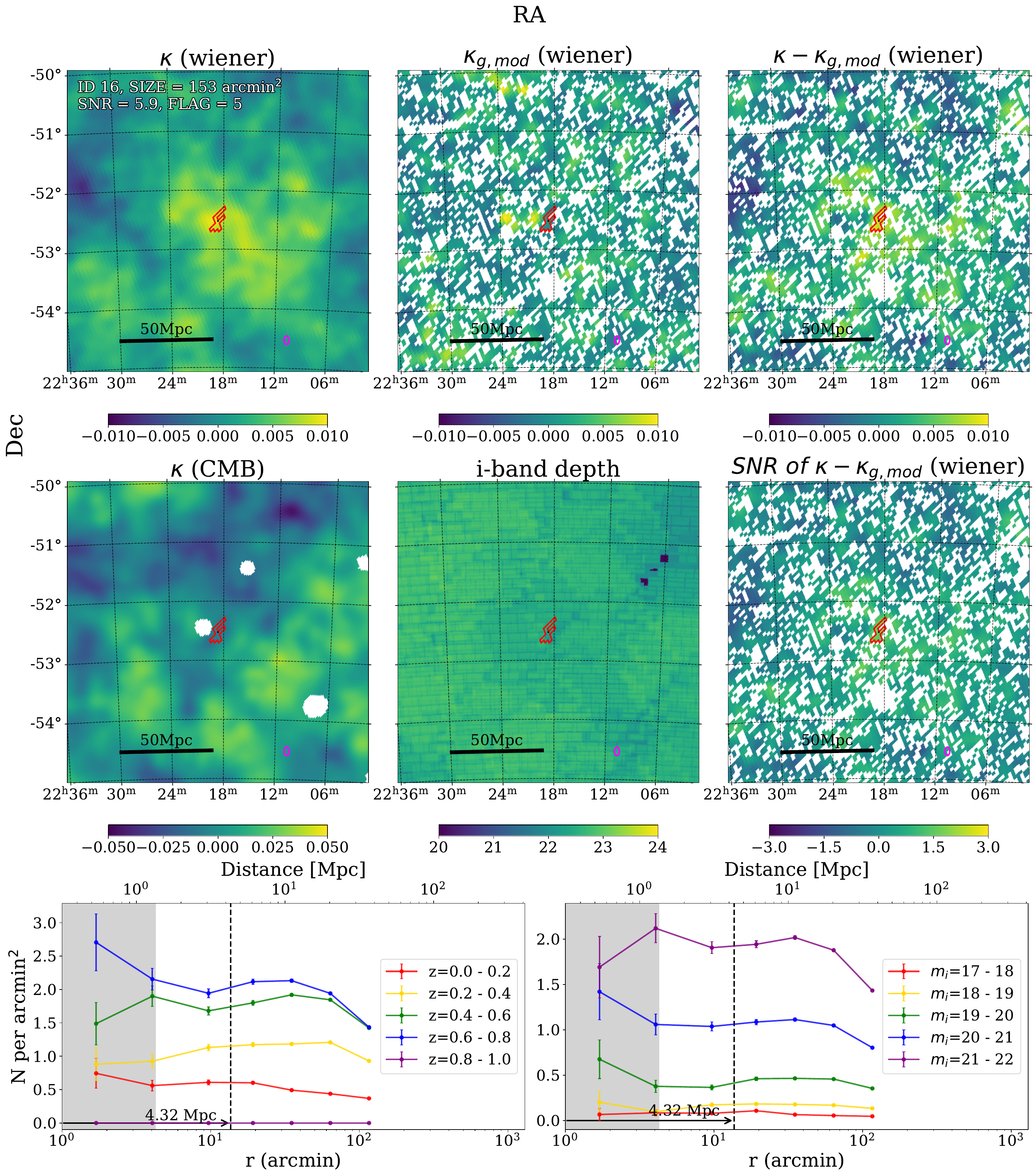}
\figsetgrpnote{The 5$^\circ$$\times$5$^\circ$ cutout maps of the dark structure candidate ID 16. The red lines outline the the boundary of the candidate in each map. The small magenta circle in the bottom-right corner of each panel represents the 10 arcmin smoothing scale, while the small black dot at the center of each panel marks the centroid of the candidate region. (top-left) Wiener convergence map. (top-center) Galaxy convergence map scaled to the Wiener weak lensing convergence. (top-right) Residual map obtained by subtracting the scaled galaxy convergence from the Wiener convergence. (middle-left) CMB lensing convergence map. (middle-center) Observation depth map for i-band. (middle-right) S/N map of the residual. (bottom-left) Radial galaxy surface number density profile around candidate ID 16, shown as a function of redshift bins and (bottom-right) the same profile shown as a function of magnitude bins.}
\figsetgrpend

\figsetgrpstart
\figsetgrpnum{8.17}
\figsetgrptitle{Candidate ID 17}
\figsetplot{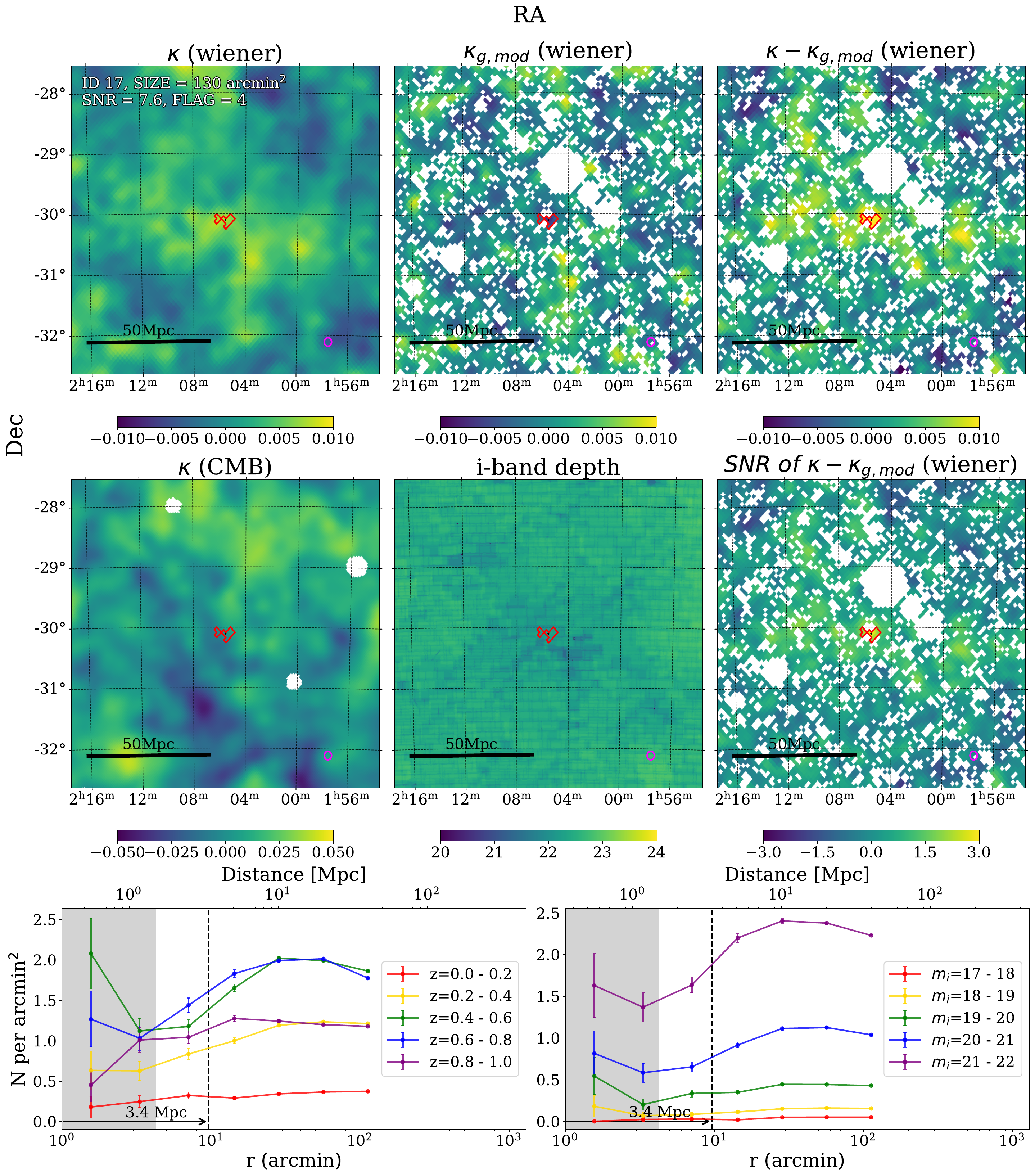}
\figsetgrpnote{The 5$^\circ$$\times$5$^\circ$ cutout maps of the dark structure candidate ID 17. The red lines outline the the boundary of the candidate in each map. The small magenta circle in the bottom-right corner of each panel represents the 10 arcmin smoothing scale, while the small black dot at the center of each panel marks the centroid of the candidate region. (top-left) Wiener convergence map. (top-center) Galaxy convergence map scaled to the Wiener weak lensing convergence. (top-right) Residual map obtained by subtracting the scaled galaxy convergence from the Wiener convergence. (middle-left) CMB lensing convergence map. (middle-center) Observation depth map for i-band. (middle-right) S/N map of the residual. (bottom-left) Radial galaxy surface number density profile around candidate ID 17, shown as a function of redshift bins and (bottom-right) the same profile shown as a function of magnitude bins.}
\figsetgrpend

\figsetgrpstart
\figsetgrpnum{8.18}
\figsetgrptitle{Candidate ID 18}
\figsetplot{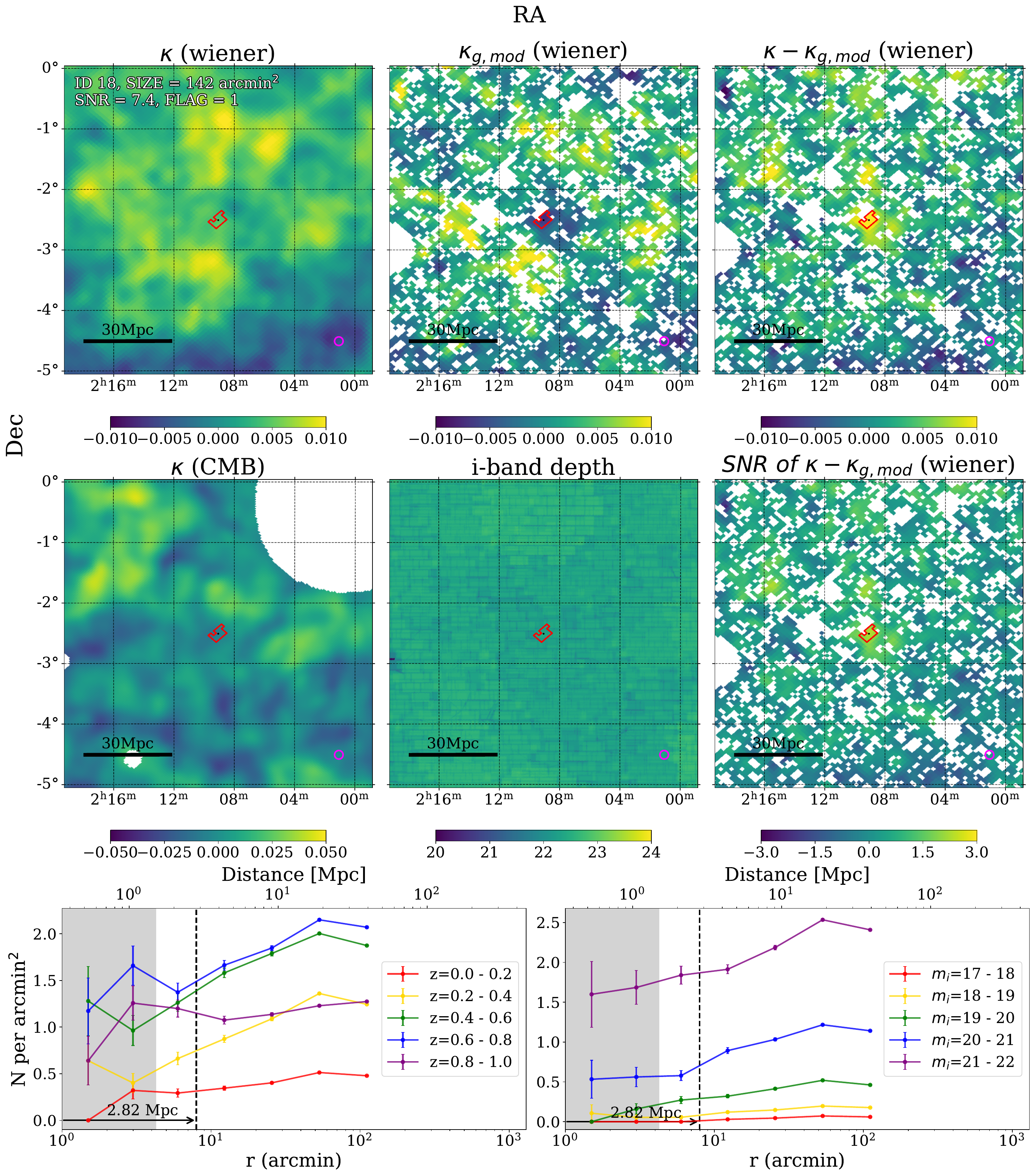}
\figsetgrpnote{The 5$^\circ$$\times$5$^\circ$ cutout maps of the dark structure candidate ID 18. The red lines outline the the boundary of the candidate in each map. The small magenta circle in the bottom-right corner of each panel represents the 10 arcmin smoothing scale, while the small black dot at the center of each panel marks the centroid of the candidate region. (top-left) Wiener convergence map. (top-center) Galaxy convergence map scaled to the Wiener weak lensing convergence. (top-right) Residual map obtained by subtracting the scaled galaxy convergence from the Wiener convergence. (middle-left) CMB lensing convergence map. (middle-center) Observation depth map for i-band. (middle-right) S/N map of the residual. (bottom-left) Radial galaxy surface number density profile around candidate ID 18, shown as a function of redshift bins and (bottom-right) the same profile shown as a function of magnitude bins.}
\figsetgrpend

\figsetgrpstart
\figsetgrpnum{8.19}
\figsetgrptitle{Candidate ID 19}
\figsetplot{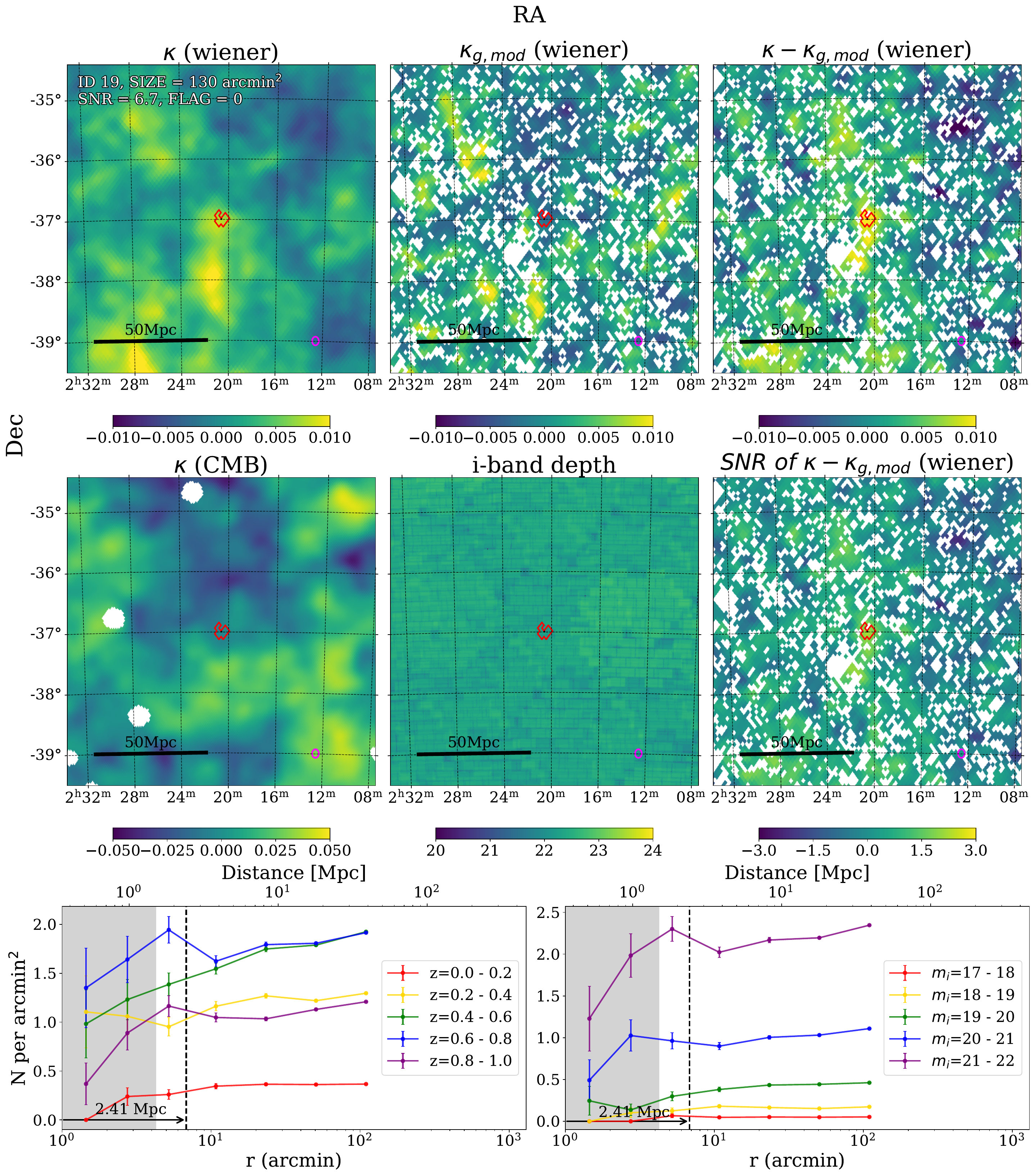}
\figsetgrpnote{The 5$^\circ$$\times$5$^\circ$ cutout maps of the dark structure candidate ID 19. The red lines outline the the boundary of the candidate in each map. The small magenta circle in the bottom-right corner of each panel represents the 10 arcmin smoothing scale, while the small black dot at the center of each panel marks the centroid of the candidate region. (top-left) Wiener convergence map. (top-center) Galaxy convergence map scaled to the Wiener weak lensing convergence. (top-right) Residual map obtained by subtracting the scaled galaxy convergence from the Wiener convergence. (middle-left) CMB lensing convergence map. (middle-center) Observation depth map for i-band. (middle-right) S/N map of the residual. (bottom-left) Radial galaxy surface number density profile around candidate ID 19, shown as a function of redshift bins and (bottom-right) the same profile shown as a function of magnitude bins.}
\figsetgrpend

\figsetgrpstart
\figsetgrpnum{8.20}
\figsetgrptitle{Candidate ID 20}
\figsetplot{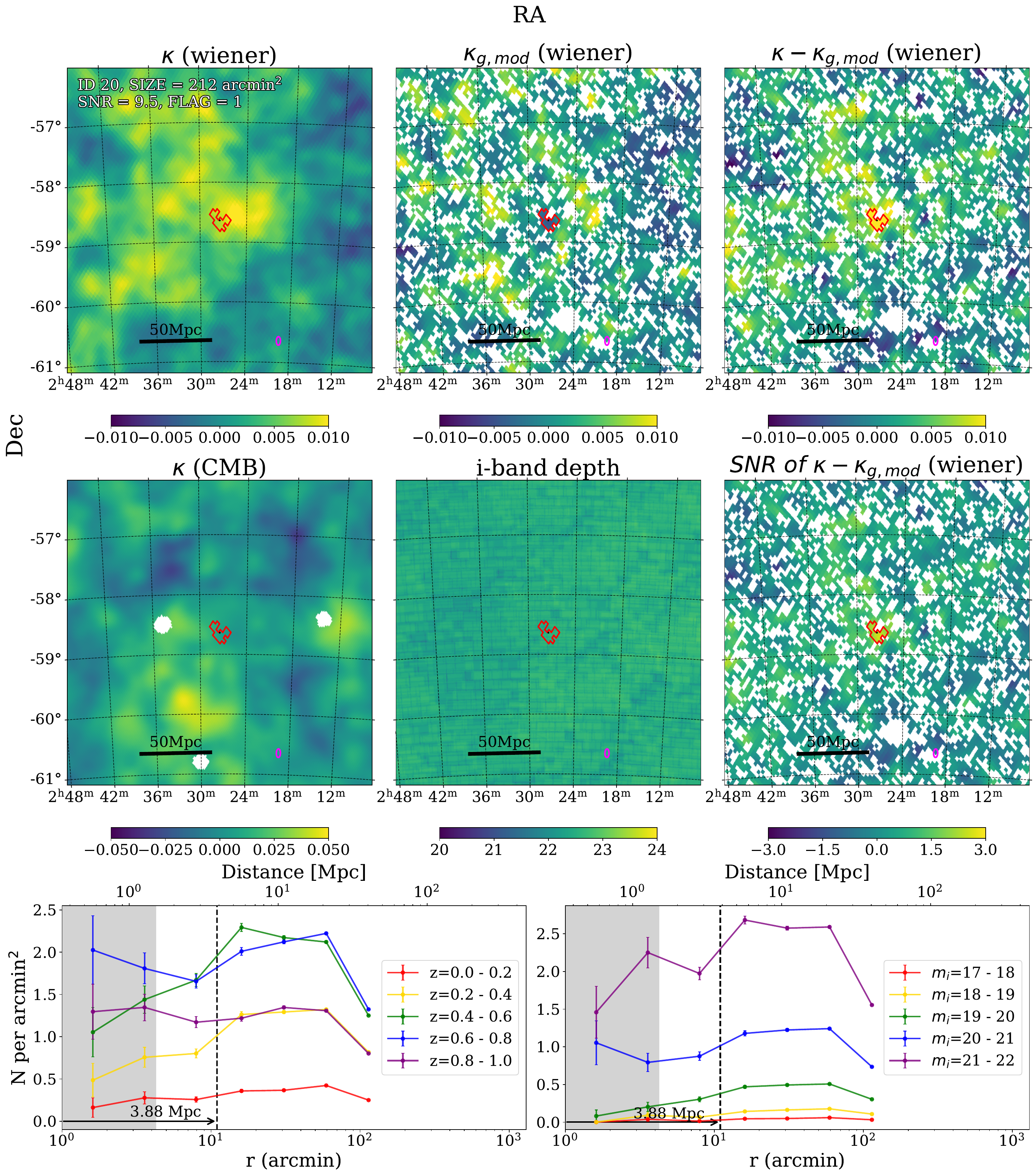}
\figsetgrpnote{The 5$^\circ$$\times$5$^\circ$ cutout maps of the dark structure candidate ID 20. The red lines outline the the boundary of the candidate in each map. The small magenta circle in the bottom-right corner of each panel represents the 10 arcmin smoothing scale, while the small black dot at the center of each panel marks the centroid of the candidate region. (top-left) Wiener convergence map. (top-center) Galaxy convergence map scaled to the Wiener weak lensing convergence. (top-right) Residual map obtained by subtracting the scaled galaxy convergence from the Wiener convergence. (middle-left) CMB lensing convergence map. (middle-center) Observation depth map for i-band. (middle-right) S/N map of the residual. (bottom-left) Radial galaxy surface number density profile around candidate ID 20, shown as a function of redshift bins and (bottom-right) the same profile shown as a function of magnitude bins.}
\figsetgrpend

\figsetgrpstart
\figsetgrpnum{8.21}
\figsetgrptitle{Candidate ID 21}
\figsetplot{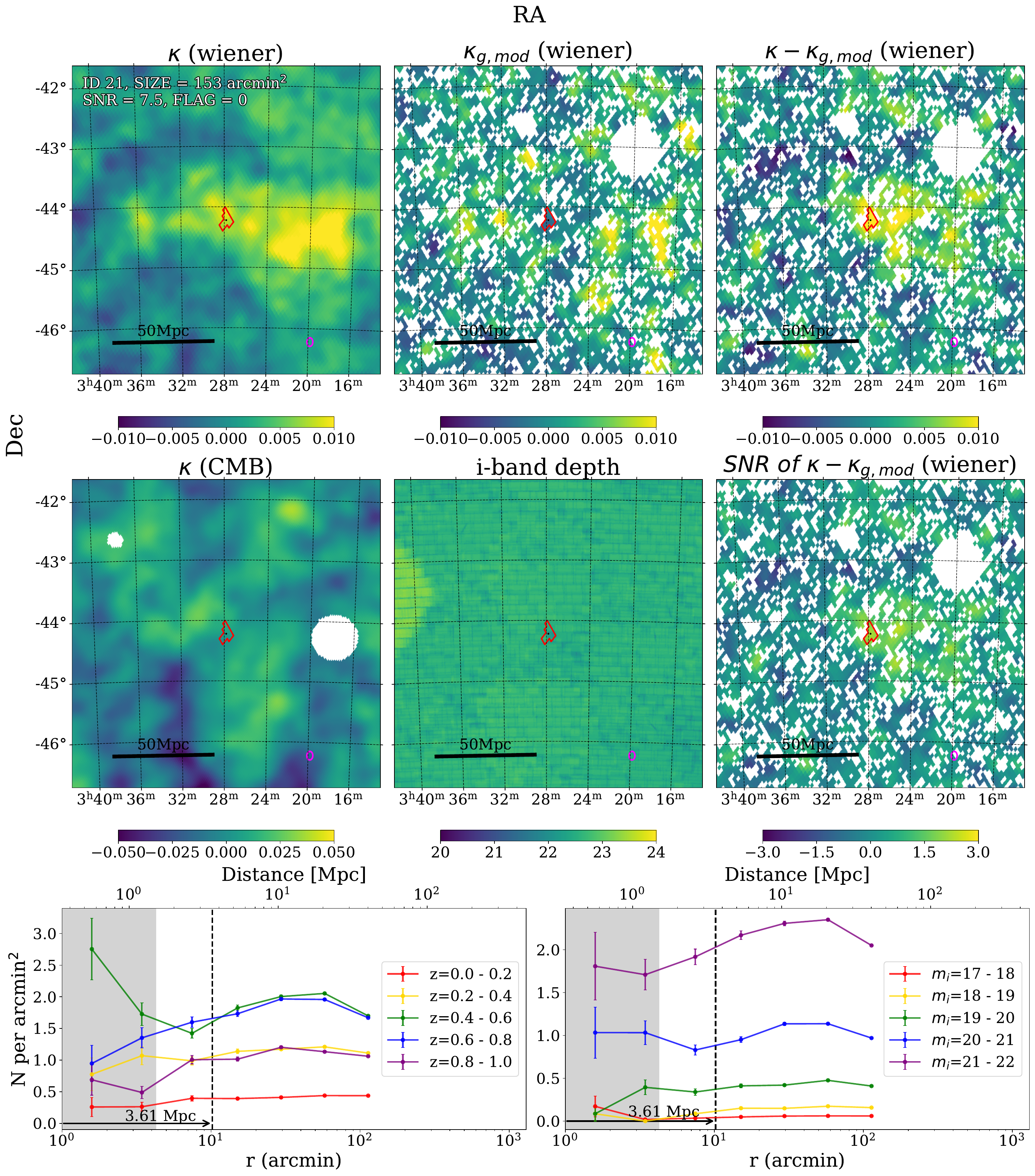}
\figsetgrpnote{The 5$^\circ$$\times$5$^\circ$ cutout maps of the dark structure candidate ID 21. The red lines outline the the boundary of the candidate in each map. The small magenta circle in the bottom-right corner of each panel represents the 10 arcmin smoothing scale, while the small black dot at the center of each panel marks the centroid of the candidate region. (top-left) Wiener convergence map. (top-center) Galaxy convergence map scaled to the Wiener weak lensing convergence. (top-right) Residual map obtained by subtracting the scaled galaxy convergence from the Wiener convergence. (middle-left) CMB lensing convergence map. (middle-center) Observation depth map for i-band. (middle-right) S/N map of the residual. (bottom-left) Radial galaxy surface number density profile around candidate ID 21, shown as a function of redshift bins and (bottom-right) the same profile shown as a function of magnitude bins.}
\figsetgrpend

\figsetgrpstart
\figsetgrpnum{8.22}
\figsetgrptitle{Candidate ID 22}
\figsetplot{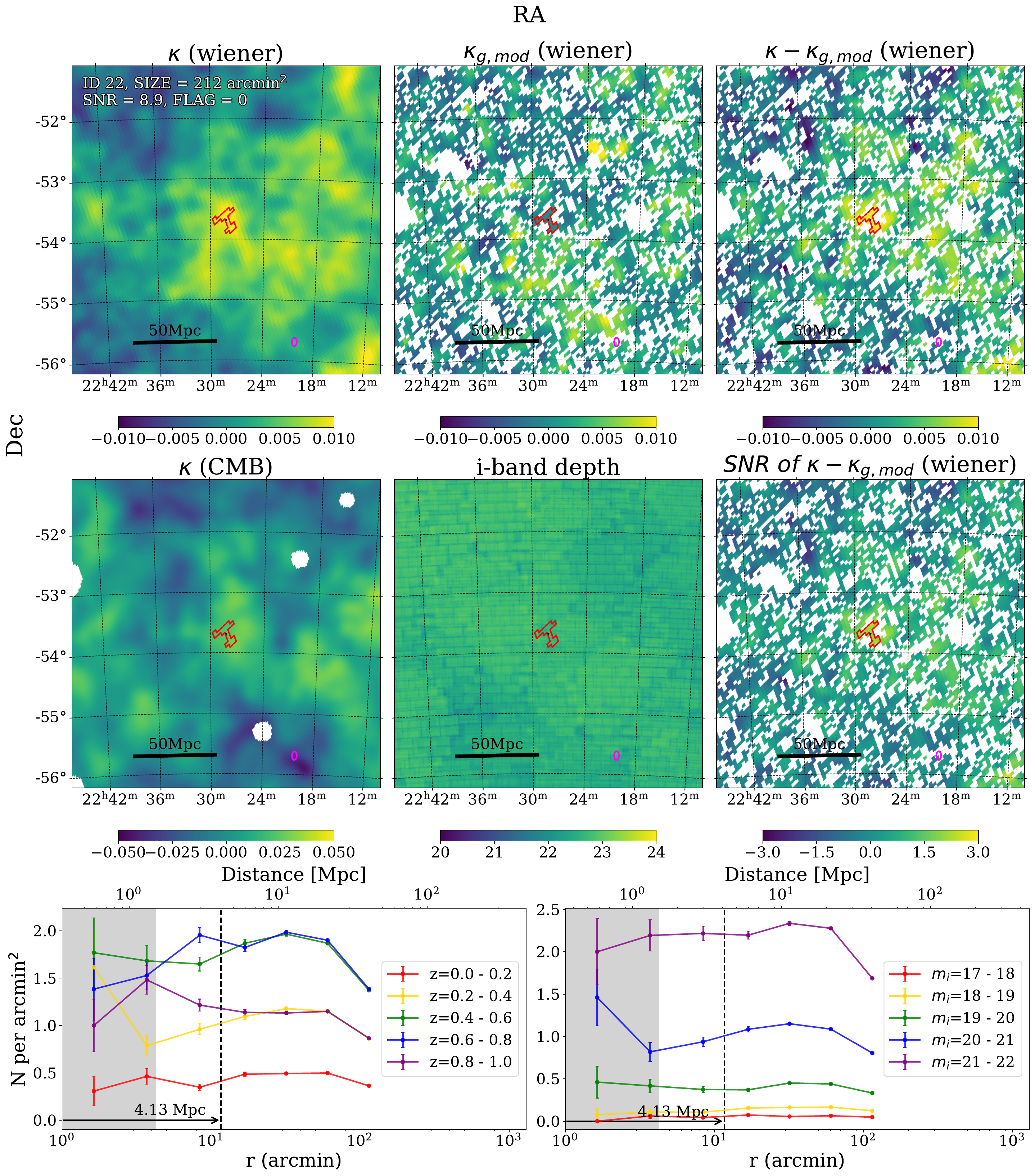}
\figsetgrpnote{The 5$^\circ$$\times$5$^\circ$ cutout maps of the dark structure candidate ID 22. The red lines outline the the boundary of the candidate in each map. The small magenta circle in the bottom-right corner of each panel represents the 10 arcmin smoothing scale, while the small black dot at the center of each panel marks the centroid of the candidate region. (top-left) Wiener convergence map. (top-center) Galaxy convergence map scaled to the Wiener weak lensing convergence. (top-right) Residual map obtained by subtracting the scaled galaxy convergence from the Wiener convergence. (middle-left) CMB lensing convergence map. (middle-center) Observation depth map for i-band. (middle-right) S/N map of the residual. (bottom-left) Radial galaxy surface number density profile around candidate ID 22, shown as a function of redshift bins and (bottom-right) the same profile shown as a function of magnitude bins.}
\figsetgrpend

\figsetend

To better examine the deficit of galaxies within the candidate region, we check the radial profile of the galaxy surface number density. We use the DES Y3 Gold catalog, and apply the same selection criteria as those used for the construction of the galaxy convergence maps. The bottom-left panel of Figure \ref{fig:exzoomin_fig8} represents the radial profile around the candidate ID 22, extending up to approximately 2.5$^\circ$. The galaxy surface density decreases toward the center for most of the redshift bins. We also check the radial profile in magnitude bins (bottom-right panel), and find a decreasing trend toward the inner region for most of the magnitude bins.

Furthermore, we examine the CMB lensing maps to investigate the dark matter content of the candidate regions. We use the CMB lensing convergence map from the Planck 2018 data release \citep{Planck2018_lensing20}. Because the raw CMB lensing convergence map exhibits significant noise when visualized directly \citep{Liu&Hill15}, we apply band filtering and Wiener filtering to improve its visual clarity \citep[e.g.,][]{Planck2018_lensing20}. The CMB lensing map at the location of the dark structure candidate ID 22 is shown in the middle-left panel of Figure~\ref{fig:exzoomin_fig8}. Although the weak-lensing and CMB lensing convergence maps are derived from source galaxies with different redshift distributions, and therefore probe the foreground mass at distinct redshift planes \citep{ACTDR6_Madhavacheril24}, several dark structure candidates show correlations between them. This correlation indicates that the CMB lensing signal also supports the presence of substantial mass in these candidates.

In summary, Figure \ref{fig:exzoomin_fig8} shows that candidate ID 22 is a very promising dark structure candidate. For each tomographic bin, the 10 arcmin smoothing scale approximately corresponds to physical scales of 1.7, 2.3, 3.2, and 3.5 Mpc, respectively. The physical size of the most extended boundary distance shown in the top-right panel of Figure \ref{fig:exzoomin_fig8} is 4.13 Mpc, which is comparable to the typical scale of a galaxy group.

\section{Discussion} \label{sec:dis}

\subsection{Further Improvements in Methodology and Data}\label{sec:6.1}

First, future deep imaging surveys will be crucial for further detecting and validating dark structure candidates. Confirming the relative paucity of galaxies for some candidates is currently limited by the shallow depth of the survey. In addition, the large fraction of masked regions hinders the identification of dark structures located within masks. To increase the resolution of our map and reduce the masked fraction, deep photometric surveys are required to mitigate the shot noise from galaxy counts in each spatial pixel.

We assess the observational effects on dark structure detection using a simple injection–recovery test. We generate a null map by shuffling the pixel values of the residual map and then inject one dark structure candidate at a random position, preserving its size and morphology. The mask and uncertainty of the original residual map are conserved, meaning that the injected candidates cannot always be detected by our dark structure detection pipeline. From 100 realizations of the injection test for each candidate, we obtain the detection rate from 35\% to 72\% when we vary the constraint on the masked fraction of the injected region. This result suggests that approximately 30-64 dark structure candidates may exist in the DES, of which about 8-42 remain undetected in this work due to observational limitations. With larger galaxy samples from future deep imaging surveys, many of these undetected candidates could potentially be revealed.

Also, the source density in the DES Y3 weak-lensing convergence map can lose the small-scale information. The DES Y3 shear catalog \citep{DESshear} contains 100,204,026 objects, with an effective source density of $n_{eff} = 5.59$ galaxies per arcmin$^{2}$. This source density is quite low compared to that of weak lensing studies that focus on smaller regions, such as galaxy clusters. We apply the 10 arcmin Gaussian smoothing and select only the candidate regions consisting of more than 10 HEALPix pixels, because the small-scale information from the WL mass map may not be well reconstructed with low source densities. To ensure a relative deficit of galaxies with small fraction of masked regions and identify smaller-scale candidates, future studies should use galaxy catalogs and weak-lensing mass maps with higher source densities from deeper observations, like the Vera C. Rubin Observatory’s Legacy Survey of Space and Time (LSST; \citealt{LSSToverview}).

Second, a more refined uncertainty estimate for weak-lensing convergence maps is necessary to ensure the reliability of weak-lensing signals in candidate regions. In this work, we estimate convergence uncertainties using the standard deviation of 29 samples from Wiener reconstruction based on the full set of source galaxies. Regardless of the convergence map reconstruction method and the tomographic bin of the source galaxies, we use identical datasets for the uncertainty estimation. These issues can potentially bias our candidate selection, which can be tested with better computing power.

In addition, the galaxy convergence maps should ideally be derived from galaxy masses, not galaxy counts. The correct definition of $\delta_g$ is the matter density fluctuation inferred from galaxies. However, in this study, we rely solely on galaxy counts and apply a simple Gaussian scaling as shown in Figure \ref{fig:scaling_w_fig5}. Developing a more sophisticated method to determine a better proxy for galaxy mass will improve the galaxy convergence maps, and show a higher correlation with the weak-lensing convergence maps. 

Differences in the photometric redshift estimation methods for the source galaxies used in weak-lensing convergence and for the lens galaxies used in galaxy convergence maps can also lead to discrepancies. \cite{DESWLmap} used the full redshift probability distributions from \cite{SOMPZ} for each tomographic bin in the weak-lensing analysis, whereas the galaxy convergence is constructed using point estimates from the DNF photometric redshift catalog. We compare the redshift distributions of all source galaxies in each bin with the fiducial distribution from \cite{SOMPZ}. While the overall shapes and peak positions are similar, small-scale differences remain. Using consistent photometric redshift estimates for both maps would be desirable in future studies.

Furthermore, large uncertainties in photometric redshifts may introduce biases into the galaxy convergence map. To assess the impact of using photometric instead of spectroscopic redshifts, we compare results based on the Dark Energy Spectroscopic Instrument (DESI) Data Release 1 spectroscopic redshift catalog and the DESI Legacy Survey Data Release 9 photometric redshift catalog. We construct the galaxy convergence maps over a $10^\circ \times 10^\circ$ region located within the most completely observed area. Galaxies are counted in each three-dimensional spatial bin according to their photometric or spectroscopic redshifts.

The galaxy convergence maps constructed from spectroscopic and photometric redshifts show no significant difference. This indicates that uncertainties in photometric redshifts do not introduce a severe bias into the galaxy convergence maps. We attribute this to the use of a relatively wide redshift bin ($\Delta z=0.1$) for galaxy counting and lensing efficiency estimation, such that redshift uncertainties smaller than this interval have little impact on the results. The integrative nature of the galaxy convergence along the line of sight further mitigates the effect of redshift uncertainties.

\subsection{The Nature of Dark Structures and Its Implications}
Dark structures share similarities with dark galaxies in that both have a star deficiency. \cite{Gain24} studied dark galaxies using the IllustrisTNG cosmological simulation and found that these galaxies tend to lack star-forming gas due to their initial residence in underdense regions of the Universe (see also \citealt{Kwon25} for the corresponding observational study). This study also suggested that dark galaxies are less likely to merge, leading to higher angular momentum throughout their evolution. This high angular momentum results in larger sizes and lower densities than those of normal galaxies. In addition, cosmic reionization heats the gas, resulting in a significant loss of star-forming gas for dark galaxies. Dark structures may form in similar low-density environments. They may be unable to accumulate sufficient baryon from their surroundings, leading to the formation of dark matter-dominated structures.

\citet{Jee12,Jee14} discussed about the nature of the dark core identified in the galaxy cluster A520. The dark core shares similarities with dark structure candidates, as both are identified by comparing weak-lensing convergence maps with galaxy distributions. \cite{Jee12} proposed several possible explanations for the presence of a dark core. They hypothesized the existence of a compact high mass-to-light ratio galaxy group at the dark core, a distant background cluster, the ejection of bright galaxies, a filament along the line-of-sight direction, collisional dark matter, and the contribution of neighboring structures. 

The dark structure candidates in this study are on the scale of galaxy groups or small clusters. Given their low galaxy number densities, one possible explanation is that they host a population of high mass-to-light ratio groups. To test this hypothesis, independent weak-lensing analyses are required to measure the total enclosed lensing mass of these candidates. By comparing the derived lensing masses with the luminosities of the galaxies, we can determine the precise mass-to-light ratios and assess whether they align with expectations. Additionally, some dark structure candidates are surrounded by galaxy overdensities. The ejection of bright galaxies may have led to certain features. Infalling satellite galaxies to the dark matter concentration may have been ejected to the surroundings via three-body encounters, depleting the galaxy population within the candidates \citep{Sales07}. Another possible scenario is that these candidates correspond to sparse filaments with no significant galaxy concentration. A filament aligned along the line of sight, without a sufficient amount of baryon, could exhibit similar characteristics to those observed in dark structure candidates.

In future studies, hydrodynamical cosmological simulations with ray tracing can be used to generate mock weak-lensing and galaxy convergence maps, enabling tests of our methodology and the identification of spurious dark structure-like signals. Furthermore, exploring simulations with varied cosmological parameters will be crucial for probing the physical origins of dark matter-dominated structures that are not easily explained within the standard \lcdm\ framework.

Our search for large-scale dark structures has both cosmological and methodological significance. If dark structures do exist, determining their abundance and precise locations would be crucial for advancing our understanding of dark matter and its role in structure formation. Follow-up multiwavelength observations, especially of star-forming gas, could shed light on the physical mechanisms that make this structure remain dark. In addition, the existence of dark structures could serve as another direct probe for dark matter. The methodology developed here can be applied to other weak-lensing convergence maps, including those from upcoming surveys such as LSST \citep{LSSToverview}, Euclid \citep{Euclidoverview}, and the Nancy Grace Roman Space Telescope \citep{RomanWL}, to find more plausible dark structure candidates.

\section{Summary}\label{sec:summary}
We present a novel method for comparing the distributions of dark matter and galaxies using weak-lensing convergence and galaxy convergence maps. The galaxy convergence maps are constructed by applying a lensing-weighted kernel to the two-dimensional galaxy density fluctuation maps. Our focus is on the identification of dark structure candidates, which are dark matter-dominated structures on a few Mpc scales without a sufficient number of galaxies. The main results are as follows. 

\begin{enumerate}
    \item A comparison between the weak-lensing and galaxy convergence maps shows general agreement. However, there are several regions that exhibit a large excess of dark matter density over galaxy number density, which could be considered as dark structure candidates.
    \item From the comparison of two weak-lensing convergence maps (i.e. Wiener and GLIMPSE) for those regions, we could identify 22 dark structure candidates.
    \item A careful examination of those 22 candidates results in seven most promising candidates.
\end{enumerate}

Our method can be applied to future large-scale galaxy surveys, such as LSST \citep{LSSToverview}, Euclid \citep{Euclidoverview}, and the Nancy Grace Roman Space Telescope \citep{RomanWL}, which can allow us to identify more plausible dark structure candidates and to further investigate their physical origin.

\section*{Acknowledgments}
We thank the referee for constructive comments that helped improve the paper. We also thank Kim HyeongHan for valuable comments. H.S.H. acknowledges support from a National Research Foundation of Korea (NRF) grant funded by the Korean government (MSIT), NRF-2021R1A2C1094577; Samsung Electronic Co., Ltd. (project No. IO220811-01945-01); and Hyunsong Educational \& Cultural Foundation. N.J. acknowledges support from ERC-selected UKRI Frontier Research grant EP/Y03015X/1.
This project used public archival data from the Dark Energy Survey (DES). Funding for the DES Projects has been provided by the U.S. Department of Energy, the U.S. National Science Foundation, the Ministry of Science and Education of Spain, the Science and Technology Facilities Council of the United Kingdom, the Higher Education Funding Council for England, the National Center for Supercomputing Applications at the University of Illinois at Urbana-Champaign, the Kavli Institute of Cosmological Physics at the University of Chicago, the Center for Cosmology and Astro-Particle Physics at the Ohio State University, the Mitchell Institute for Fundamental Physics and Astronomy at Texas A\&M University, Financiadora de Estudos e Projetos, Funda{\c c}{\~a}o Carlos Chagas Filho de Amparo {\`a} Pesquisa do Estado do Rio de Janeiro, Conselho Nacional de Desenvolvimento Cient{\'i}fico e Tecnol{\'o}gico and the Minist{\'e}rio da Ci{\^e}ncia, Tecnologia e Inova{\c c}{\~a}o, the Deutsche Forschungsgemeinschaft, and the Collaborating Institutions in the Dark Energy Survey.
The Collaborating Institutions are Argonne National Laboratory, the University of California at Santa Cruz, the University of Cambridge, Centro de Investigaciones Energ{\'e}ticas, Medioambientales y Tecnol{\'o}gicas-Madrid, the University of Chicago, University College London, the DES-Brazil Consortium, the University of Edinburgh, the Eidgen{\"o}ssische Technische Hochschule (ETH) Z{\"u}rich,  Fermi National Accelerator Laboratory, the University of Illinois at Urbana-Champaign, the Institut de Ci{\`e}ncies de l'Espai (IEEC/CSIC), the Institut de F{\'i}sica d'Altes Energies, Lawrence Berkeley National Laboratory, the Ludwig-Maximilians Universit{\"a}t M{\"u}nchen and the associated Excellence Cluster Universe, the University of Michigan, the National Optical Astronomy Observatory, the University of Nottingham, The Ohio State University, the OzDES Membership Consortium, the University of Pennsylvania, the University of Portsmouth, SLAC National Accelerator Laboratory, Stanford University, the University of Sussex, and Texas A\&M University.
Based in part on observations at Cerro Tololo Inter-American Observatory, National Optical Astronomy Observatory, which is operated by the Association of Universities for Research in Astronomy (AURA) under a cooperative agreement with the National Science Foundation.
This research uses services or data provided by the Astro Data Lab, which is part of the Community Science and Data Center (CSDC) Program of NSF NOIRLab. NOIRLab is operated by the Association of Universities for Research in Astronomy (AURA), Inc. under a cooperative agreement with the U.S. National Science Foundation.

%






\section*{Appendix A: Correlation between GLIMPSE Convergence Map and Galaxy Convergence Map}
\begin{figure*}
    \centering
    \includegraphics[width=1\linewidth, trim=70 70 70 0, clip]{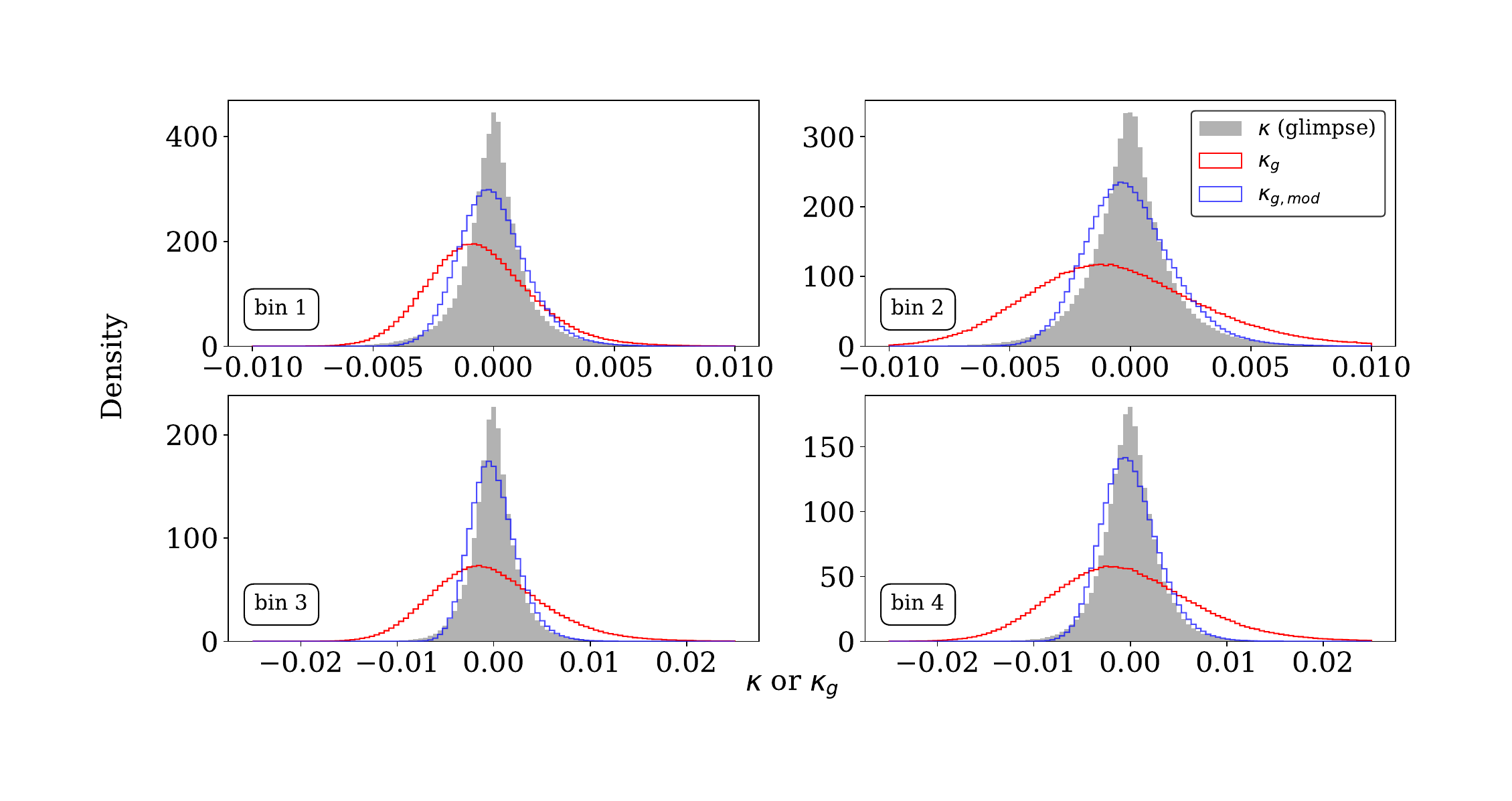}
    \caption{Similar to Figure \ref{fig:scaling_w_fig5}, but for the weak-lensing convergence from GLIMPSE reconstruction.}
    \label{fig:scaling_g}
\end{figure*}

\begin{figure*}
    \centering
    \includegraphics[width=1\linewidth,trim=70 60 70 30, clip]{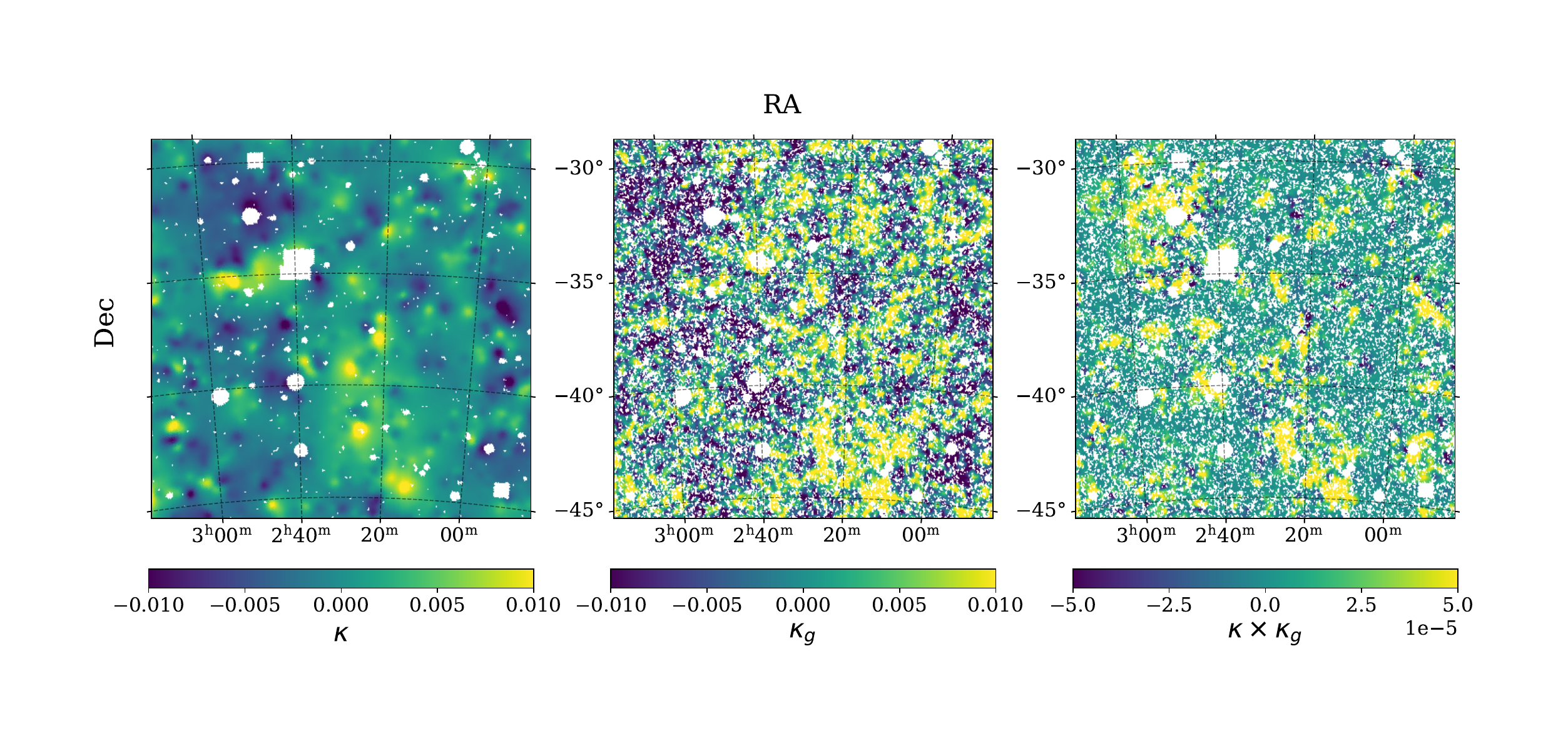}
    \caption{Similar to Figure \ref{fig:ktimeskg_fig6}, but for the GLIMPSE weak-lensing convergence maps.}
    \label{fig:glimpsecorr}
\end{figure*}

\begin{figure}
    \centering
    \includegraphics[width=1\linewidth]{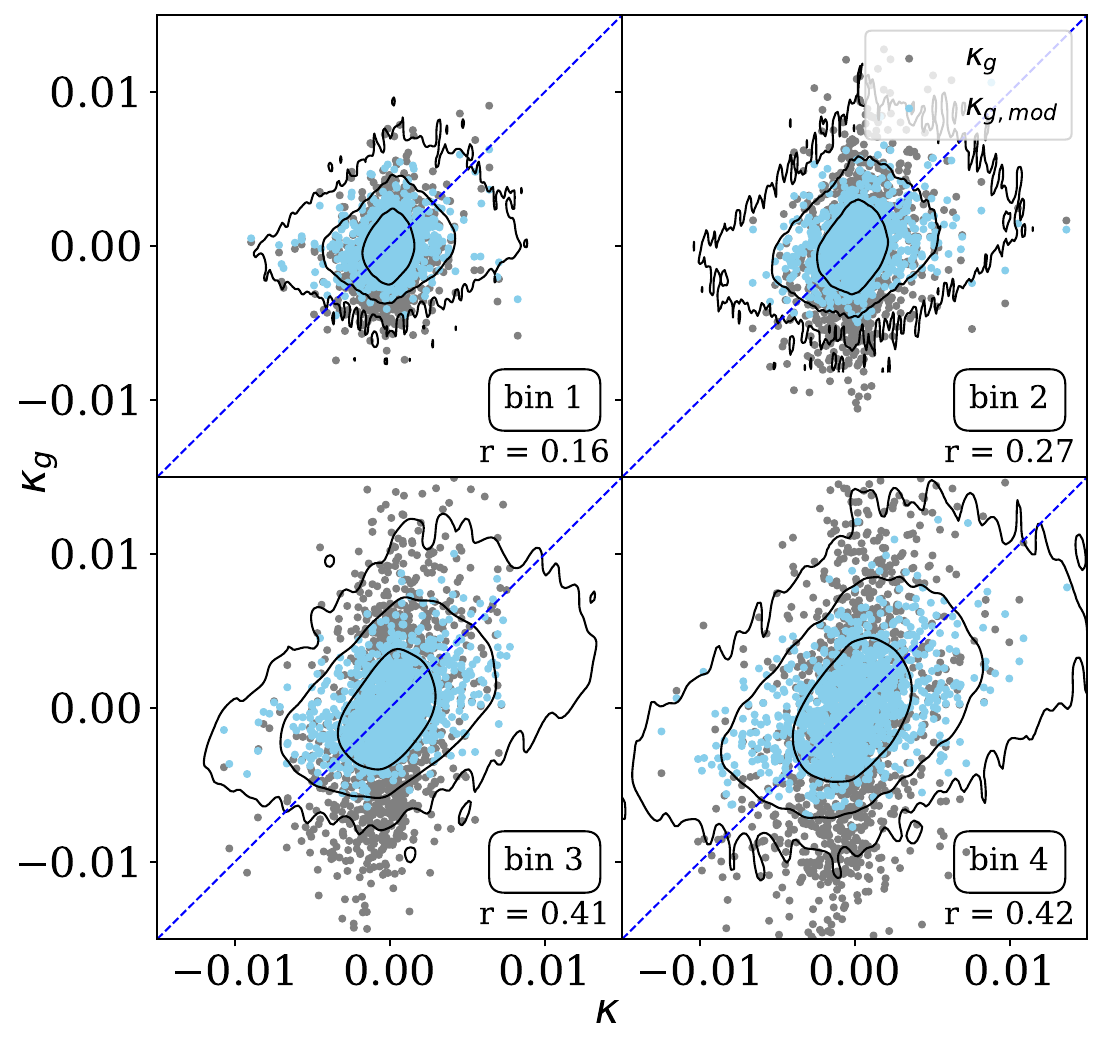}
    \caption{Similar to Figure \ref{fig:kvskg_fig7}, but for the weak-lensing convergence from the GLIMPSE method.}
    \label{fig:glimpsepix}
\end{figure}

Figure \ref{fig:scaling_g} shows the histograms of the weak-lensing convergence reconstructed using the GLIMPSE method and the galaxy convergence for each tomographic bin, similar to Figure \ref{fig:scaling_w_fig5}. We scale the galaxy convergence as described in Section \ref{sec:scaling}, to the GLIMPSE weak-lensing convergence distribution. The distribution of the scaled galaxy convergence appears well adjusted, aligning with the distribution of the GLIMPSE convergence. Figure \ref{fig:glimpsecorr} represents the same cutout as in Figure \ref{fig:ktimeskg_fig6}, but using the GLIMPSE convergence map instead of the Wiener convergence map. The correlation between weak-lensing convergence and galaxy convergence is clearly visible in the reddish regions of the third panel. Figure \ref{fig:glimpsepix} shows a pixel-by-pixel comparison between the GLIMPSE weak-lensing convergence and the galaxy convergence, analogous to Figure \ref{fig:kvskg_fig7}. As in the case of the Wiener convergence, a strong linear correlation between $\kappa$ and $\kappa_g$ is evident. These figures confirm that the high-correlation signal observed between the Wiener weak-lensing convergence and galaxy convergence is also present when using the GLIMPSE weak-lensing convergence.

\section*{Appendix B: Dark Structure Candidates in GLIMPSE Convergence Map Comparison}
\begin{figure*}
    \centering
    \includegraphics[width=1\linewidth]{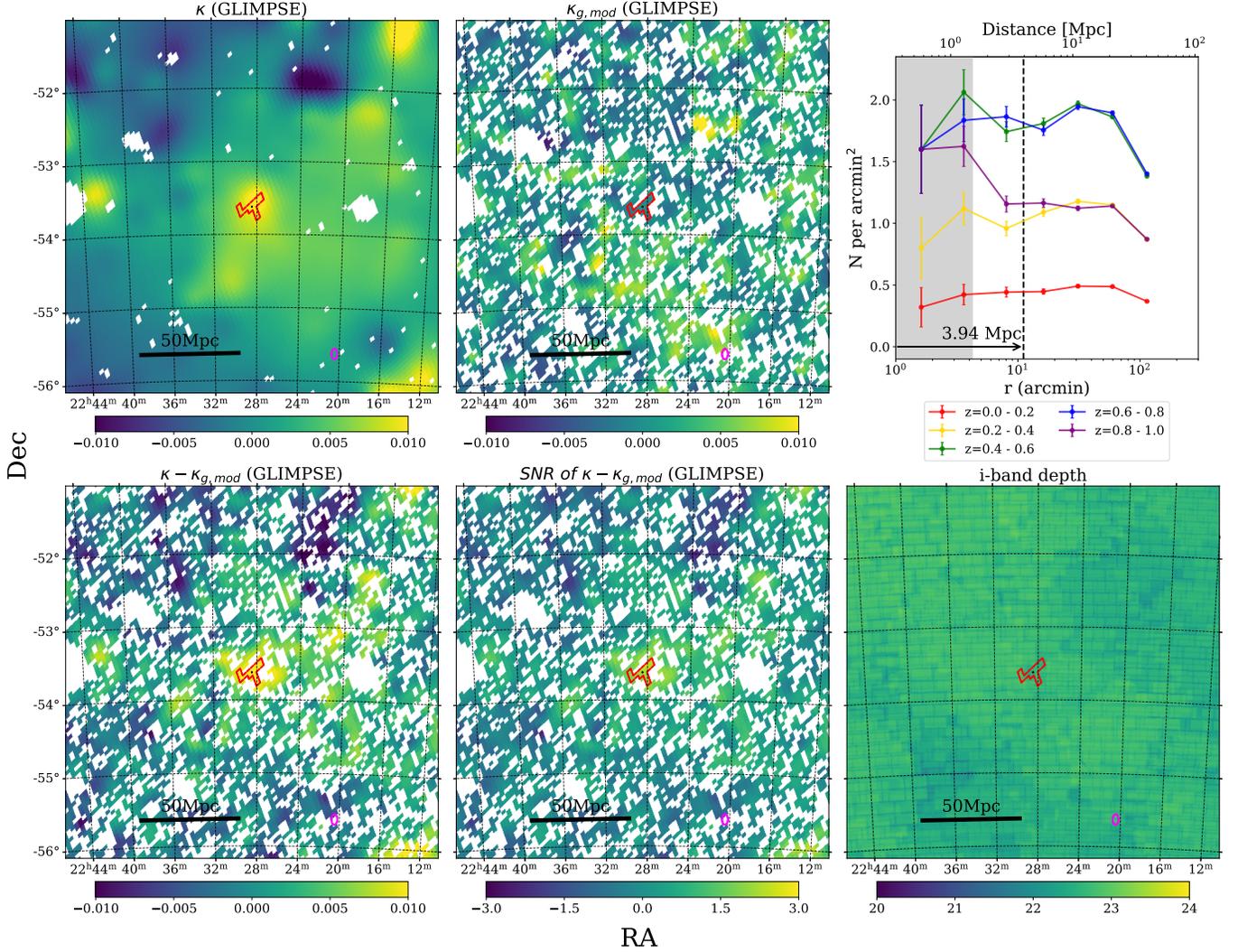}
    \caption{The 5$^\circ$$\times$5$^\circ$ cutout maps of the dark structure candidate ID 22. The red lines outline the the boundary of the candidate in each map. The small magenta circle in the bottom-right corner of each panel represents the 10 arcmin smoothing scale, while the small black dot at the center of each panel marks the centroid of the candidate region. (top-left) GLIMPSE convergence map. (top-center) Galaxy convergence map scaled to the GLIMPSE weak lensing convergence. (top-right) Residual map obtained by subtracting the scaled galaxy convergence from the GLIMPSE convergence. (middle-left) CMB lensing convergence map. (middle-center) Observation depth map for i-band. (middle-right) S/N map of the residual. (bottom-left) Radial galaxy surface number density profile around candidate ID 22, shown as a function of redshift bins and (bottom-right) shown as a function of magnitude bins. The gray shaded region represents the circular radius of the smoothing scale, while the black dashed line indicates the most extended boundary distance from the candidate centroid. The complete figure set for 22 dark structure candidates with the GLIMPSE convergence map is available in the online journal.}
    \label{fig:AppBGLIMSPE}
\end{figure*}

Figure \ref{fig:AppBGLIMSPE} represents the counterpart to Figure \ref{fig:exzoomin_fig8}, using the GLIMPSE weak-lensing convergence map instead of the Wiener convergence map. It  displays the comparison between the GLIMPSE convergence and the galaxy convergence, scaled to match the GLIMPSE convergence. The complete figure set for 22 dark structure candidates is available in the online journal.

\figsetstart
\figsetnum{13}
\figsettitle{Dark Structure Candidates in GLIMPSE Convergence Map}

\figsetgrpstart
\figsetgrpnum{12.1}
\figsetgrptitle{Candidate ID 1}
\figsetplot{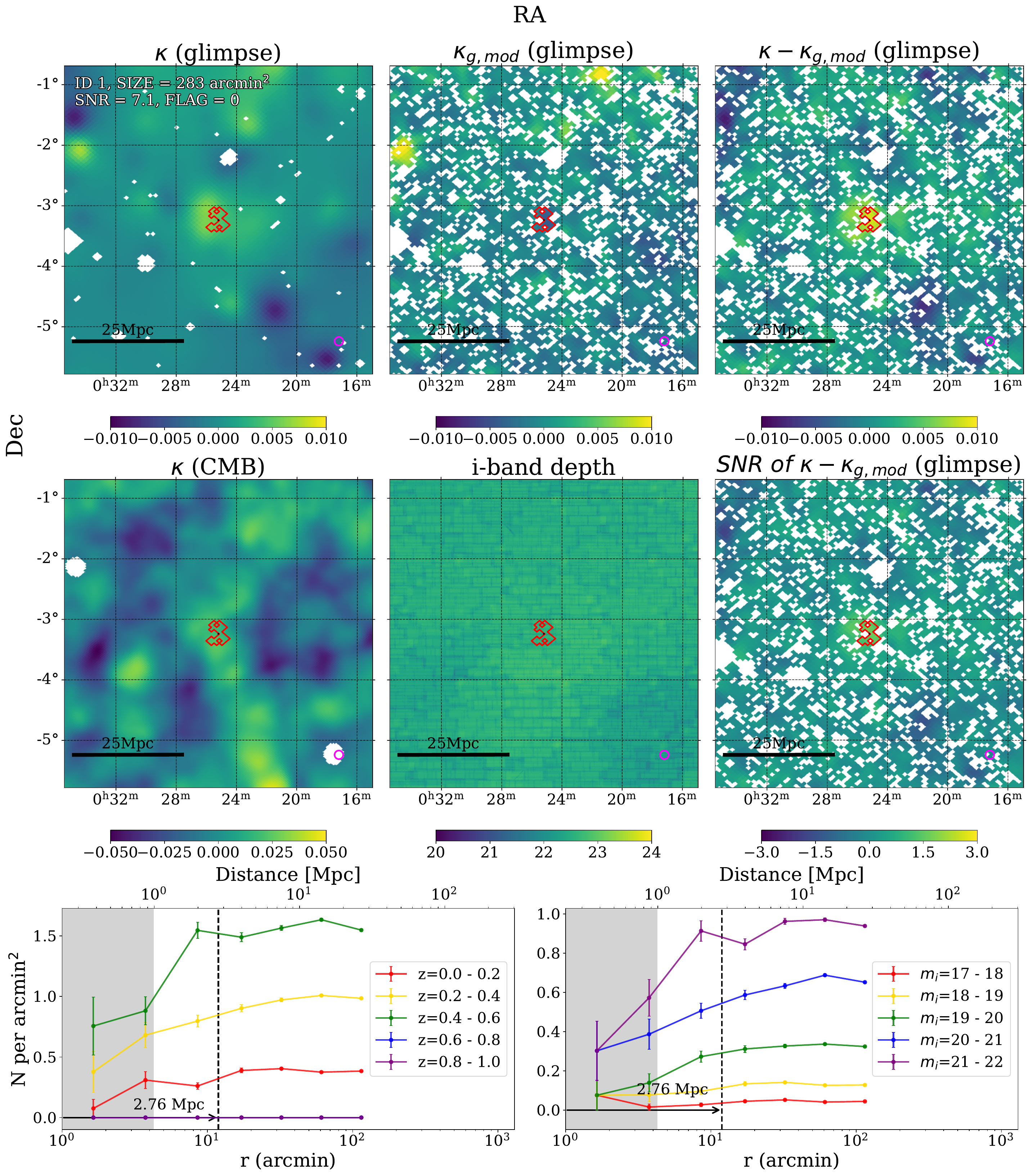}
\figsetgrpnote{The 5$^\circ$$\times$5$^\circ$ cutout maps of the dark structure candidate ID 1. The red lines outline the the boundary of the candidate in each map. The small magenta circle in the bottom-right corner of each panel represents the 10 arcmin smoothing scale, while the small black dot at the center of each panel marks the centroid of the candidate region. (top-left) GLIMPSE convergence map. (top-center) Galaxy convergence map scaled to the GLIMPSE weak lensing convergence. (top-right) Residual map obtained by subtracting the scaled galaxy convergence from the GLIMPSE convergence. (middle-left) CMB lensing convergence map. (middle-center) Observation depth map for i-band. (middle-right) S/N map of the residual. (bottom-left) Radial galaxy surface number density profile around candidate ID 1, shown as a function of redshift bins and (bottom-right) the same profile shown as a function of magnitude bins.}
\figsetgrpend

\figsetgrpstart
\figsetgrpnum{12.2}
\figsetgrptitle{Candidate ID 2}
\figsetplot{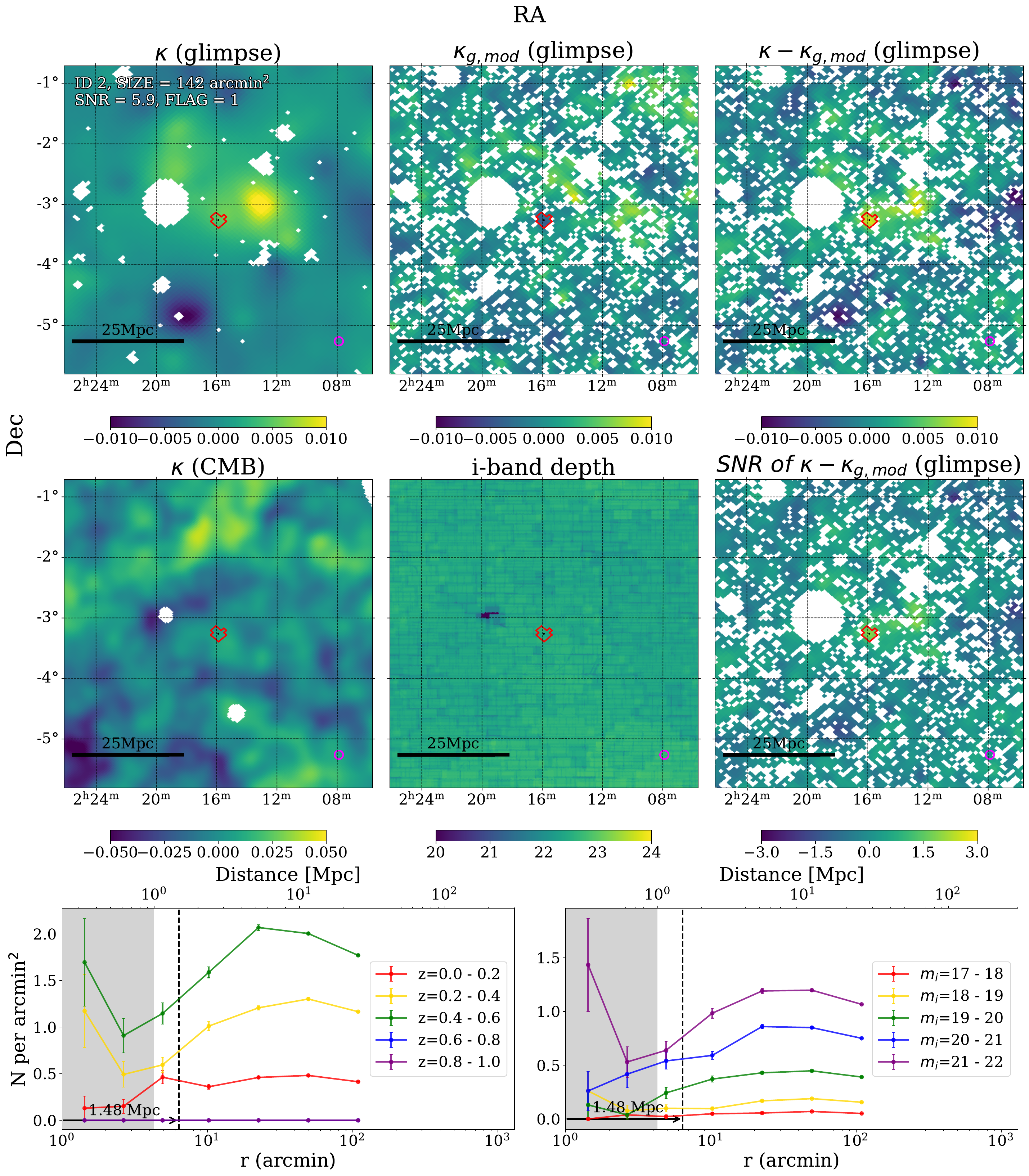}
\figsetgrpnote{The 5$^\circ$$\times$5$^\circ$ cutout maps of the dark structure candidate ID 2. The red lines outline the the boundary of the candidate in each map. The small magenta circle in the bottom-right corner of each panel represents the 10 arcmin smoothing scale, while the small black dot at the center of each panel marks the centroid of the candidate region. (top-left) GLIMPSE convergence map. (top-center) Galaxy convergence map scaled to the GLIMPSE weak lensing convergence. (top-right) Residual map obtained by subtracting the scaled galaxy convergence from the GLIMPSE convergence. (middle-left) CMB lensing convergence map. (middle-center) Observation depth map for i-band. (middle-right) S/N map of the residual. (bottom-left) Radial galaxy surface number density profile around candidate ID 2, shown as a function of redshift bins and (bottom-right) the same profile shown as a function of magnitude bins.}
\figsetgrpend

\figsetgrpstart
\figsetgrpnum{12.3}
\figsetgrptitle{Candidate ID 3}
\figsetplot{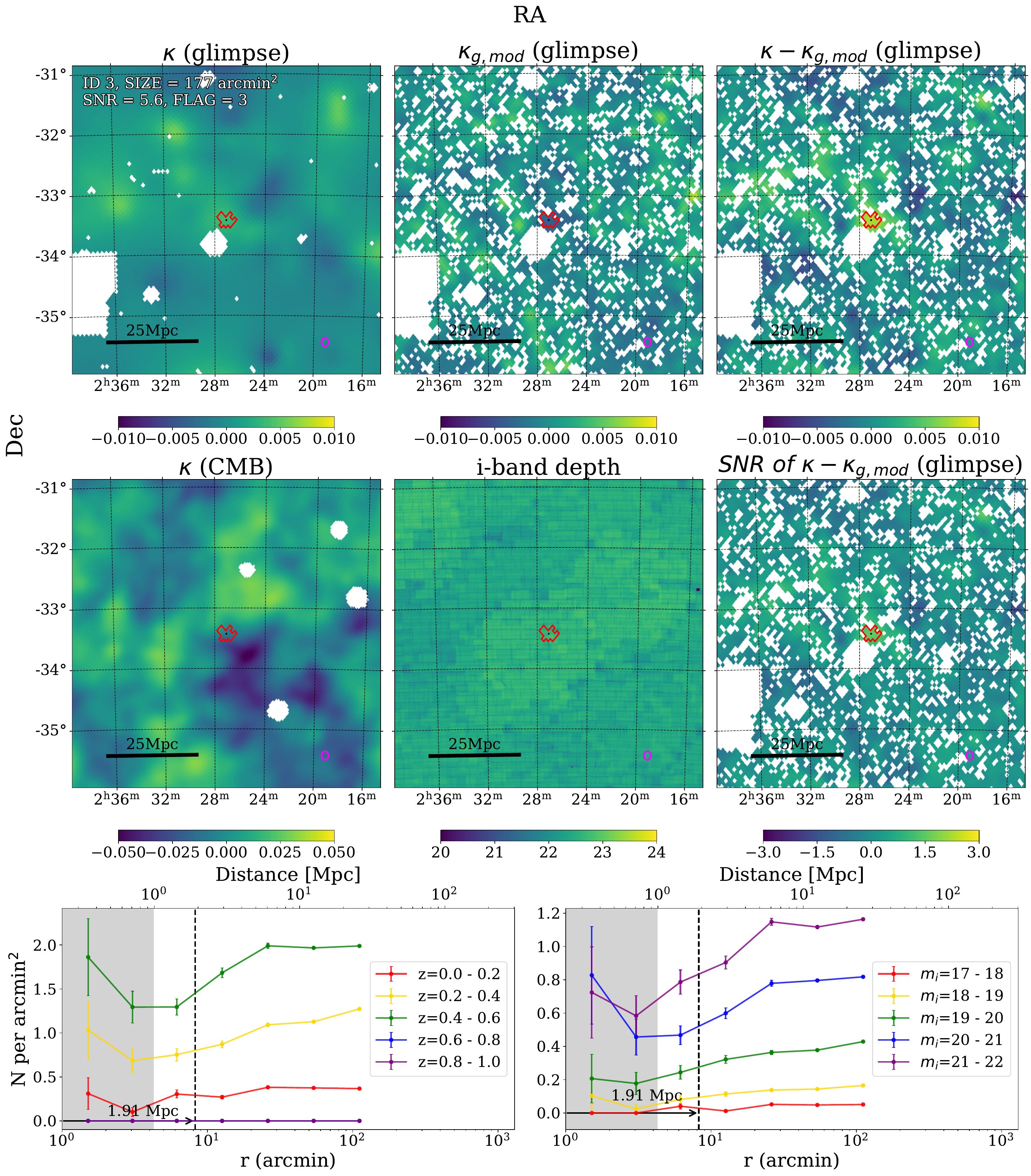}
\figsetgrpnote{The 5$^\circ$$\times$5$^\circ$ cutout maps of the dark structure candidate ID 3. The red lines outline the the boundary of the candidate in each map. The small magenta circle in the bottom-right corner of each panel represents the 10 arcmin smoothing scale, while the small black dot at the center of each panel marks the centroid of the candidate region. (top-left) GLIMPSE convergence map. (top-center) Galaxy convergence map scaled to the GLIMPSE weak lensing convergence. (top-right) Residual map obtained by subtracting the scaled galaxy convergence from the GLIMPSE convergence. (middle-left) CMB lensing convergence map. (middle-center) Observation depth map for i-band. (middle-right) S/N map of the residual. (bottom-left) Radial galaxy surface number density profile around candidate ID 3, shown as a function of redshift bins and (bottom-right) the same profile shown as a function of magnitude bins.}
\figsetgrpend

\figsetgrpstart
\figsetgrpnum{12.4}
\figsetgrptitle{Candidate ID 4}
\figsetplot{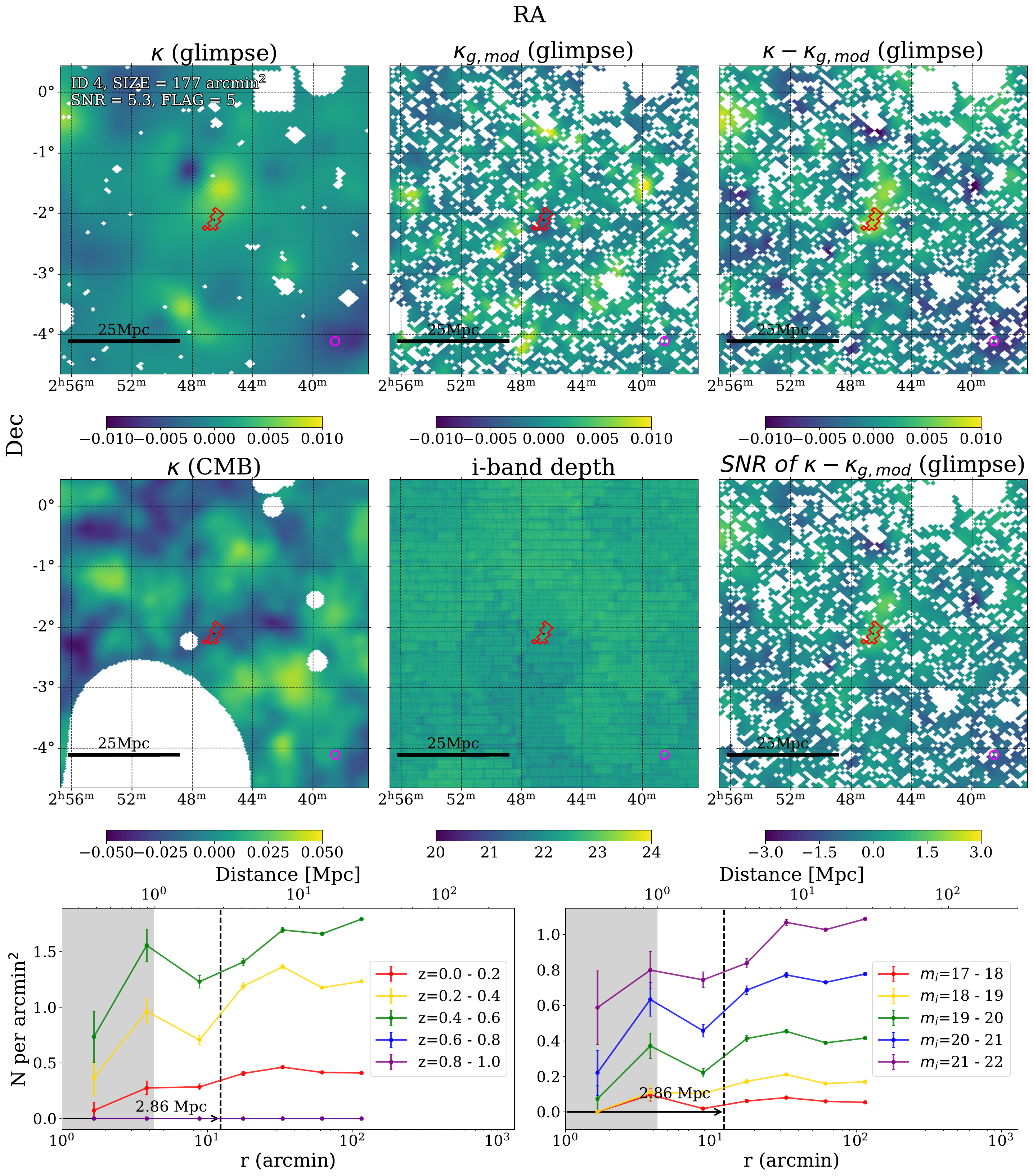}
\figsetgrpnote{The 5$^\circ$$\times$5$^\circ$ cutout maps of the dark structure candidate ID 4. The red lines outline the the boundary of the candidate in each map. The small magenta circle in the bottom-right corner of each panel represents the 10 arcmin smoothing scale, while the small black dot at the center of each panel marks the centroid of the candidate region. (top-left) GLIMPSE convergence map. (top-center) Galaxy convergence map scaled to the GLIMPSE weak lensing convergence. (top-right) Residual map obtained by subtracting the scaled galaxy convergence from the GLIMPSE convergence. (middle-left) CMB lensing convergence map. (middle-center) Observation depth map for i-band. (middle-right) S/N map of the residual. (bottom-left) Radial galaxy surface number density profile around candidate ID 4, shown as a function of redshift bins and (bottom-right) the same profile shown as a function of magnitude bins.}
\figsetgrpend

\figsetgrpstart
\figsetgrpnum{12.5}
\figsetgrptitle{Candidate ID 5}
\figsetplot{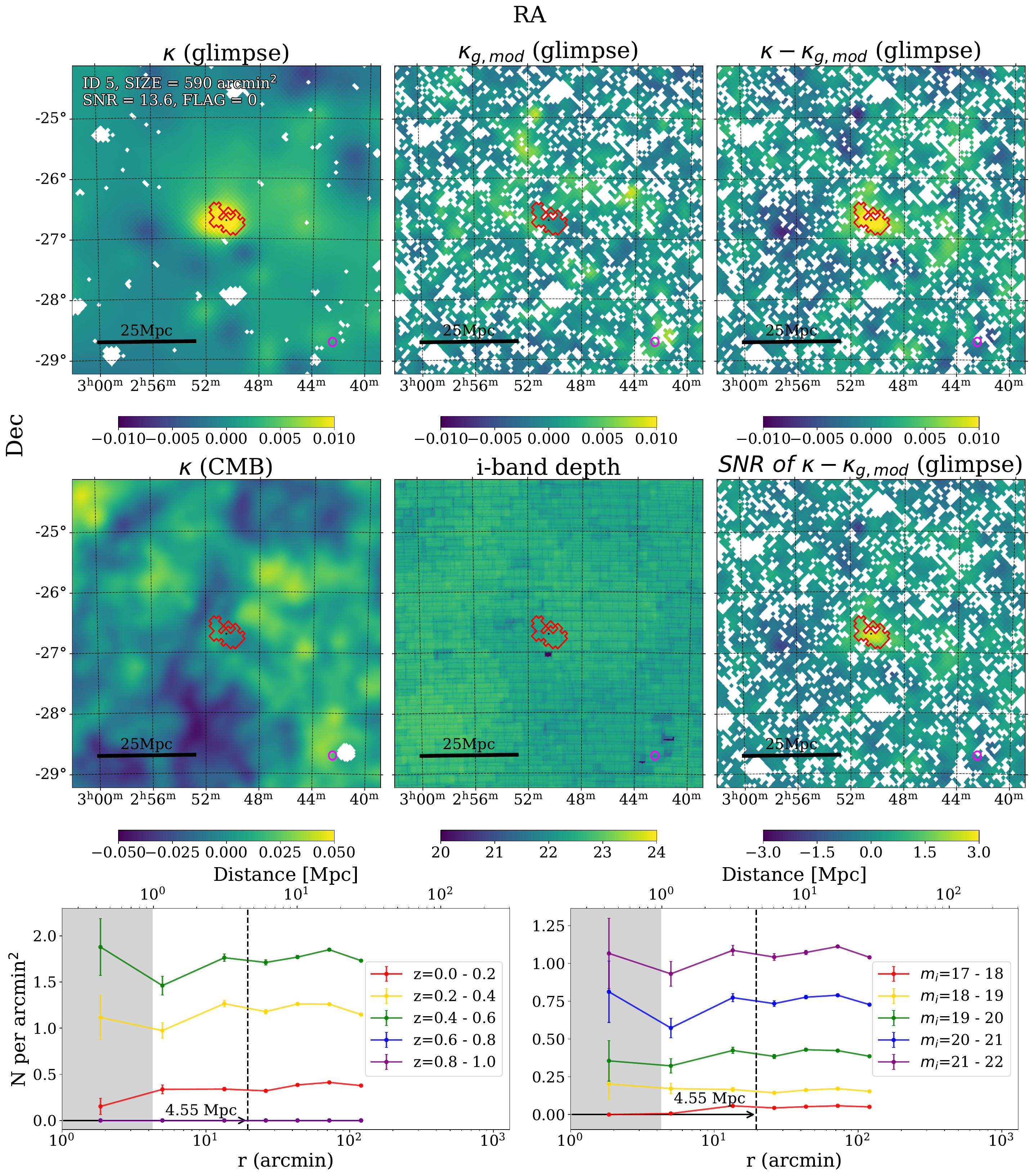}
\figsetgrpnote{The 5$^\circ$$\times$5$^\circ$ cutout maps of the dark structure candidate ID 5. The red lines outline the the boundary of the candidate in each map. The small magenta circle in the bottom-right corner of each panel represents the 10 arcmin smoothing scale, while the small black dot at the center of each panel marks the centroid of the candidate region. (top-left) GLIMPSE convergence map. (top-center) Galaxy convergence map scaled to the GLIMPSE weak lensing convergence. (top-right) Residual map obtained by subtracting the scaled galaxy convergence from the GLIMPSE convergence. (middle-left) CMB lensing convergence map. (middle-center) Observation depth map for i-band. (middle-right) S/N map of the residual. (bottom-left) Radial galaxy surface number density profile around candidate ID 5, shown as a function of redshift bins and (bottom-right) the same profile shown as a function of magnitude bins.}
\figsetgrpend

\figsetgrpstart
\figsetgrpnum{12.6}
\figsetgrptitle{Candidate ID 6}
\figsetplot{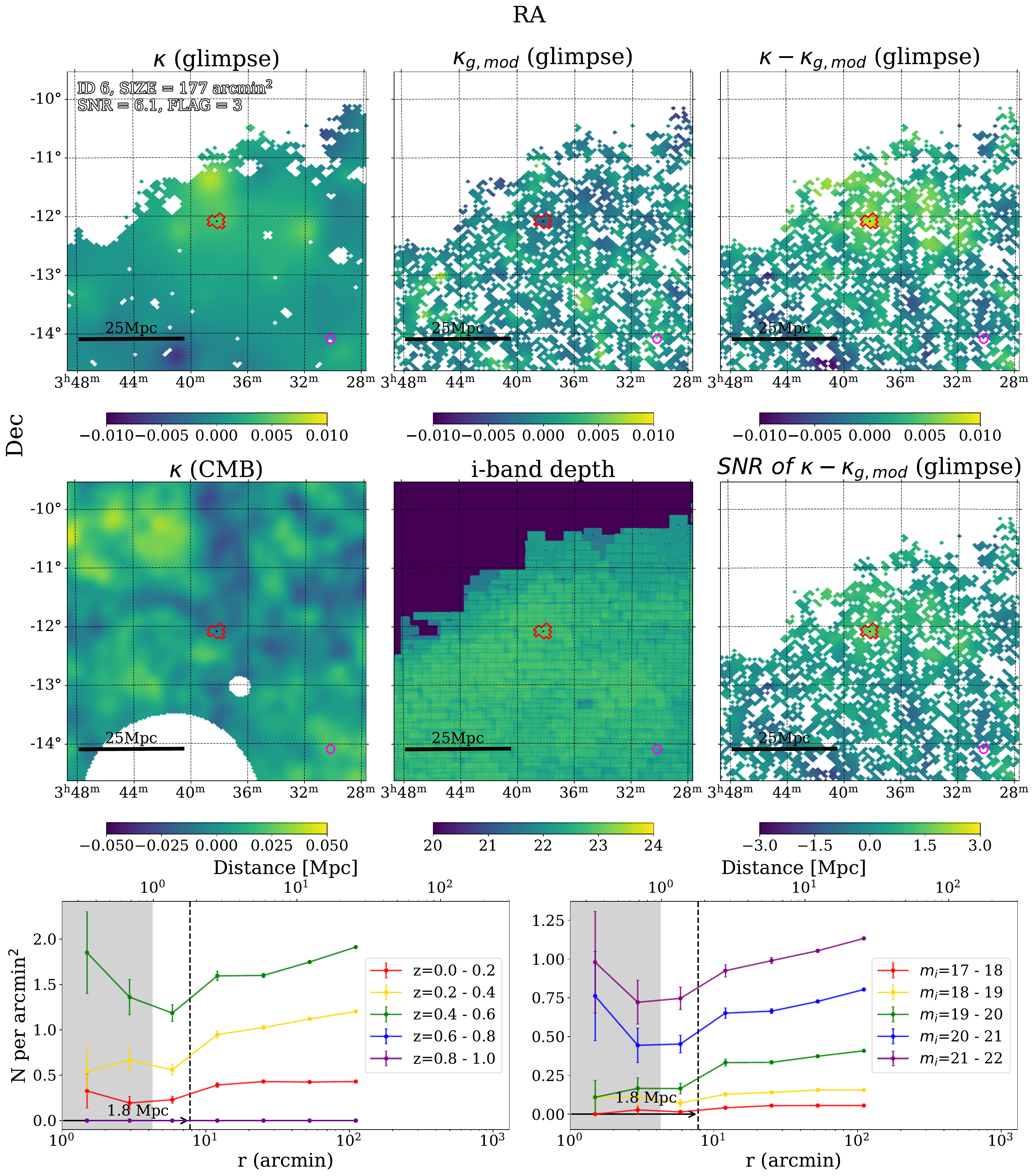}
\figsetgrpnote{The 5$^\circ$$\times$5$^\circ$ cutout maps of the dark structure candidate ID 6. The red lines outline the the boundary of the candidate in each map. The small magenta circle in the bottom-right corner of each panel represents the 10 arcmin smoothing scale, while the small black dot at the center of each panel marks the centroid of the candidate region. (top-left) GLIMPSE convergence map. (top-center) Galaxy convergence map scaled to the GLIMPSE weak lensing convergence. (top-right) Residual map obtained by subtracting the scaled galaxy convergence from the GLIMPSE convergence. (middle-left) CMB lensing convergence map. (middle-center) Observation depth map for i-band. (middle-right) S/N map of the residual. (bottom-left) Radial galaxy surface number density profile around candidate ID 6, shown as a function of redshift bins and (bottom-right) the same profile shown as a function of magnitude bins.}
\figsetgrpend

\figsetgrpstart
\figsetgrpnum{12.7}
\figsetgrptitle{Candidate ID 7}
\figsetplot{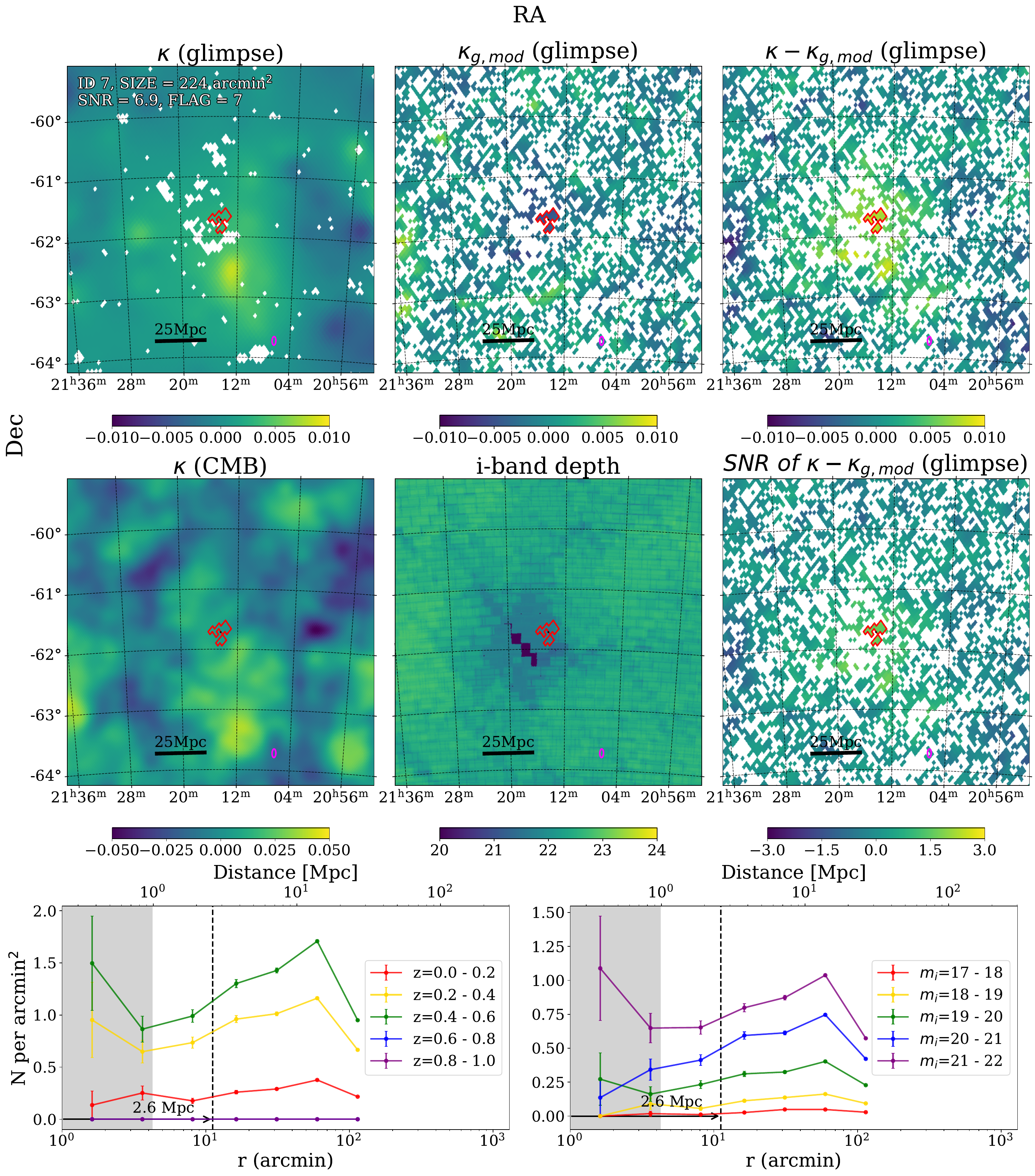}
\figsetgrpnote{The 5$^\circ$$\times$5$^\circ$ cutout maps of the dark structure candidate ID 7. The red lines outline the the boundary of the candidate in each map. The small magenta circle in the bottom-right corner of each panel represents the 10 arcmin smoothing scale, while the small black dot at the center of each panel marks the centroid of the candidate region. (top-left) GLIMPSE convergence map. (top-center) Galaxy convergence map scaled to the GLIMPSE weak lensing convergence. (top-right) Residual map obtained by subtracting the scaled galaxy convergence from the GLIMPSE convergence. (middle-left) CMB lensing convergence map. (middle-center) Observation depth map for i-band. (middle-right) S/N map of the residual. (bottom-left) Radial galaxy surface number density profile around candidate ID 7, shown as a function of redshift bins and (bottom-right) the same profile shown as a function of magnitude bins.}
\figsetgrpend

\figsetgrpstart
\figsetgrpnum{12.8}
\figsetgrptitle{Candidate ID 8}
\figsetplot{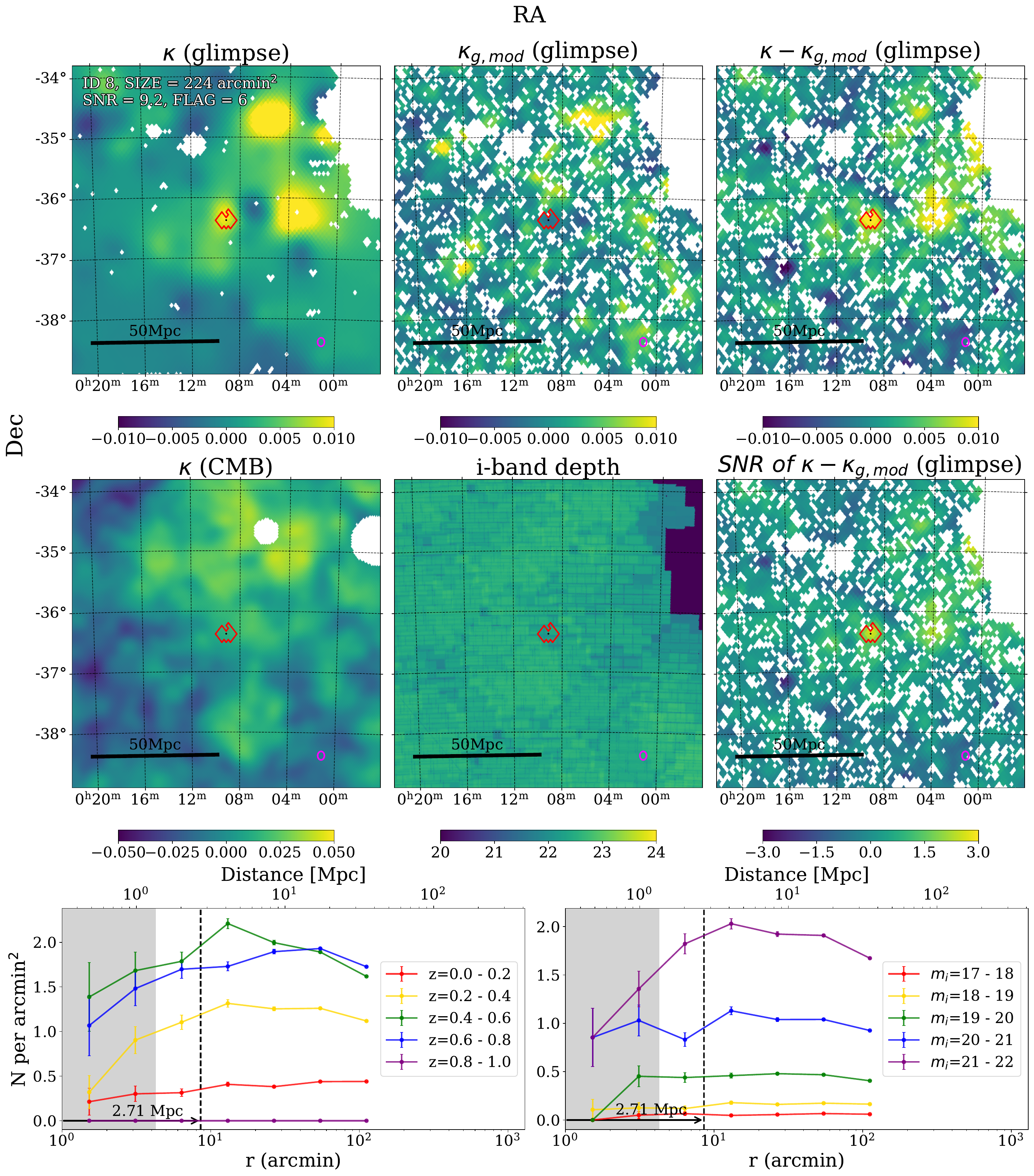}
\figsetgrpnote{The 5$^\circ$$\times$5$^\circ$ cutout maps of the dark structure candidate ID 8. The red lines outline the the boundary of the candidate in each map. The small magenta circle in the bottom-right corner of each panel represents the 10 arcmin smoothing scale, while the small black dot at the center of each panel marks the centroid of the candidate region. (top-left) GLIMPSE convergence map. (top-center) Galaxy convergence map scaled to the GLIMPSE weak lensing convergence. (top-right) Residual map obtained by subtracting the scaled galaxy convergence from the GLIMPSE convergence. (middle-left) CMB lensing convergence map. (middle-center) Observation depth map for i-band. (middle-right) S/N map of the residual. (bottom-left) Radial galaxy surface number density profile around candidate ID 8, shown as a function of redshift bins and (bottom-right) the same profile shown as a function of magnitude bins.}
\figsetgrpend

\figsetgrpstart
\figsetgrpnum{12.9}
\figsetgrptitle{Candidate ID 9}
\figsetplot{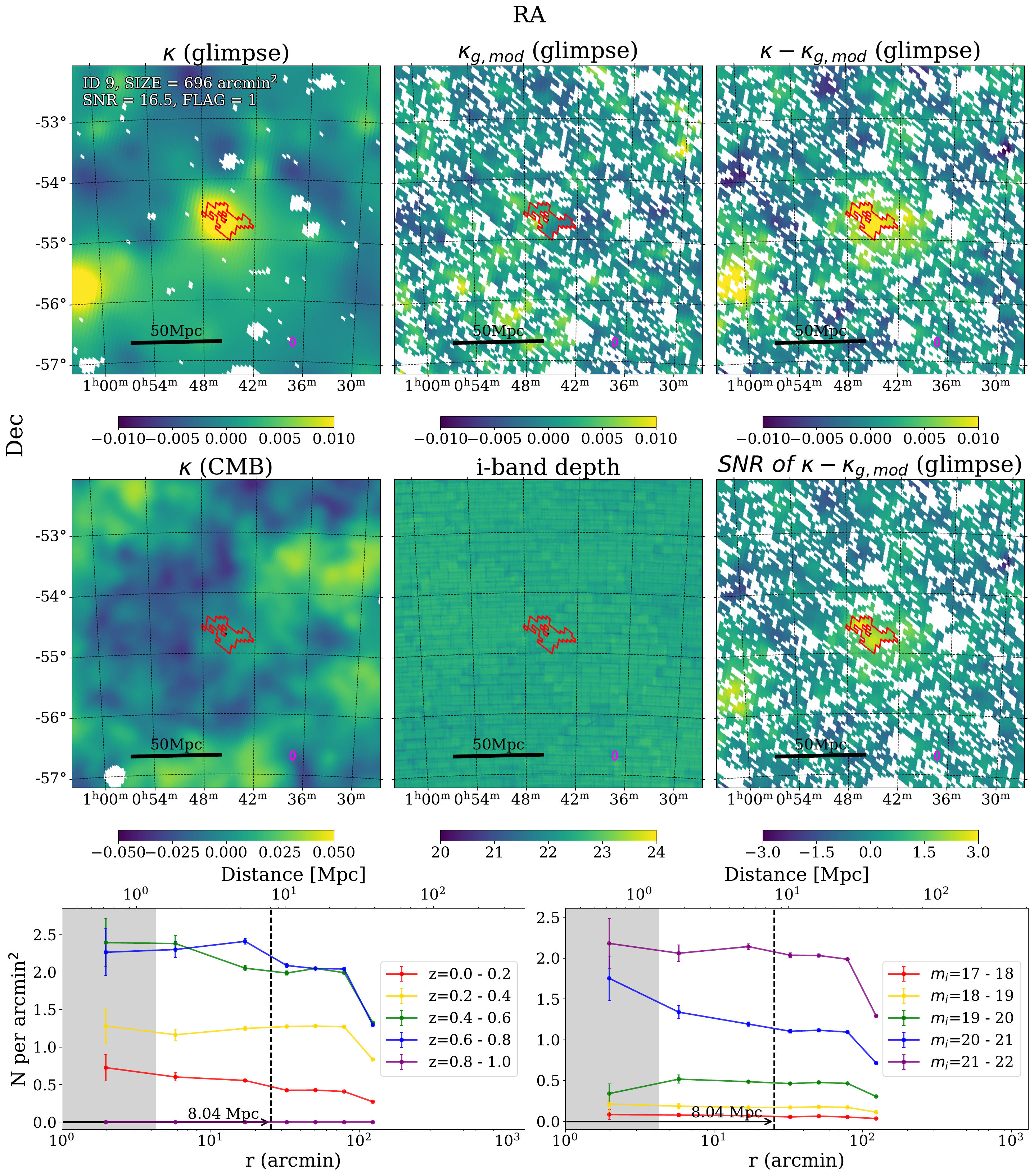}
\figsetgrpnote{The 5$^\circ$$\times$5$^\circ$ cutout maps of the dark structure candidate ID 9. The red lines outline the the boundary of the candidate in each map. The small magenta circle in the bottom-right corner of each panel represents the 10 arcmin smoothing scale, while the small black dot at the center of each panel marks the centroid of the candidate region. (top-left) GLIMPSE convergence map. (top-center) Galaxy convergence map scaled to the GLIMPSE weak lensing convergence. (top-right) Residual map obtained by subtracting the scaled galaxy convergence from the GLIMPSE convergence. (middle-left) CMB lensing convergence map. (middle-center) Observation depth map for i-band. (middle-right) S/N map of the residual. (bottom-left) Radial galaxy surface number density profile around candidate ID 9, shown as a function of redshift bins and (bottom-right) the same profile shown as a function of magnitude bins.}
\figsetgrpend

\figsetgrpstart
\figsetgrpnum{12.10}
\figsetgrptitle{Candidate ID 10}
\figsetplot{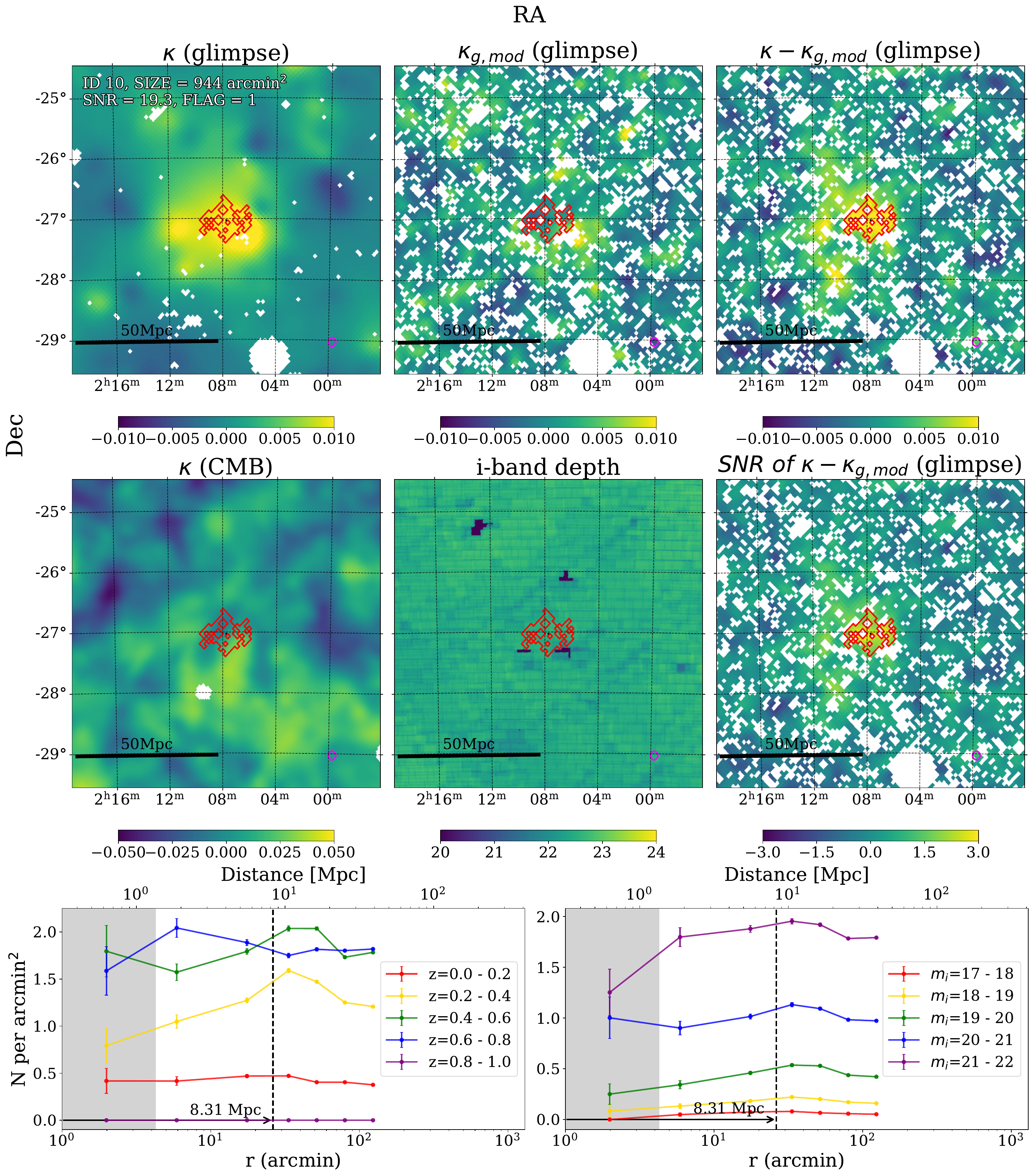}
\figsetgrpnote{The 5$^\circ$$\times$5$^\circ$ cutout maps of the dark structure candidate ID 10. The red lines outline the the boundary of the candidate in each map. The small magenta circle in the bottom-right corner of each panel represents the 10 arcmin smoothing scale, while the small black dot at the center of each panel marks the centroid of the candidate region. (top-left) GLIMPSE convergence map. (top-center) Galaxy convergence map scaled to the GLIMPSE weak lensing convergence. (top-right) Residual map obtained by subtracting the scaled galaxy convergence from the GLIMPSE convergence. (middle-left) CMB lensing convergence map. (middle-center) Observation depth map for i-band. (middle-right) S/N map of the residual. (bottom-left) Radial galaxy surface number density profile around candidate ID 10, shown as a function of redshift bins and (bottom-right) the same profile shown as a function of magnitude bins.}
\figsetgrpend

\figsetgrpstart
\figsetgrpnum{12.11}
\figsetgrptitle{Candidate ID 11}
\figsetplot{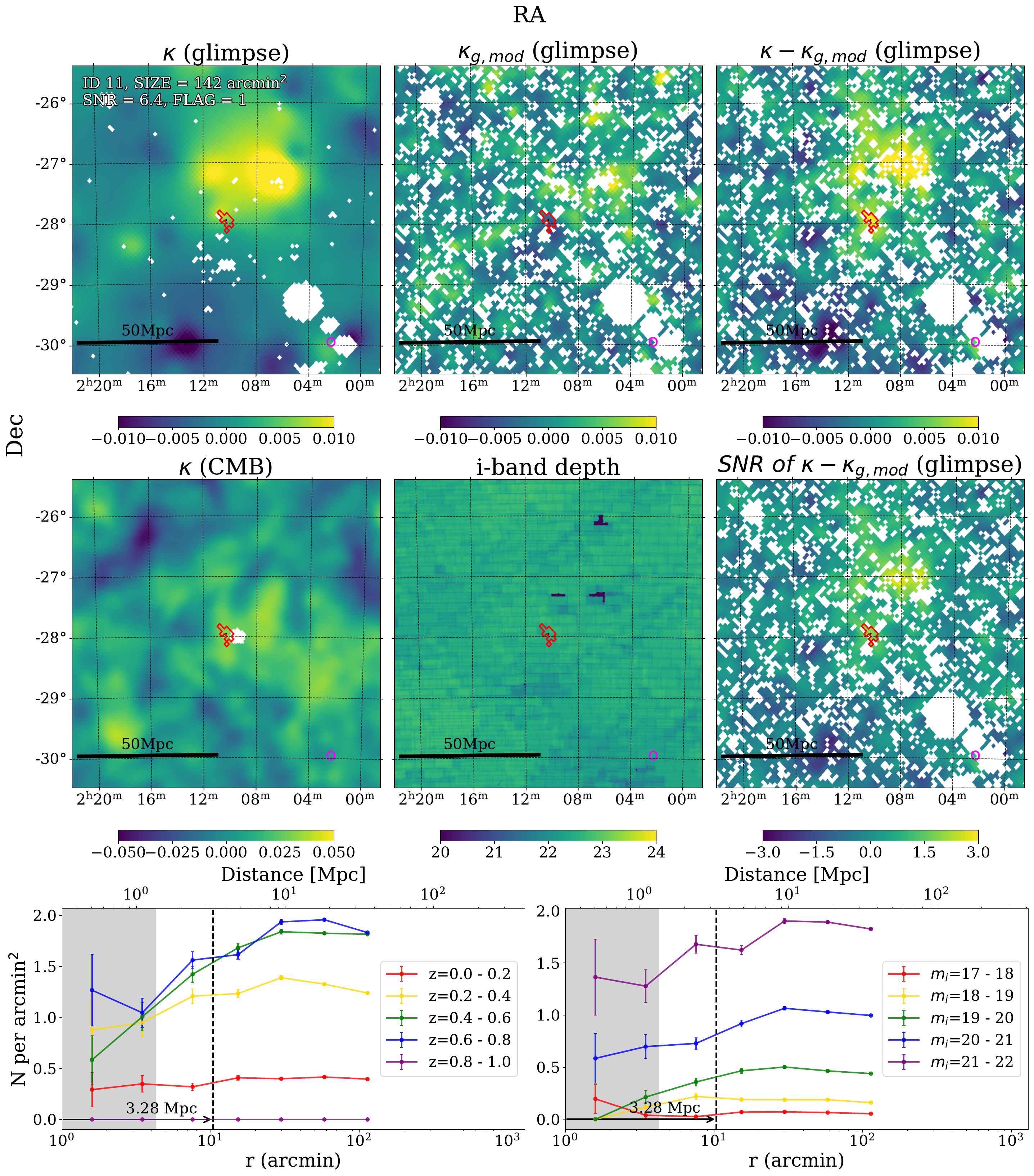}
\figsetgrpnote{The 5$^\circ$$\times$5$^\circ$ cutout maps of the dark structure candidate ID 11. The red lines outline the the boundary of the candidate in each map. The small magenta circle in the bottom-right corner of each panel represents the 10 arcmin smoothing scale, while the small black dot at the center of each panel marks the centroid of the candidate region. (top-left) GLIMPSE convergence map. (top-center) Galaxy convergence map scaled to the GLIMPSE weak lensing convergence. (top-right) Residual map obtained by subtracting the scaled galaxy convergence from the GLIMPSE convergence. (middle-left) CMB lensing convergence map. (middle-center) Observation depth map for i-band. (middle-right) S/N map of the residual. (bottom-left) Radial galaxy surface number density profile around candidate ID 11, shown as a function of redshift bins and (bottom-right) the same profile shown as a function of magnitude bins.}
\figsetgrpend

\figsetgrpstart
\figsetgrpnum{12.12}
\figsetgrptitle{Candidate ID 12}
\figsetplot{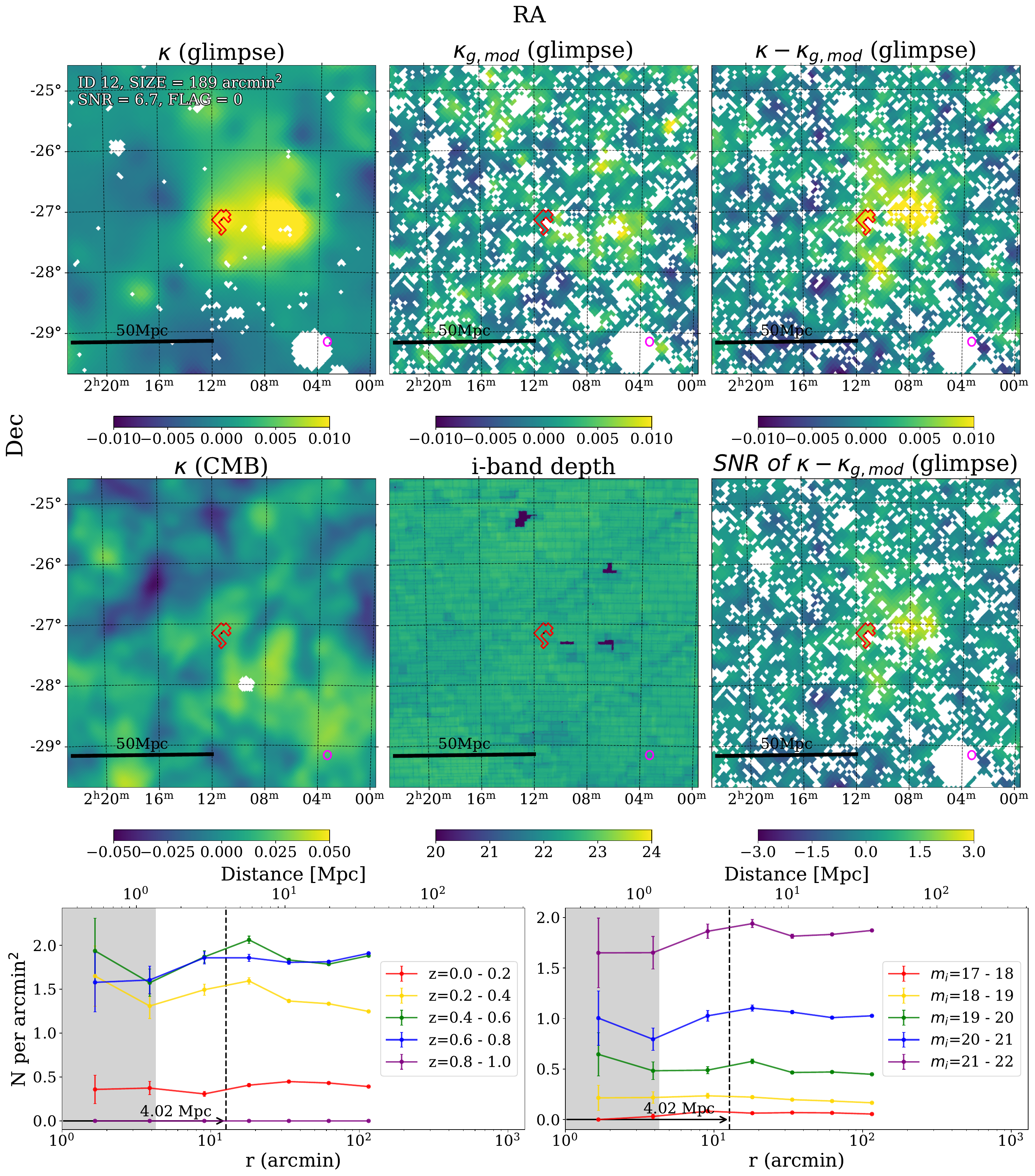}
\figsetgrpnote{The 5$^\circ$$\times$5$^\circ$ cutout maps of the dark structure candidate ID 12. The red lines outline the the boundary of the candidate in each map. The small magenta circle in the bottom-right corner of each panel represents the 10 arcmin smoothing scale, while the small black dot at the center of each panel marks the centroid of the candidate region. (top-left) GLIMPSE convergence map. (top-center) Galaxy convergence map scaled to the GLIMPSE weak lensing convergence. (top-right) Residual map obtained by subtracting the scaled galaxy convergence from the GLIMPSE convergence. (middle-left) CMB lensing convergence map. (middle-center) Observation depth map for i-band. (middle-right) S/N map of the residual. (bottom-left) Radial galaxy surface number density profile around candidate ID 12, shown as a function of redshift bins and (bottom-right) the same profile shown as a function of magnitude bins.}
\figsetgrpend

\figsetgrpstart
\figsetgrpnum{12.13}
\figsetgrptitle{Candidate ID 13}
\figsetplot{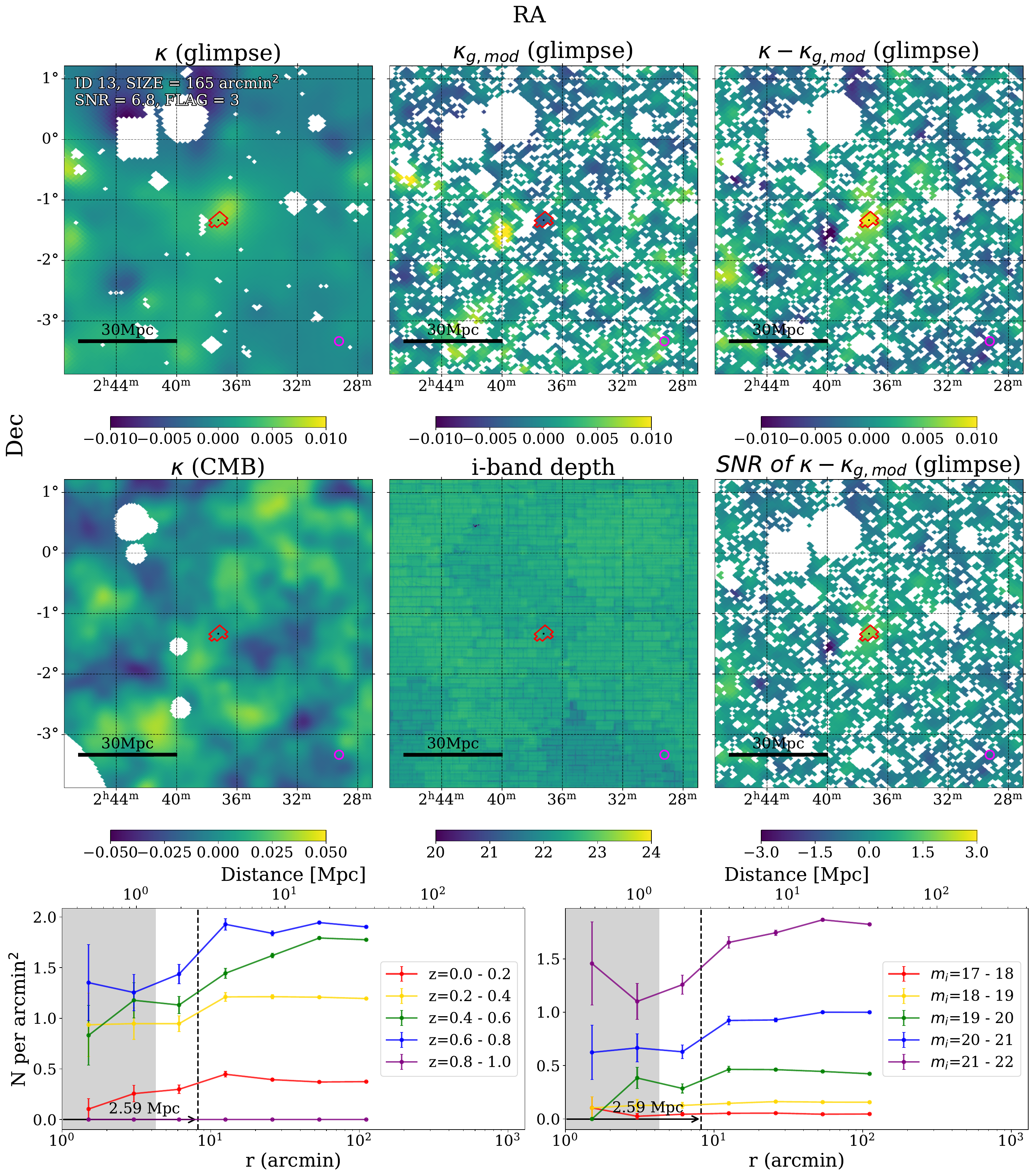}
\figsetgrpnote{The 5$^\circ$$\times$5$^\circ$ cutout maps of the dark structure candidate ID 13. The red lines outline the the boundary of the candidate in each map. The small magenta circle in the bottom-right corner of each panel represents the 10 arcmin smoothing scale, while the small black dot at the center of each panel marks the centroid of the candidate region. (top-left) GLIMPSE convergence map. (top-center) Galaxy convergence map scaled to the GLIMPSE weak lensing convergence. (top-right) Residual map obtained by subtracting the scaled galaxy convergence from the GLIMPSE convergence. (middle-left) CMB lensing convergence map. (middle-center) Observation depth map for i-band. (middle-right) S/N map of the residual. (bottom-left) Radial galaxy surface number density profile around candidate ID 13, shown as a function of redshift bins and (bottom-right) the same profile shown as a function of magnitude bins.}
\figsetgrpend

\figsetgrpstart
\figsetgrpnum{12.14}
\figsetgrptitle{Candidate ID 14}
\figsetplot{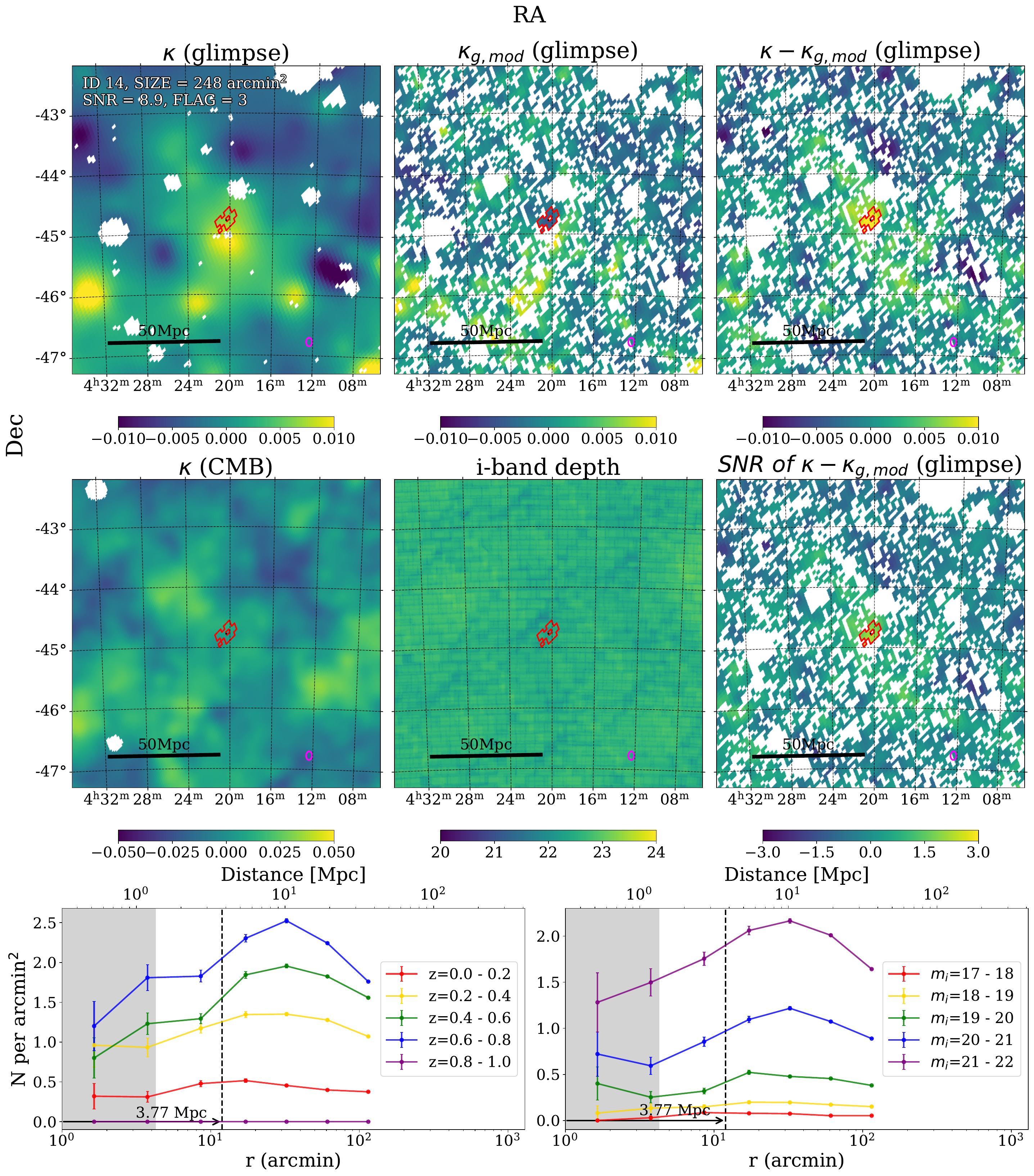}
\figsetgrpnote{The 5$^\circ$$\times$5$^\circ$ cutout maps of the dark structure candidate ID 14. The red lines outline the the boundary of the candidate in each map. The small magenta circle in the bottom-right corner of each panel represents the 10 arcmin smoothing scale, while the small black dot at the center of each panel marks the centroid of the candidate region. (top-left) GLIMPSE convergence map. (top-center) Galaxy convergence map scaled to the GLIMPSE weak lensing convergence. (top-right) Residual map obtained by subtracting the scaled galaxy convergence from the GLIMPSE convergence. (middle-left) CMB lensing convergence map. (middle-center) Observation depth map for i-band. (middle-right) S/N map of the residual. (bottom-left) Radial galaxy surface number density profile around candidate ID 14, shown as a function of redshift bins and (bottom-right) the same profile shown as a function of magnitude bins.}
\figsetgrpend

\figsetgrpstart
\figsetgrpnum{12.15}
\figsetgrptitle{Candidate ID 15}
\figsetplot{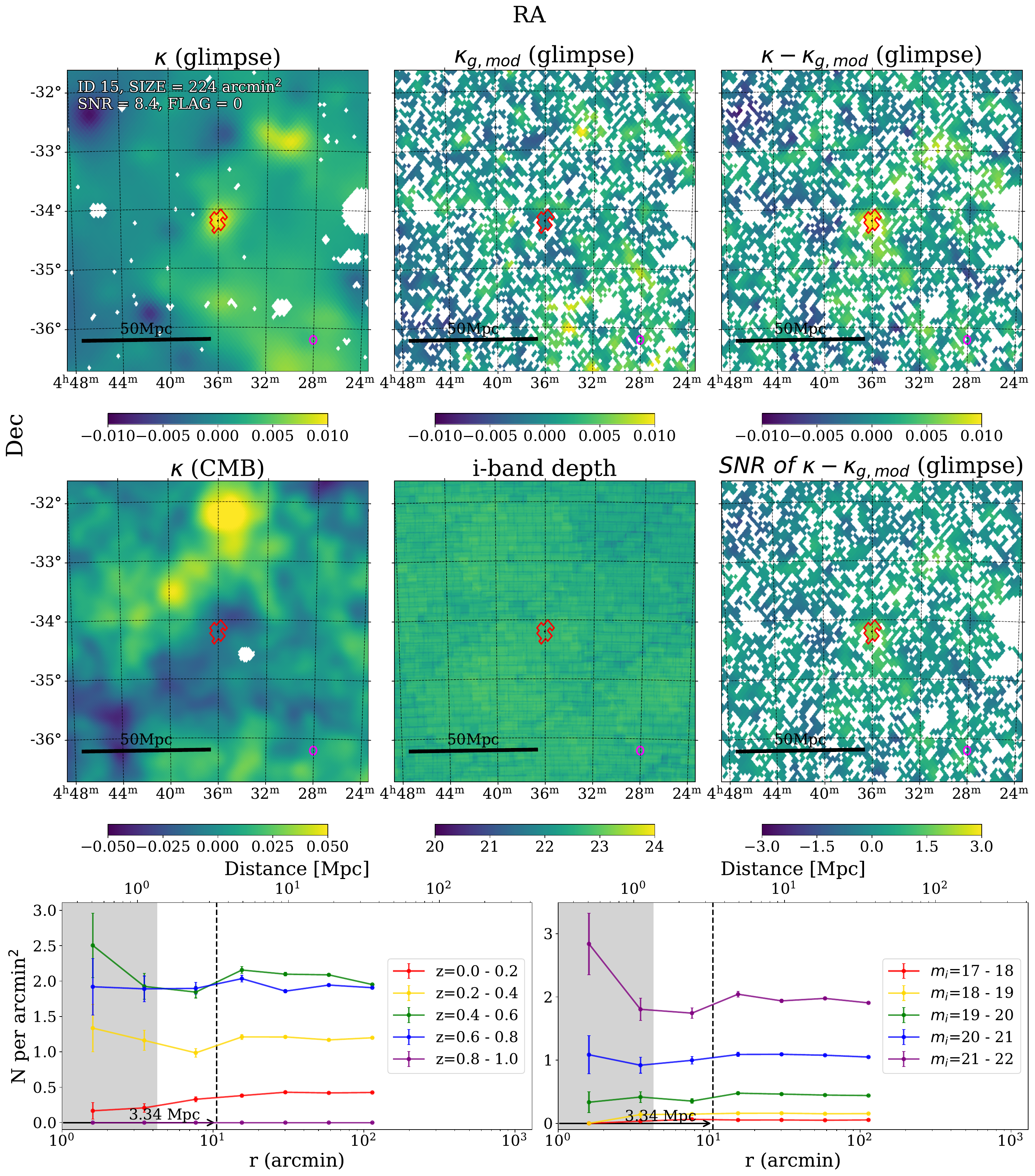}
\figsetgrpnote{The 5$^\circ$$\times$5$^\circ$ cutout maps of the dark structure candidate ID 15. The red lines outline the the boundary of the candidate in each map. The small magenta circle in the bottom-right corner of each panel represents the 10 arcmin smoothing scale, while the small black dot at the center of each panel marks the centroid of the candidate region. (top-left) GLIMPSE convergence map. (top-center) Galaxy convergence map scaled to the GLIMPSE weak lensing convergence. (top-right) Residual map obtained by subtracting the scaled galaxy convergence from the GLIMPSE convergence. (middle-left) CMB lensing convergence map. (middle-center) Observation depth map for i-band. (middle-right) S/N map of the residual. (bottom-left) Radial galaxy surface number density profile around candidate ID 15, shown as a function of redshift bins and (bottom-right) the same profile shown as a function of magnitude bins.}
\figsetgrpend

\figsetgrpstart
\figsetgrpnum{12.16}
\figsetgrptitle{Candidate ID 16}
\figsetplot{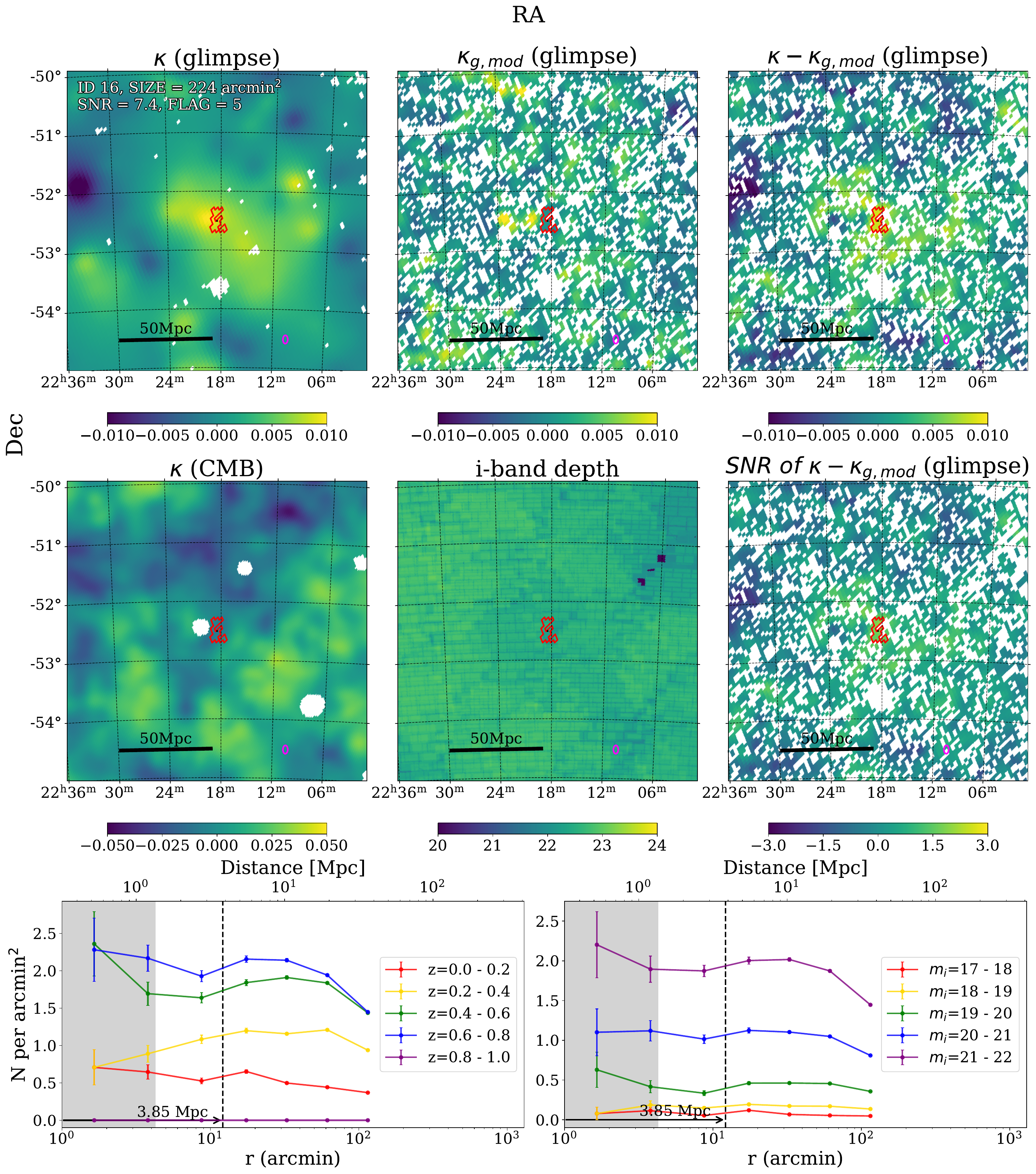}
\figsetgrpnote{The 5$^\circ$$\times$5$^\circ$ cutout maps of the dark structure candidate ID 16. The red lines outline the the boundary of the candidate in each map. The small magenta circle in the bottom-right corner of each panel represents the 10 arcmin smoothing scale, while the small black dot at the center of each panel marks the centroid of the candidate region. (top-left) GLIMPSE convergence map. (top-center) Galaxy convergence map scaled to the GLIMPSE weak lensing convergence. (top-right) Residual map obtained by subtracting the scaled galaxy convergence from the GLIMPSE convergence. (middle-left) CMB lensing convergence map. (middle-center) Observation depth map for i-band. (middle-right) S/N map of the residual. (bottom-left) Radial galaxy surface number density profile around candidate ID 16, shown as a function of redshift bins and (bottom-right) the same profile shown as a function of magnitude bins.}
\figsetgrpend

\figsetgrpstart
\figsetgrpnum{12.17}
\figsetgrptitle{Candidate ID 17}
\figsetplot{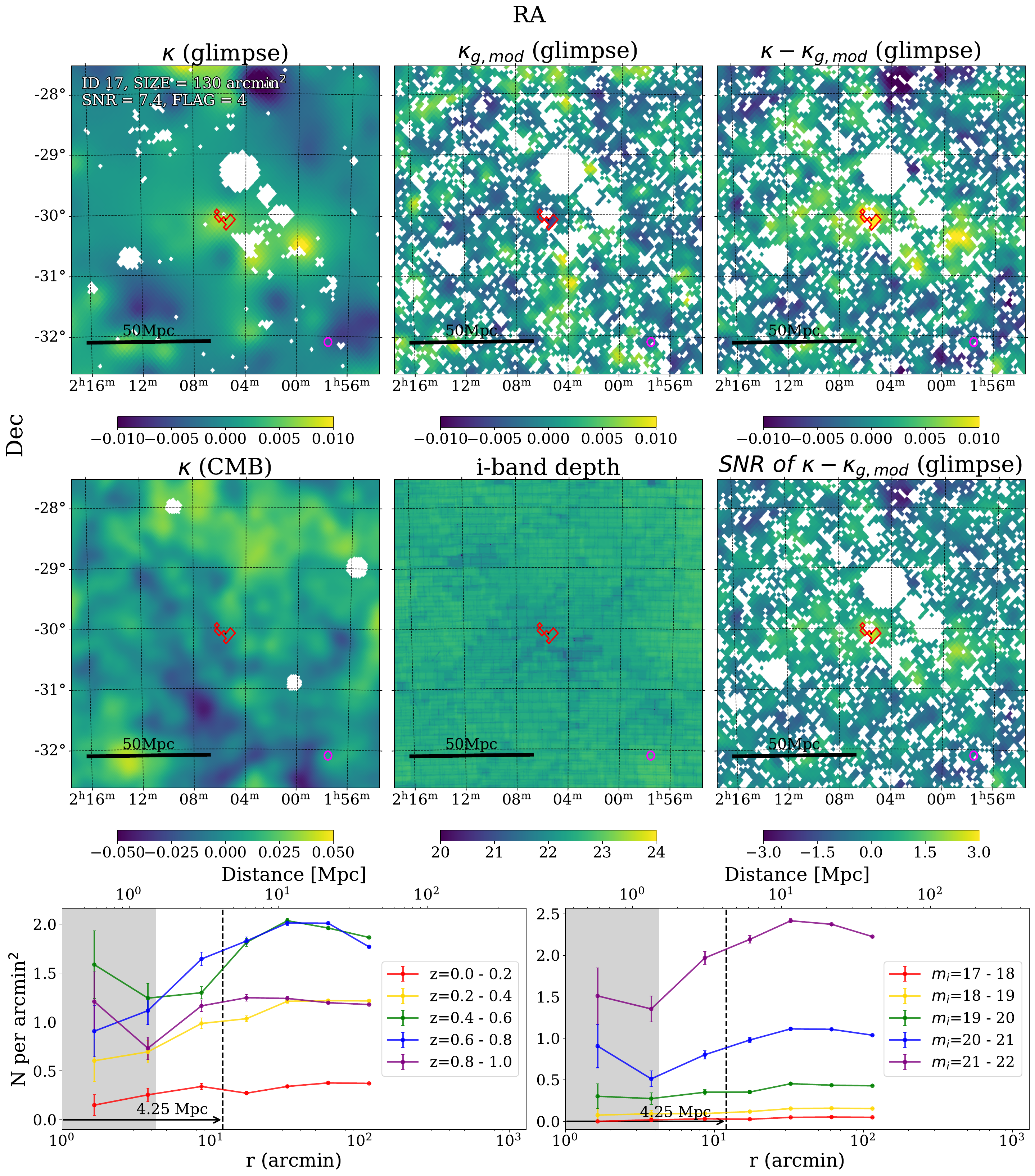}
\figsetgrpnote{The 5$^\circ$$\times$5$^\circ$ cutout maps of the dark structure candidate ID 17. The red lines outline the the boundary of the candidate in each map. The small magenta circle in the bottom-right corner of each panel represents the 10 arcmin smoothing scale, while the small black dot at the center of each panel marks the centroid of the candidate region. (top-left) GLIMPSE convergence map. (top-center) Galaxy convergence map scaled to the GLIMPSE weak lensing convergence. (top-right) Residual map obtained by subtracting the scaled galaxy convergence from the GLIMPSE convergence. (middle-left) CMB lensing convergence map. (middle-center) Observation depth map for i-band. (middle-right) S/N map of the residual. (bottom-left) Radial galaxy surface number density profile around candidate ID 17, shown as a function of redshift bins and (bottom-right) the same profile shown as a function of magnitude bins.}
\figsetgrpend

\figsetgrpstart
\figsetgrpnum{12.18}
\figsetgrptitle{Candidate ID 18}
\figsetplot{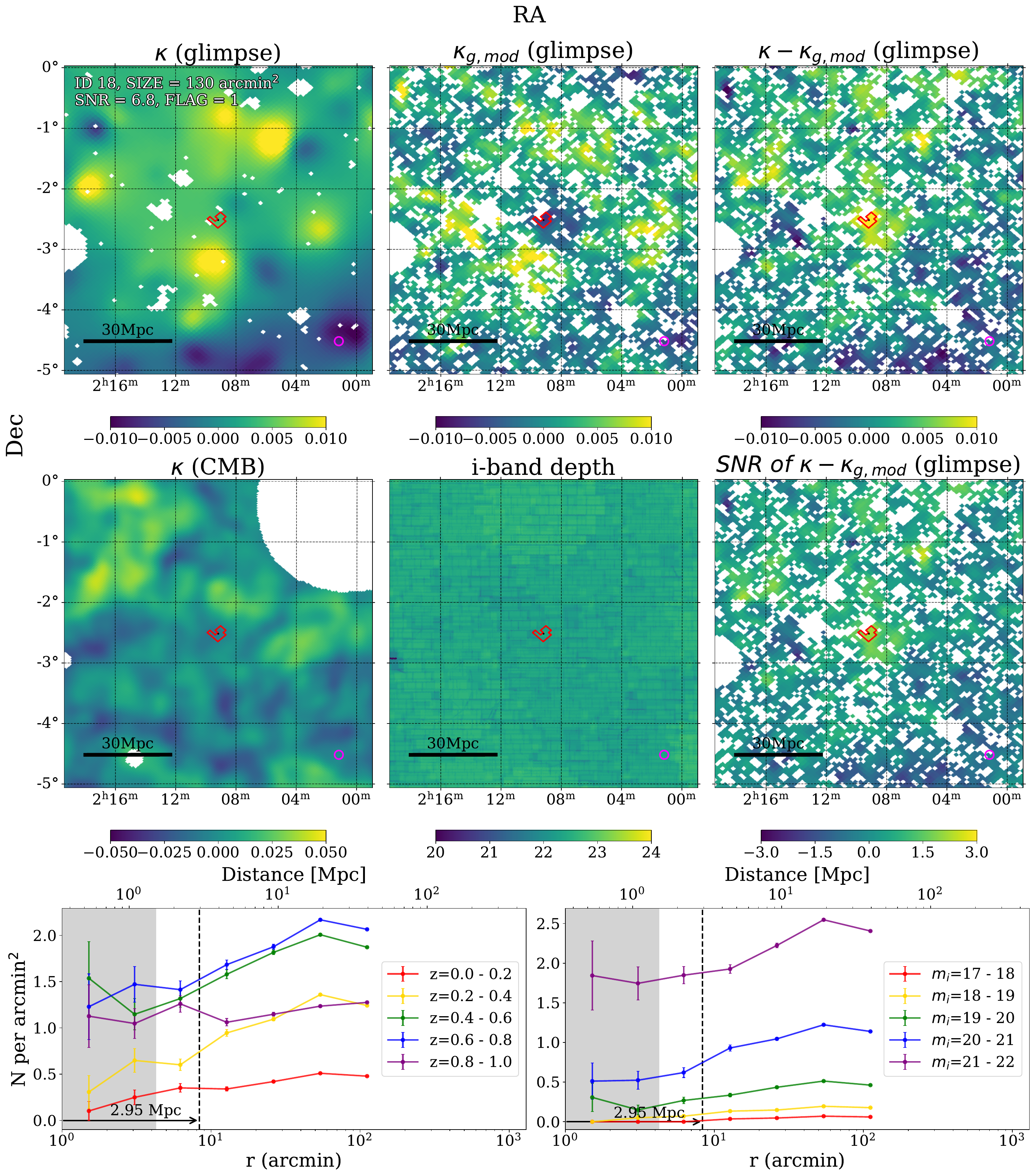}
\figsetgrpnote{The 5$^\circ$$\times$5$^\circ$ cutout maps of the dark structure candidate ID 18. The red lines outline the the boundary of the candidate in each map. The small magenta circle in the bottom-right corner of each panel represents the 10 arcmin smoothing scale, while the small black dot at the center of each panel marks the centroid of the candidate region. (top-left) GLIMPSE convergence map. (top-center) Galaxy convergence map scaled to the GLIMPSE weak lensing convergence. (top-right) Residual map obtained by subtracting the scaled galaxy convergence from the GLIMPSE convergence. (middle-left) CMB lensing convergence map. (middle-center) Observation depth map for i-band. (middle-right) S/N map of the residual. (bottom-left) Radial galaxy surface number density profile around candidate ID 18, shown as a function of redshift bins and (bottom-right) the same profile shown as a function of magnitude bins.}
\figsetgrpend

\figsetgrpstart
\figsetgrpnum{12.19}
\figsetgrptitle{Candidate ID 19}
\figsetplot{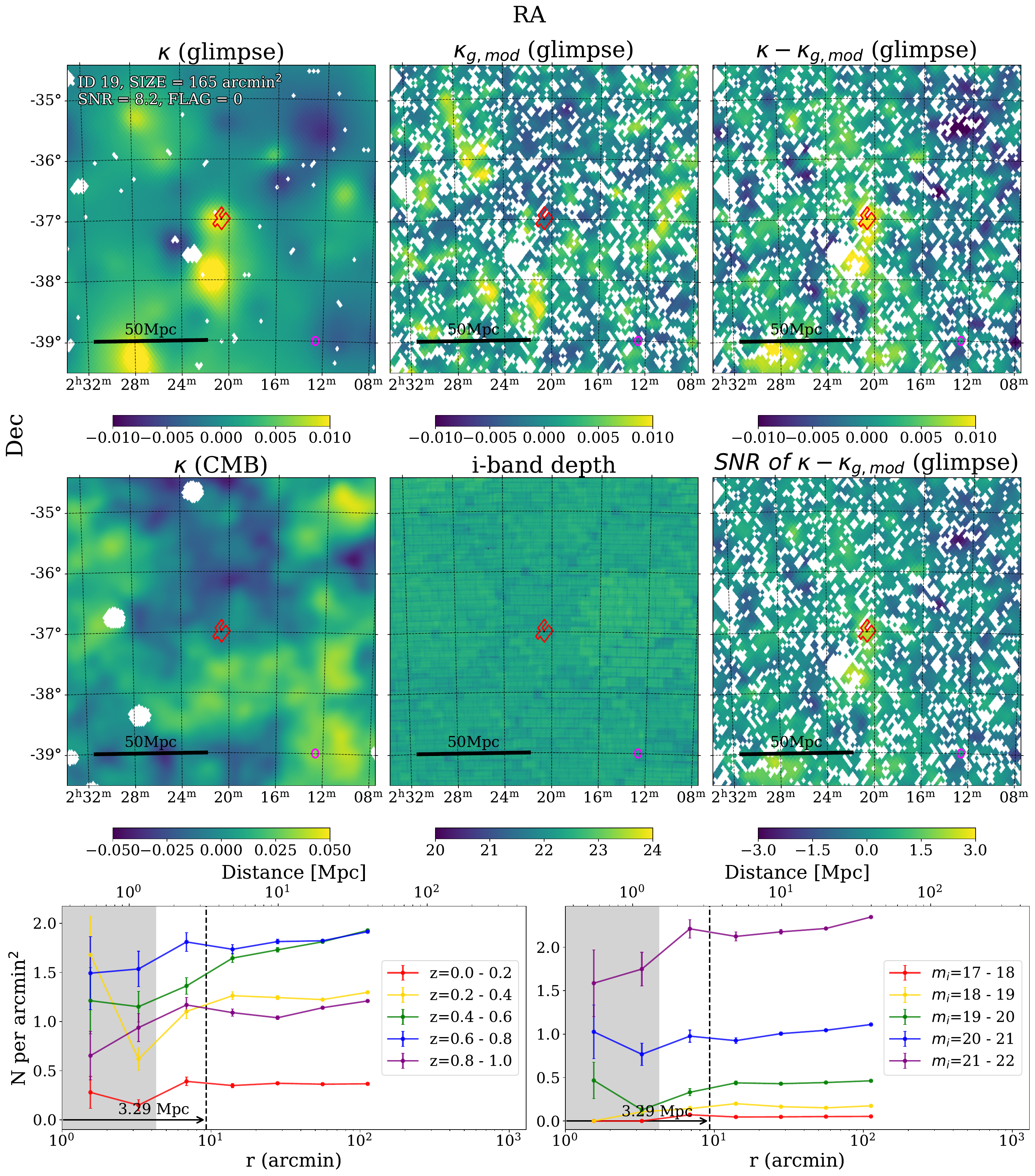}
\figsetgrpnote{The 5$^\circ$$\times$5$^\circ$ cutout maps of the dark structure candidate ID 19. The red lines outline the the boundary of the candidate in each map. The small magenta circle in the bottom-right corner of each panel represents the 10 arcmin smoothing scale, while the small black dot at the center of each panel marks the centroid of the candidate region. (top-left) GLIMPSE convergence map. (top-center) Galaxy convergence map scaled to the GLIMPSE weak lensing convergence. (top-right) Residual map obtained by subtracting the scaled galaxy convergence from the GLIMPSE convergence. (middle-left) CMB lensing convergence map. (middle-center) Observation depth map for i-band. (middle-right) S/N map of the residual. (bottom-left) Radial galaxy surface number density profile around candidate ID 19, shown as a function of redshift bins and (bottom-right) the same profile shown as a function of magnitude bins.}
\figsetgrpend

\figsetgrpstart
\figsetgrpnum{12.20}
\figsetgrptitle{Candidate ID 20}
\figsetplot{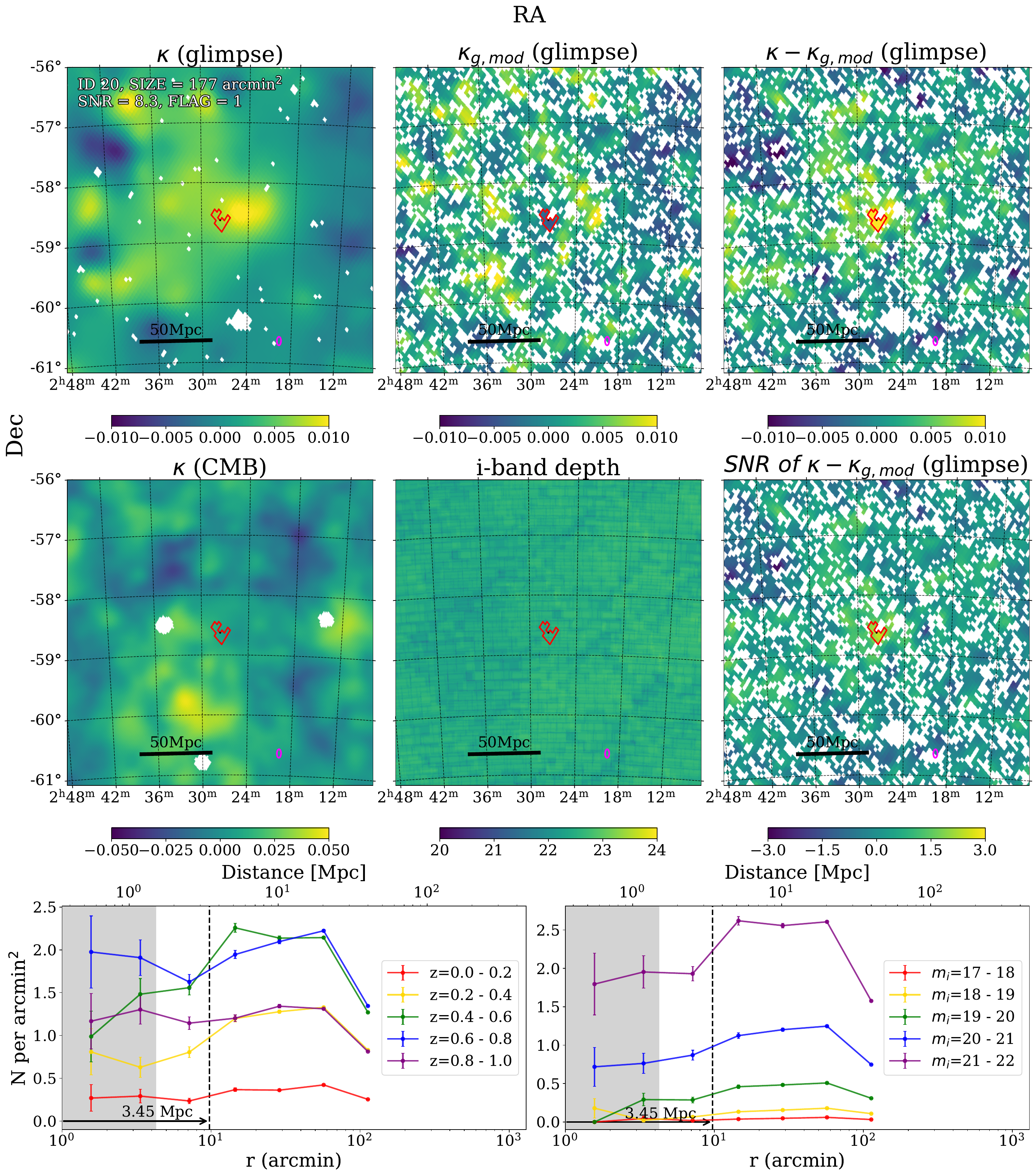}
\figsetgrpnote{The 5$^\circ$$\times$5$^\circ$ cutout maps of the dark structure candidate ID 20. The red lines outline the the boundary of the candidate in each map. The small magenta circle in the bottom-right corner of each panel represents the 10 arcmin smoothing scale, while the small black dot at the center of each panel marks the centroid of the candidate region. (top-left) GLIMPSE convergence map. (top-center) Galaxy convergence map scaled to the GLIMPSE weak lensing convergence. (top-right) Residual map obtained by subtracting the scaled galaxy convergence from the GLIMPSE convergence. (middle-left) CMB lensing convergence map. (middle-center) Observation depth map for i-band. (middle-right) S/N map of the residual. (bottom-left) Radial galaxy surface number density profile around candidate ID 20, shown as a function of redshift bins and (bottom-right) the same profile shown as a function of magnitude bins.}
\figsetgrpend

\figsetgrpstart
\figsetgrpnum{12.21}
\figsetgrptitle{Candidate ID 21}
\figsetplot{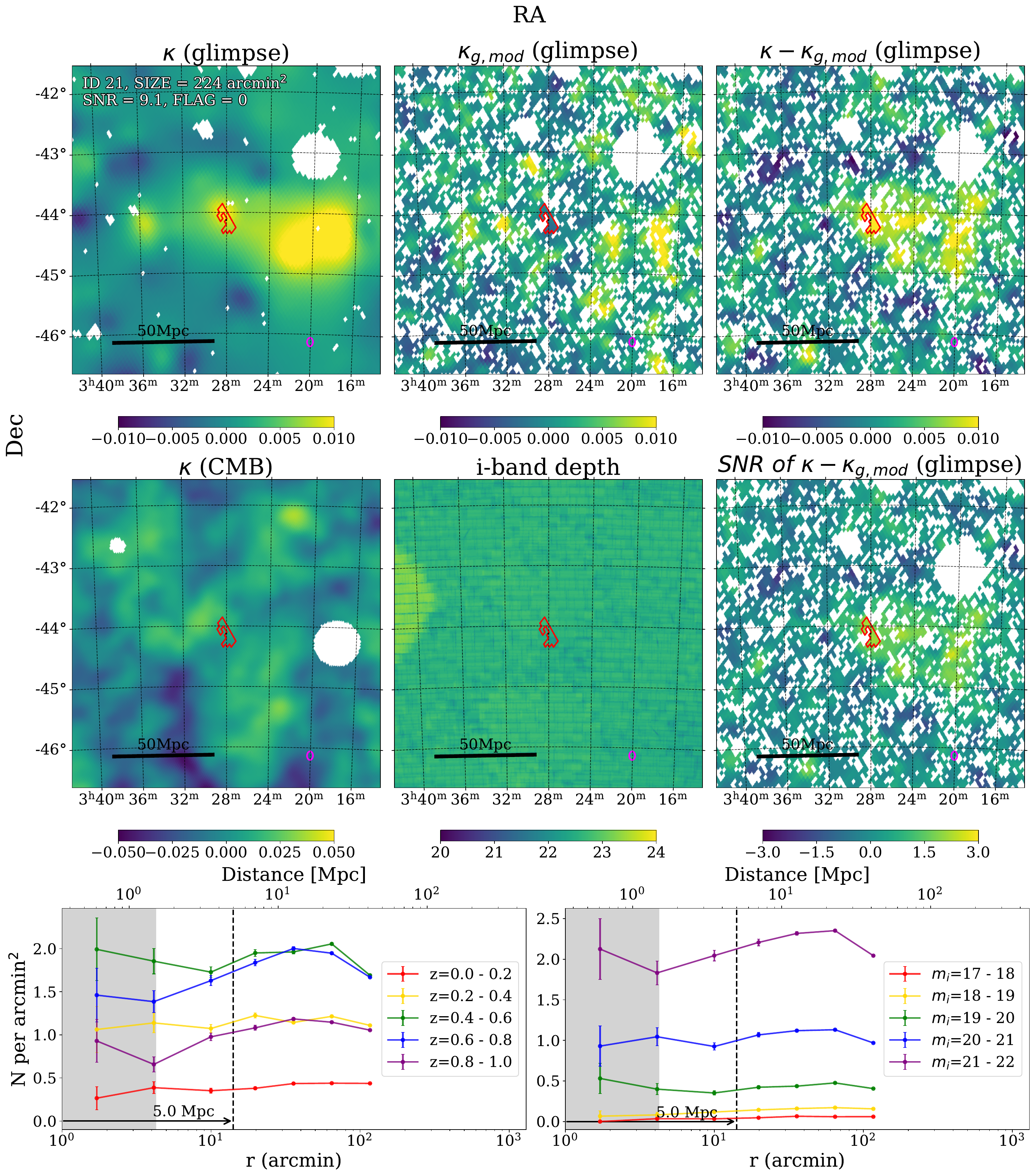}
\figsetgrpnote{The 5$^\circ$$\times$5$^\circ$ cutout maps of the dark structure candidate ID 21. The red lines outline the the boundary of the candidate in each map. The small magenta circle in the bottom-right corner of each panel represents the 10 arcmin smoothing scale, while the small black dot at the center of each panel marks the centroid of the candidate region. (top-left) GLIMPSE convergence map. (top-center) Galaxy convergence map scaled to the GLIMPSE weak lensing convergence. (top-right) Residual map obtained by subtracting the scaled galaxy convergence from the GLIMPSE convergence. (middle-left) CMB lensing convergence map. (middle-center) Observation depth map for i-band. (middle-right) S/N map of the residual. (bottom-left) Radial galaxy surface number density profile around candidate ID 21, shown as a function of redshift bins and (bottom-right) the same profile shown as a function of magnitude bins.}
\figsetgrpend

\figsetgrpstart
\figsetgrpnum{12.22}
\figsetgrptitle{Candidate ID 22}
\figsetplot{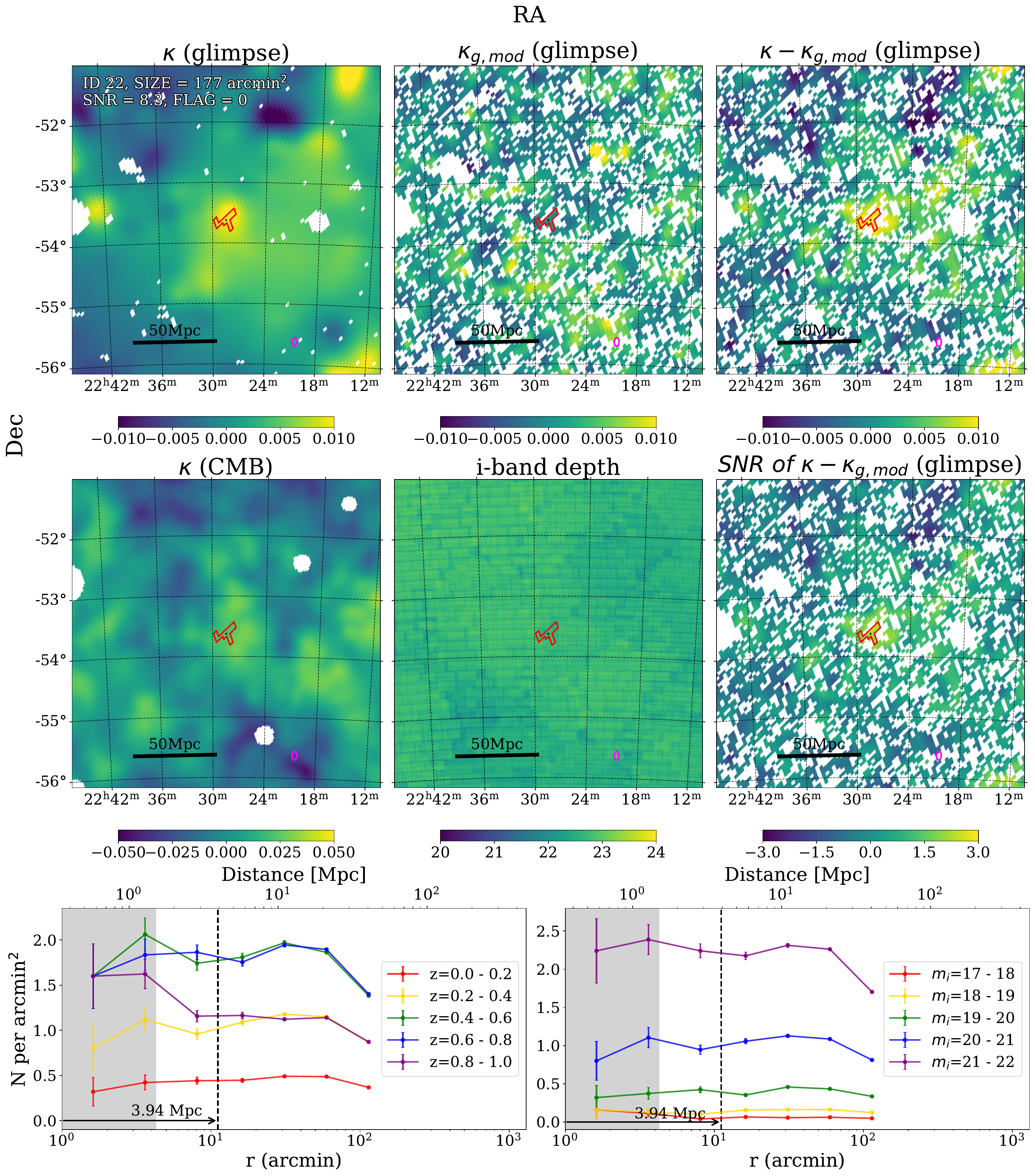}
\figsetgrpnote{The 5$^\circ$$\times$5$^\circ$ cutout maps of the dark structure candidate ID 22. The red lines outline the the boundary of the candidate in each map. The small magenta circle in the bottom-right corner of each panel represents the 10 arcmin smoothing scale, while the small black dot at the center of each panel marks the centroid of the candidate region. (top-left) GLIMPSE convergence map. (top-center) Galaxy convergence map scaled to the GLIMPSE weak lensing convergence. (top-right) Residual map obtained by subtracting the scaled galaxy convergence from the GLIMPSE convergence. (middle-left) CMB lensing convergence map. (middle-center) Observation depth map for i-band. (middle-right) S/N map of the residual. (bottom-left) Radial galaxy surface number density profile around candidate ID 22, shown as a function of redshift bins and (bottom-right) the same profile shown as a function of magnitude bins.}
\figsetgrpend

\figsetend

\bibliography{ms}{}
\bibliographystyle{aasjournal}


\end{document}